\documentclass[singlespace]{ccw_chithesis}

\usepackage{graphics}
\usepackage{natbib}

\def\bfv{{\bf v}}
\def\bfu{{\bf u}}
\def\bfj{{\bf J}}
\def\bfnabla{{\bf \nabla}}
\def\bfzero{{\bf 0}}

\begin{document}

\title{SINGULARITIES, UNIVERSALITY, AND SCALING \\ IN EVAPORATIVE DEPOSITION PATTERNS}
\author{Yuri O. Popov}
\department{Physics}
\division{Physical Sciences}
\degree{Doctor of Philosophy}
\date{December 2003}
\maketitle

\makecopyright

\dedication
\begin{center}
\emph{To Denis}
\end{center}

\acknowledgments{I am greatly indebted to Tom Witten for his advice, everlasting support, and infinite patience.  He has been an advisor, a teacher, and a friend during all these years; it is doubtful this work would be completed without him.  I am deeply grateful.

I am very thankful to Denis who was a source of inspiration during the latest years of this work, and who provided all the needed love and support.

My parents gave me their trust and no-nonsense judgment that helped me to get to this stage; I would like to thank them.

The University of Chicago was a wonderful place to work at during these years, and I would like to express my gratitude to many members of its community who made this experience truly unforgettable.  Todd Dupont, David Grier, and Leo Kadanoff provided invaluable help during various stages of this work; other faculty members of the James Frank Institute contributed to the unique scientific atmosphere I was surrounded by.  Postdocs and graduate students --- Itai Cohen, Robert Deegan, Vladimir Belyi, Ajay Gopinathan, Toan Nguyen, Jung-Ren Huang, Ajay Gopal, Alexey Tkachenko, Joe Plewa, Brian DiDonna, Hiam Diamant, Swan Behrens, Matthias M\"obius, Jennifer Curtis, Adrian Muresan, Dorel Buta, Li-Sheng Tseng, Shunji Egusa, Tao Liang --- helped in a countless number of ways and generally made my experience at the university enjoyable, both academically and otherwise.

My personal friends, both within and outside the university, --- Bruno Carneiro da Cunha, Crist\'{\i}an Garc\'{\i}a, Pedro Amaral, Alexander Karaivanov, Mariana Gatzeva, Andrea Tiseno, Denis Suprun, Leonid Malyshkin, Andrey Tsvetkov, Andrey Tokarev, Denis Shpakov, Pavel Danilov, Boris Baryshnikov, Maria Razumova, Vladislav Gav\-ri\-lets --- made this journey bright and exciting; my life without them would be only a fraction of the life I really had during these years.  Thanks to all of them.

Finally, I would like to thank Nobuko McNeill, Van Bistrow, Rosemary Garrison, and Melva Smith for easing both the formal and the informal parts of the university experience, for their care and indispensable assistance.}

\tableofcontents

\listoffigures

\begin{abstract}
The theory of solute transfer and deposit growth in evaporating sessile drops on a plane substrate is presented.  The main ideas and the principal equations are formulated.  The problem is solved analytically for two important geometries: round drops (drops with circular boundary) and pointed drops (drops with angular boundary).  The surface shape, the evaporation rate, the flow field, and the mass of the solute deposited at the drop perimeter are obtained as functions of the drying time and the drop geometry.  In addition, a model accounting for the spatial extent of the deposit arising from the non-zero volume of the solute particles is solved for round drops.  The geometrical characteristics of the deposition patterns as functions of the drying time, the drop geometry, and the initial concentration of the solute are found analytically for small initial concentrations of solute and numerically for arbitrary initial concentrations of solute.  The universality of the theoretical results is emphasized, and comparison to the experimental data is made.
\end{abstract}

\mainmatter

\chapter{Introduction}

The problem of the so-called ``coffee-drop deposit'' has recently aroused great interest.  The residue left when coffee dries on the countertop is usually darkest and hence most concentrated along the perimeter of the stain.  Ring-like stains, with the solute segregated to the edge of a drying drop, are not particular to coffee.  Mineral rings left on washed glassware, banded deposits of salt on the sidewalk during winter, and enhanced edges in water color paintings are all examples of the variety of physical systems displaying similar behavior and understood by coffee-drop deposit terminology.

Understanding the process of drying of such solutions is important for many scientific and industrial applications, where ability to control the distribution of the solute during drying process is at stake.  For instance, in the paint industry, the pigment should be evenly dispersed after drying, and the segregation effects are highly undesirable.  Also, in the protein crystallography~\cite{pre1, pre2}, attempts are made to assemble the two-dimensional crystals by using evaporation driven convection, and hence solute concentration gradients should be avoided.  On the other hand, in the production of nanowires~\cite{pre3} or in surface patterning~\cite{pre4} perimeter-concentrated deposits may be of advantage.  Recent important applications of this phenomenon related to DNA stretching in a flow have emerged as well~\cite{jpcb2}.  For instance, a high-throughput automatic DNA mapping was suggested~\cite{jpcb1}, where fluid flow induced by evaporation is used for both stretching DNA molecules and depositing them onto a substrate.  Droplet drying is also important in the attempts to create arrays of DNA spots for gene expression analysis.

Ring-like deposit patterns have been studied experimentally by a number of groups.  Difficulties of obtaining a uniform deposit~\cite{pre5}, deformation of sessile drops due to a sol-gel transition of the solute at the contact line~\cite{pre6, pre7}, stick-slip motion of the contact line of colloidal liquids~\cite{pre8, pre9}, and the effect of ring formation on the evaporation of the sessile drops~\cite{pre0} were all reported.  The evaporation of the sessile drops (regardless of solute presence) has also been investigated extensively.  Constancy of the evaporation flux was demonstrated~\cite{jpcb3, jpcb4}, and the change of the geometrical characteristics (contact angle, drop height, contact-line radius) during drying was measured in detail~\cite{jpcb5, jpcb6, jpcb7, jpcb8}. 

The most recent and complete experimental effort to date on coffee-drop deposits was conducted by Robert Deegan {\em et al.}~\cite{deegan1, deegan2, deegan3, deegan4}.  Most experimental data referred to in this work originate from observations and measurements of this group.  They reported extensive results on ring formation and demonstrated that these could be quantitatively accounted for.  The main ideas of the theory of solute transfer in such physical systems have also been developed in their work~\cite{deegan1}.  It was observed that the contact line of a drop of liquid remains pinned during most of the drying process.  While the highest evaporation occurs at the edges, the bulk of the solvent is concentrated closer to the center of the drop.  In order to replenish the liquid removed by evaporation at the edge, a flow from the inner to the outer regions must exist inside the drop.  This flow is capable of transferring all of the solute to the contact line and thus accounts for the strong contact-line concentration of the residue left after complete drying.  The idea of this theory is very robust since it is independent of the nature of the solute and only requires the pinning of the edge during drying (which can occur in a number of possible ways: surface roughness, chemical heterogeneities {\em etc}).  This theory accounts quantitatively for many phenomena observed experimentally; among other things, we will reproduce its main conclusions in this work.

Mathematically, the most complicated part of this problem is related to determining the evaporation rate from the surface of the drop.  An analogy between the diffusive concentration fields and the electrostatic potential fields was suggested~\cite{lebedev, jpcb0}, so that an equivalent electrostatic problem can be solved instead of the problem of evaporation of a sessile drop.  Important analytical solutions to this equivalent electrostatic problem in various geometries were first derived by Lebedev~\cite{lebedev}.  A number of useful consequences from these analytical results were later reported in Ref.~\cite{hu}.

In this work, we discuss the theory of solute transfer and deposit growth in evaporating sessile droplets on a substrate and provide quantitative account for many observed phenomena and measurement results.  Chapter~2 discusses the main ideas and the general theory; all principal equations are derived in that chapter.  While most of its equations have been reported previously, the derivation presented here is original and deals with some mathematical issues never fully addressed before in the context of the current problem.  Chapter~2 is the basis for all the following chapters of this work, and its content is required for all the other chapters.

While the principal equations are fully derived and presented in Chapter~2, their solution depends heavily on the geometry of the drop.  The flow pattern discussed in this work is a type of hydrodynamic flow that is sensitive to the perimeter shape, {\em i.e.}\ the shape of the contact line.  Mathematically, solution to the differential equations depends on the boundary conditions.  Chapters~3 and 4 discuss the analytical solution to this problem in two important geometries.  The two geometries are the drops with circular boundary (round drops) and the drops with angular boundary (pointed drops).  The choice of these two geometries is not accidental.  An arbitrary boundary line can be represented as a sequence of smooth segments, which can be approximated by circular arcs, and fractures, which can be approximated by angular regions.  Thus, knowledge of analytical solution for both circular (Chapter~3) and angular (Chapter~4) boundary shapes fills out the quantitative picture of solute transfer and deposit growth for an arbitrary drop boundary.

The case of the round drops is the most important from the practical point of view and the easiest to deal with mathematically.  This case allows for a full analytical solution, and this solution has been obtained earlier.  Here, in Chapter~3, we reproduce concisely the earlier results and report some new ones.  Its content is a prerequisite to Chapter~5, but is not essential for understanding Chapter~4 (although it is used for drawing some parallels and for comparison of the results in the two geometries).

The case of the pointed drops (Chapter~4), while also important, is much more complicated mathematically than the round-drop solution is.  Presence of the vertex of the angle introduces a singularity at this vertex in addition to the weaker singularity at the contact line.  Singularities govern the solutions to differential equations, and thus presence of the angle and its vertex changes the results substantially.  Also, an angular region, as a mathematical object, is infinite, while a circular region is always bounded.  The real drops with a fracture must always have a third (the furthest from the vertex) side of its contact line, and therefore the overall solution depends on the shape of that furthest part of the boundary.  At the same time, we are interested only in the universal features of the solution that are independent of that furthest part.  Keeping in mind this lust for universality and the mathematical complexity mentioned above, we specify only one boundary of the drop (the vertex and the two sides of the angle) leaving the remainder of the boundary curve unspecified.  Such an approach turns out to be sufficient to determine the universal features of the solution, and it allows us to find all the important singularities as power laws of distance from the vertex of the angle.  Most of the results of Chapter~4 were originally obtained in our earlier works~\cite{popov1, popov2}.  Chapters~4 and 5 are completely independent of each other.

Chapters~2--4 address the issue of the deposit mass accumulation at the drop boundary, however, they treat the solute particles as if they do not occupy any volume, and hence all the solute can be accommodated at the one-dimensional singularity of the contact line.  In reality, the solute deposit accumulated at the perimeter of a drying drop has some thickness, for instance, the shape of the solute residue in a round drop is more like a ring rather than an infinitely thin circumference of the circle.  We attribute this finite volume of the solute deposit simply to the finite size of the solute particles, {\em i.e.}\ we assume the particles do occupy some volume and hence cannot be packed denser than certain mass per unit volume.  A model accounting for the finite size of the deposit and determining its geometric characteristics (height, width) is considered in Chapter~5.  The model is solved for the simplest case of circular geometry, and its results are compared to the zero-volume results of Chapter~3 and to the experimental results of Refs.~\cite{deegan3, deegan4}.  Both the analytical and the numerical solutions are provided, and both compare well with the experimental data (and with each other).  Results of Chapter~5 are presented for the first time.  A chapter of conclusions completes this work.

Before proceeding to the main matter, we would like to point out some features common to all our results and giving rise to the title of this work.  Flows found here are capable of transferring 100\% of the solute to the contact line, and thus account for the strong perimeter concentration of the solute in all cases.  Many quantities scale as power laws; some others follow different functional dependencies.  However, both the exponents of the power laws and any elements of the other functional dependencies turn out to be {\em universal\/} within our model.  They do not depend on any parameters of the system other than the system geometry.  Within the range of applicability of our theory, there are no fitting parameters, no undetermined constants, and no unknown coefficients.  Thus, for instance, exponents of the power laws are as universal as the exponent of distance $-2$ in the Coulomb's law.  We view this universality as one of the main advantages and one of the most exciting features of this theory.

\chapter{Main ideas and general theory of solute transfer and deposit growth in evaporating drops}

In this chapter we will present the general theory of solute transfer to the contact line that will subsequently be solved analytically for two important geometries.

\section{System, geometry, coordinates, and assumptions}

We consider a sessile droplet of solution on a horizontal surface (substrate).  The nature of the solute is not essential for the mechanism.  The typical diameter of the solute particles in Deegan's experiments was of the order of 0.1--1~$\mu$m; we will assume a similar order of magnitude throughout this work.  For smaller particles diffusion becomes more important compared to the hydrodynamic flows of this work.  For larger particles sedimentation may become important when particles exceed certain size.

The droplet is bounded by the contact line in the plane of the substrate.  The (macroscopic) contact line is defined as the common one-dimensional boundary of all three phases (liquid, air and solid substrate).  We will not specify the shape of the contact line in this chapter; in the later chapters we will make this selection and describe the two qualitatively different limits --- circular drops and angular drops.  As we explained in the introduction, these shapes account for the smooth and fractured segments of an arbitrary contact line.  Main equations are not sensitive to the particular geometry, only the boundary conditions are.

We assume that the droplet is sufficiently small so that the surface tension is dominant, and the gravitational effects can be safely neglected.  Mathematically, this is controlled by the Bond number $Bo = \rho g R h_{max} / \sigma$, which accounts for the balance of surface tension and gravitational force on the surface shape.  Here $\rho$ is the density of fluid, $g$ is the gravitational constant, $R$ is a typical size of the drop in the plane of the substrate, $h_{max}$ is the maximal height of the drop at the beginning of the drying process, and $\sigma$ is the surface tension at the liquid-air interface.  For the typical experimental conditions the Bond number is of the order of 0.02--0.05, and thus gravity indeed is unimportant and surface shape is governed mostly by the surface tension.

\begin{figure}
\begin{center}
\includegraphics{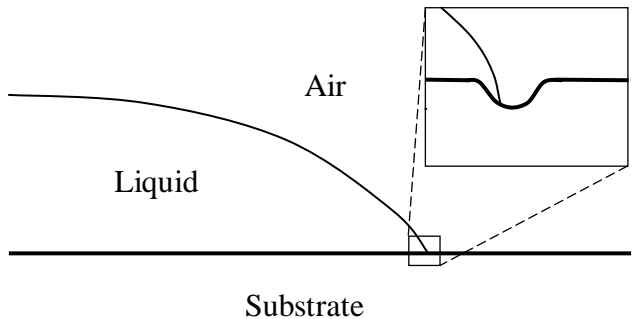}
\caption{Illustration of the possibility of a wide range of contact angles in the presence of a groove or another inhomogeneity.}
\label{contactangleeps}
\end{center}
\end{figure}

At the same time, we do {\em not\/} assume that the contact angle $\theta$ between the liquid-air interface and the plane of the substrate is constant along the contact line on the substrate, nor do we assume it is constant in time.  To achieve a prescribed boundary shape (other than a perfect circle on an ideal plane), the substrate must have scratches, grooves or other inhomogeneities (sufficiently small compared to the dimensions of the droplet), which {\em pin\/} the contact line.  A strongly pinned contact line can sustain a wide range of (macroscopic) contact angles.  The contact angle is not fixed by the interfacial tensions as it is on a uniform surface (Fig.~\ref{contactangleeps}).  Throughout this work we will deal with small contact angles ($\theta \ll 1$) as is almost always the case in the experimental realizations (typically, $\theta_{max} < 0.1$--0.3); however, the general equations of this chapter do not rely on the smallness of the contact angle.

We will use the cylindrical coordinates $(r,\phi,z)$ throughout this work, as most natural for both geometries of interest.  Coordinate $z$ is always normal to the plane of the substrate, and the plane itself is described by $z = 0$, with $z$ being positive on the droplet side of the space.  Coordinates $(r,\phi)$ are the polar radius and the azimuthal angle, respectively, in the plane of the substrate.  The origin is chosen at the center of the circle in the circular geometry and at the vertex of the angle in the angular geometry.  The geometry of the problem is quite complicated despite the visible simplicity.  In both geometries of interest, we consider an object, whose symmetry does not match the symmetry of any simple orthogonal coordinate system of the three-dimensional space.  For instance, solution of the Laplace equation (needed below) requires introduction of the special coordinate systems (the toroidal coordinates and the conical coordinates) with heavy use of various special functions.  Similar difficulties related to the geometry arise in the other parts of the problem as well. 

We describe the surface shape of the drop $h(r,\phi)$ by local mean curvature $K$ that is spatially uniform at any given moment of time, but changes with time as droplet dries.  Ideally, the surface shape should be considered dynamically together with the flow field inside the drop.  However, as we show below, for flow velocities much lower than the characteristic velocity $v^* = \sigma/3\eta$ (where $\sigma$ is surface tension and $\eta$ is dynamic viscosity), which is about 24~m/s for water under normal conditions, one can consider the surface shape independently of the flow and use the equilibrium result at any given moment of time for finding the flow at that time.  Another way of expressing the same condition is to refer to the capillary number $Ca = \eta \tilde v / \sigma$ (where $\tilde v$ is the characteristic value of the flow velocity of the order of 1--10~$\mu$m/s), which is the ratio of viscous to capillary forces.  This ratio is of the order of $10^{-8}$--$10^{-7}$ under typical experimental conditions, clearly demonstrating that capillary forces are by far the dominant ones in this system and that surface shape is practically equilibrium and depends on time adiabatically. 

We consider {\em slow\/} flows, {\em i.e.}\ flows with low Reynolds numbers (also known as ``creeping flows'').  This amounts to the neglect of the inertial terms in the Navier-Stokes equation.  As all the conditions above are, this is well justified by the real experimental conditions.  We also employ the so-called ``lubrication approximation''.  It is essentially based on the two conditions reflecting the thinness of the drop and resulting from the separation of the vertical and horizontal scales.  One is that the pressure inside the drop $p$ does not depend on the coordinate $z$ normal to the substrate:  $\partial_z p = 0$.  The other is related to the small slope of the free surface $|\bfnabla h| \ll 1$, which is equivalent to the dominance of the $z$-derivatives of any component $u_i$ of flow velocity $\bfu$: $\partial_z u_i \gg \partial_s u_i$ (index $s$ refers to the derivatives with respect to any coordinate in the plane of the substrate).  The lubrication approximation is a standard simplifying procedure for this class of hydrodynamic problems~\cite{greenspan, cameron, brenner}.

Before proceeding to the main section of this chapter and formulating the main ideas of the theory, we will make a brief note on evaporation rate.

\section{On evaporation rate}

In order to determine the flow caused by evaporation, one needs to know the flux profile of liquid leaving each point of the surface by evaporation.  This quantity will be seen to be independent of the processes going on inside the drop and must be determined prior to considering any such processes.

\begin{figure}
\begin{center}
\includegraphics{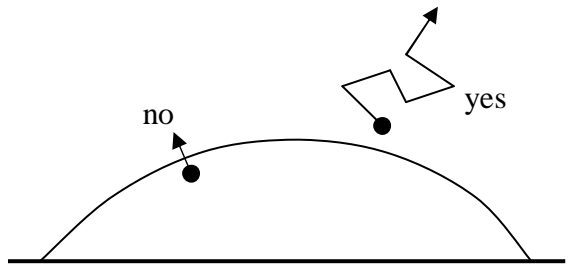}
\caption{The rate-limiting process for evaporative mass loss.  It is the diffusion of saturated vapor just above the interface rather than the transfer across the interface.}
\label{rlp}
\end{center}
\end{figure}

The functional form of the evaporation rate $J(r,\phi)$ (defined as evaporative mass loss per unit surface area per unit time) depends on the rate-limiting step, which can, in principle, be either the transfer rate across the liquid-vapor interface or the diffusive relaxation of the saturated vapor layer immediately above the drop.  We assume everywhere that the rate-limiting step is diffusion of liquid vapor (Fig.~\ref{rlp}) and that evaporation rapidly attains a steady state.  Indeed, the transfer rate across the liquid-vapor interface is characterized by the time scale of the order of $10^{-10}$~s, while the diffusion process has characteristic times of the order of $R^2/D$ (where $D$ is the diffusion constant for vapor in air and $R$ is a characteristic size of the drop), which is of the order of seconds for water drops under typical drying conditions.  Also, the ratio of the time required for the vapor-phase water concentration to adjust to the changes in the droplet shape ($R^2/D$) to the droplet evaporation time $t_f$ is of the order of $(n_s - n_\infty)/\rho \approx 10^{-5}$, where $n_s$ is the density of saturated vapor just above the liquid-air interface and $n_\infty$ is the ambient vapor density~\cite{hu}.  Thus, indeed, vapor concentration adjusts rapidly compared to the time required for water evaporation, and the evaporation process can be considered quasi-steady.

As the rate-limiting process is the diffusion, the density of vapor above the liquid-vapor interface $n$ obeys the diffusion equation.  Since diffusion rapidly attains a steady state, this diffusion equation reduces to the Laplace equation
\begin{equation}
\nabla^2 n = 0.
\end{equation}
This equation is to be solved together with the following boundary conditions dependent on the geometry of the drop: (a) along the surface of the drop the air is saturated with vapor and hence $n$ at the interface is the constant density of saturated vapor $n_s$, (b) far away from the drop the density approaches the constant ambient vapor density $n_\infty$, and (c) vapor cannot penetrate the substrate and hence $\partial_z n = 0$ at the substrate outside of the drop.  Having found density of vapor, one can obtain the evaporation rate $\bfj = - D \nabla n$, where $D$ is the diffusion constant.

\begin{figure}
\begin{center}
\includegraphics{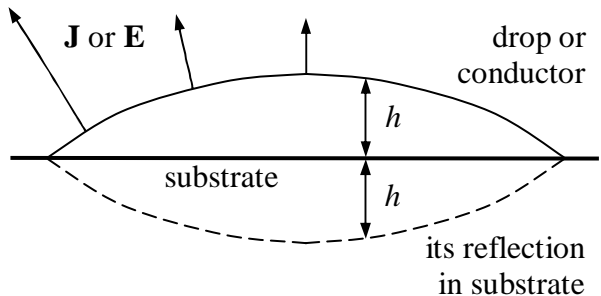}
\caption{Illustration of analogy between evaporation rate {\bf J} for a liquid drop and electric field {\bf E} for a conductor.  Consideration of the drop (or conductor) and its reflection in the plane of the substrate significantly simplifies the boundary problem.}
\label{evaprateeps}
\end{center}
\end{figure}

This boundary problem is mathematically equivalent to that of a charged conductor of the same geometry at constant potential if we identify $n$ with the electrostatic potential and $\bfj$ with the electric field.  Moreover, since there is no component of $\bfj$ normal to the substrate, we can further simplify the boundary problem by considering a conductor of the shape of our drop plus its reflection in the plane of the substrate in the full space instead of viewing only the semi-infinite space bounded by the substrate (Fig.~\ref{evaprateeps}).  This reduces the number of boundary conditions to only two: (a) $n = n_s$ on the surface of the conductor, and (b) $n = n_\infty$ at infinity.  The shape of the conductor (the drop and its reflection in the substrate) is now symmetric with respect to the plane of the substrate.  This plane-symmetric problem of finding the electric field around the conductor at constant potential in infinite space is much simpler than the original problem in the semi-infinite space, and this will be the problem we will actually be solving in order to find the evaporation rate for the two geometries of interest.  Particular solution in each of the geometries will be presented in the next two chapters.

Having formulated physical assumptions intrinsic to the theory, we are now in position to state its main ideas.

\section{The full system of equations for hydrodynamic flow: conservation of mass, Darcy's law, and Young-Laplace equation}

The essential idea behind the theory has been developed in works of Deegan {\em et al.}~\cite{deegan1, deegan2, deegan3}.  It is an experimental observation that the contact line of a drop of liquid is pinned during most of the drying process.  While the highest evaporation occurs at the edges, the bulk of the solvent is concentrated closer to the center of the drop.  In order to replenish the liquid removed by evaporation at the edge, a flow from the inner to the outer regions must exist inside the drop.  This flow is capable of transferring all of the solute to the contact line and thus accounts for the strong contact-line concentration of the residue left after complete drying.  Thus, a pinned contact line entails fluid flow toward that contact line.  The ``elasticity'' of the liquid-air interface fixed at the contact line provides the force driving this flow.

To develop this idea mathematically, we ignore for a moment any solute in the liquid.  Once the flow is found, one can track the motion of the suspended particles, since they are just carried along by the flow.  The purpose of this section is to describe a generic method for finding the hydrodynamic flow of the liquid inside the drop, which is a prerequisite to knowing the details of solute transfer.  

\begin{figure}
\begin{center}
\includegraphics{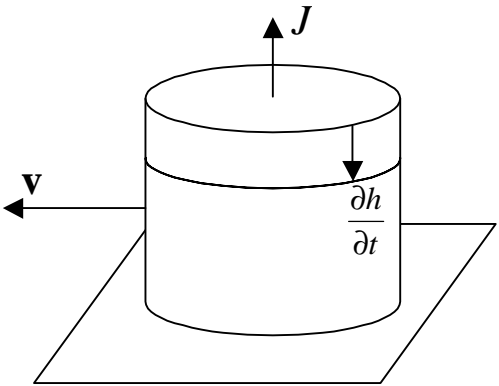}
\caption{Conservation of mass: the liquid-vapor interface lowers exactly by the amount of fluid evaporated from the surface plus the difference between the outflow and the influx of fluid from the adjacent regions.}
\label{consmasseps}
\end{center}
\end{figure}

We define depth-averaged flow velocity by
\begin{equation}
\bfv = \frac 1h \int_0^h \bfu_s \, dz,
\label{defv}\end{equation} 
where $\bfu_s$ is the in-plane component of the local three-dimensional velocity $\bfu$.  Then we write the conservation of fluid mass in the form
\begin{equation}
\bfnabla\cdot(h\bfv) + \frac{J}{\rho}\sqrt{1+(\nabla h)^2} + \partial_t h = 0,
\label{consmass}\end{equation}
where $t$ is the time, $\rho$ is the density of the fluid, and each of the quantities $h$, $J$ and $\bfv$ is a function of $r$, $\phi$ and $t$.  (We will drop the $(\nabla h)^2$ part of the second term everywhere in the following since it is always small compared to unity, as will be seen below.)  This equation represents the fact that the rate of change of the amount of fluid in a volume element (column) above an infinitesimal area on the substrate (third term) is equal to the negative of the sum of the net flux of liquid out of the column (first term) and the amount of mass evaporated from the surface element on top of that column (second term); Fig.~\ref{consmasseps} illustrates this idea.  Thus, this expression relates the depth-averaged velocity field $\bfv (r,\phi,t)$ to the liquid-vapor interface position $h(r,\phi,t)$ and the evaporation rate $J(r,\phi,t)$.  However, this is only one equation for two variables since vector $\bfv$ has generally two components in the plane of the substrate.  Moreover, while the evaporation rate $J$ is indeed independent of flow $\bfv$, the free-surface shape $h$ should in general be determined simultaneously with $\bfv$.  Thus, there are actually three unknowns to be determined together ($h$ and two components of velocity, say, $v_r$ and $v_\phi$), and hence two more equations are needed.  These additional equations will be of the hydrodynamic origin.

We start with the Navier-Stokes equation with inertial terms omitted (low Rey\-nolds numbers):
\begin{equation}
\bfnabla p = \eta \nabla^2 \bfu,
\end{equation}
where $p$ is the fluid pressure, $\eta$ is the dynamic viscosity, and $\bfu$ is the velocity.  Applying lubrication-approximation conditions $\partial_z p = 0$ and $\partial_z u_i \gg \partial_s u_i$, we arrive at the simplified form of this equation
\begin{equation}
\bfnabla_s\,p = \eta\,\partial_{zz} \bfu_s,
\end{equation}
where index $s$ again refers to the vector components along the substrate.  From now on we will suppress the subscript $s$ at the symbol of nabla-operator, and will assume for the rest of this work that this operator refers to the two-dimensional vector operations in the plane of the substrate.  Solution to the above equation with boundary conditions
\begin{equation}
\left. \bfu_s \right|_{z = 0} = \bfzero\qquad\qquad\mbox{and}\qquad\qquad\left. \partial_z \bfu_s \right|_{z = h} = \bfzero
\end{equation}
(no slip at the substrate and no stress at the liquid-air interface) yields
\begin{equation}
\bfu_s = \frac{\bfnabla p}{\eta} \left(\frac{z^2}2 - h z\right),
\label{velocity-profile}\end{equation}
or, after vertical averaging~(\ref{defv}),
\begin{equation}
\bfv = - \frac{h^2}{3\eta} \bfnabla p.
\label{darcy}\end{equation}
This result is a variant of the Darcy's law~\cite{brenner, bensimon}.  Note that since a curl of a gradient is always zero and $\eta$ is a constant, the preceding equation can be re-written as
\begin{equation}
\bfnabla\times\left(\frac{\bfv}{h^2}\right) = \bfzero.
\label{curl}\end{equation}
This condition is analogous to the condition of the potential flow ($\bfnabla\times\bfv = \bfzero$), but with a quite unusual combination of the velocity and the surface height $\bfv/h^2$ in place of the usual velocity $\bfv$.

Relation~(\ref{darcy}) provides the two sought equations in addition to the conservation of mass~(\ref{consmass}).  However, it contains one new variable, pressure $p$, and hence another equation is needed.  This last equation is provided by the condition of the mechanical equilibrium of the liquid-air interface (also known as the Young-Laplace equation) relating the pressure and the surface shape:
\begin{equation}
p = - 2 K \sigma + p_{atm}.
\label{mechequil}\end{equation}
Here $p_{atm}$ is the atmospheric pressure, $\sigma$ is the surface tension, and $K$ is the mean curvature of the surface, uniquely related to the surface shape $h$ by differential geometry.  Note that this expression is independent of both the conservation of mass~(\ref{consmass}) and the Darcy's law~(\ref{darcy}).  Thus, the complete set of equations required to fully determine the four dynamic variables $h$, $p$, $v_r$, and $v_\phi$ consists of four differential equations (together with the appropriate boundary conditions at the contact line, which are dependent on the particular geometry of the drop):  one equation of the conservation of mass~(\ref{consmass}), two equations of the Darcy's law~(\ref{darcy}), and one equation of the mechanical equilibrium of the interface~(\ref{mechequil}).  They provide all the necessary conditions to solve the problem at least in principle.

In practice, however, solution of these four {\em coupled\/} differential equations is not possible in most geometries of practical interest.  At the same time, under normal drying conditions the viscous stress is negligible, or, equivalently, the typical velocities are much smaller than $v^* = \sigma/3\eta \approx 24\mbox{ m}/\mbox{s}$ (for water under normal conditions).  As we show in the Appendix, the four equations {\em decouple\/} under these conditions.  As a result, one can employ the equilibrium result for the surface shape $h$ at any given moment of time, and then determine the pressure and the velocity fields for this fixed functional form of $h$.  Mathematically, the original system of equations can be rewritten as:
\begin{equation}
2 K = - \frac{\Delta p}\sigma,
\label{laplace}\end{equation}
\begin{equation}
\bfnabla\cdot(h^3 \bfnabla\psi) = - \frac{J}{\rho} - \partial_t h,
\label{psi}\end{equation}
\begin{equation}
\bfv = h^2 \bfnabla\psi,
\label{vpsi}\end{equation}
where $\Delta p = p_0 - p_{atm}$, $\psi = - \epsilon p_1 / 3\eta$, and $p_0$ and $p_1$ are the leading and the first-order terms in the expansion of pressure $p = p_0 + \epsilon p_1 + \cdots$ in a small parameter $\epsilon$ inversely proportional to $v^*$ (see the Appendix for details).  Note that $p_0$ is independent of $(r,\phi)$, although it does depend on time (this time dependence will be determined later in this work).  Therefore, there is a profound difference between equations~(\ref{mechequil}) and (\ref{laplace}):  the former is a local statement, with the right-hand side depending on the coordinates of a point within the drop, while the latter is a global condition of spatial constancy of the mean curvature throughout the drop.  Equation~(\ref{laplace}) defines the {\em equilibrium\/} surface shape for any given value of $p_0$ at any given moment of time, and moreover, can be solved independently of the other equations.  Thus, the procedure for finding the solution becomes significantly simplified:  first find the equilibrium surface shape $h$ from condition~(\ref{laplace}) and independently specify the functional form of the evaporation rate $J$ from an equivalent electrostatic problem, then solve equation~(\ref{psi}) for the reduced pressure $\psi$, and finally obtain the flow field $\bfv$ according to prescription~(\ref{vpsi}).  The next two chapters will be devoted to the particular steps of this procedure for the two geometries of interest.

In the next section, we will describe how knowledge of flow inside the drop allows one to find the rate of solute transfer to the contact line and determine the laws of deposit growth.

\section{Solute transfer and deposit growth}

With the velocity field inside the drop in hand, we can compute the rate of the deposit growth at the contact line.  We assume that the suspended particles are carried along by the flow with velocity equal to the fluid velocity.  Integrating the velocity field:
\begin{equation}
\frac{dr}{r \, d\phi} = \frac{v_r(r,\phi)}{v_\phi(r,\phi)},
\label{vel-ratio}\end{equation}
we find the streamline equation $r(\phi)$ or $\phi(r)$, {\em i.e.}\ the trajectory of each particle as it moves with the fluid.  This streamline equation is independent of the overall intensity of evaporation [since both $v_r$ and $v_\phi$ depend on it as a multiplicative factor dropping out of Eq.~(\ref{vel-ratio})], and thus the shape of the streamlines is universal for each geometry of the drop.  Physically, this indicates that solute particles move along the same trajectories independently of how fast evaporation occurs and hence how fast the flow is.

Given the shape of the streamlines, we can compute the time it takes an element of fluid (having started from some initial point $(r_i,\phi_i)$ and moving along a streamline) to reach the contact line.  This time can be found by integrating both sides of either $v_r(r,\phi,t) dt = dr$ or $ v_\phi(r,\phi,t) dt = r \, d\phi$ with known dependences of $v_r$ or $v_\phi$ on the variables and known relation between $r$ and $\phi$ on the streamline.  The integrations are to be conducted from $r_i$ or $\phi_i$ (at time 0) to $r_f$ or $\phi_f$ (at time $t$), respectively, where $(r_f,\phi_f)$ is the terminal endpoint of the trajectory on the contact line.  Thus, the initial location of particles that reach the contact line at time $t$ is characterized by $(r_i,\phi_i)$.  First, only particles initially located near the contact line reach that contact line.  As time goes by, particles initially located further away from the contact line and in the inner parts of the drop reach the contact line.  Finally, particles initially located in the innermost parts of the drop ({\em e.g.}\ at the center of a round drop or near the bisector of an angular drop) reach the contact line as well.  The more time elapsed, the more particles reached the contact line and the larger the area is where they were spread around initially.  One can view this process as inward propagation of the inner boundary of the set of initial locations of the particles that have reached the contact line by time $t$.  As is easy to understand, the velocity of this front is equal to the negative of the vector of fluid velocity at each point (the fluid and the particles move towards the contact line while this front moves away from it, hence a minus sign).

Within the time computed in the preceding paragraph {\em all\/} the solute that lays on the way of an element of fluid as it moves toward the contact line becomes part of the deposit.  Now, we use our knowledge of the initial distribution of the solute, namely, that the solute has constant concentration $c$ everywhere in the drop at time $t = 0$, and compute the mass of the deposit accumulated at the contact line by that time.  The mass of the deposit $dm$ accumulated at the contact-line element of length $dl$ can be found by integrating $h(r,\phi,0)$ over area $dA$ between two infinitesimally close streamlines (terminating $dl$ apart on the contact line) swept by the infinitesimal element of fluid on its way to the contact line and multiplying the result by the initial concentration $c$ of the solute:
\begin{equation}
dm = c \int_{dA} h(r,\phi,0) \, r dr d\phi.
\label{mass-def}\end{equation}
Obviously, this mass will depend of the initial location $(r_i,\phi_i)$ of the element of fluid.  Thus, both the time elapsed from the beginning of the drying process ($t$) and the deposit mass accumulated at the contact line ($dm$) depend on the initial coordinates of the arriving element of fluid (only one of the coordinates is actually an independent variable --- the other is constrained by the streamline equation).  Eliminating these coordinates from the expressions for time and mass, one can finally obtain the deposit mass accumulated at the contact line as a function of time elapsed from the beginning of the drying process.

Thus, the procedure described allows one to find the rate of solute transfer and the laws of deposit growth.  Since we use depth-averaged velocity throughout this work, we implicitly assume that there is no vertical segregation of the solute.

\chapter{Deposit growth for zero-volume particles in circular evaporating drops}

In this chapter we will provide the full solution to the problem of solute transfer in the case of round drops.  In this geometry, the contact line is the circumference of a circle (and the origin of the cylindrical coordinates is located at the center of that circle).  This case is of most importance from the point of view of practical applications and at the same time is the easiest to treat analytically.  Most results can be derived in a closed analytical form, and many of them have been obtained in earlier works~\cite{deegan1, hu}.  Some of them, however, are presented for the first time, and some correct earlier expressions.

\section{Surface shape}

We will follow the procedure explained in great detail in the preceding chapter.  First, one needs to determine the equilibrium shape of the liquid-air interface.  In circular geometry this task is particularly easy.  Let $R$ be the radius of the circular projection of the drop onto the plane of the substrate and $\theta$ be the (macroscopic) contact angle.  Then the solution to Eq.~(\ref{laplace}) with boundary condition $h(R,\phi,t) = 0$ is just a spherical cap.  In cylindrical coordinates function $h(r,\phi,t)$ is independent of $\phi$ and can be written as
\begin{equation}
h(r,t) = \sqrt{\frac{R^2}{\sin^2 \theta} - r^2} - R \cot\theta,
\label{sphericalcap}\end{equation}
where the radius of the footprint of this cap $R$ and the contact angle $\theta$ are related via the right-hand side of Eq.~(\ref{laplace}):
\begin{equation}
R = \frac{2\sigma}{\Delta p} \sin\theta.  
\end{equation}
Both $\theta$ and $\sigma/\Delta p$ change with time during drying process; however, the radius of the footprint $R$ stays constant.  In most experimental realizations $\theta \ll 1$ and the preceding expression takes even simpler form:
\begin{equation}
h(r,t) = \frac{R^2 - r^2}{2R} \theta(t) + O(\theta^3).
\label{h-circular}\end{equation}
Thus, for small $\theta$, quantity $\nabla h$ is indeed small (since $r < R$) and can be safely neglected with respect to unity ({\em i.e.}\ the free surface of the drop is nearly horizontal) as was asserted in the preceding chapter.

Knowledge of the surface shape allows one to find all the necessary geometrical characteristics of the drop, for instance, its volume
\begin{equation}
V = \int_0^R h(r,t) \, 2 \pi r dr = \pi R^3 \bar V(\theta)
\end{equation}
or the total mass of the water (or any other fluid the drop is comprised of)
\begin{equation}
M = \rho V = \pi \rho R^3 \bar V(\theta),
\label{watermass-circular}\end{equation}
where $\rho$ is the density of the water and $\bar V(\theta)$ is a function of the contact angle:
\begin{equation}
\bar V(\theta) = \frac{\cos^3 \theta - 3 \cos\theta + 2}{3 \sin^3 \theta} = \frac\theta{4} + O(\theta^3)
\end{equation}
(the last equality is an expansion in limit $\theta \ll 1$).

\section{Evaporation rate}

The other prerequisite to determining the flow inside the drop is the evaporation rate from the free surface of the drop.  This task involves solution of the equivalent electrostatic problem (the Laplace equation) for the conductor of the shape of the drop plus its reflection in the plane of the substrate (kept at constant potential, as a boundary condition).  In the case of the round drop the shape of this conductor resembles a symmetrical double-convex lens comprised of two spherical caps.  The system of orthogonal coordinates that matches the symmetry of this object (so that one of the coordinate surfaces coincides with the surface of the lens) is called the toroidal coordinates $(\alpha,\beta,\phi)$, where coordinates $\alpha$ and $\beta$ are related to the cylindrical coordinates $r$ and $z$ by
\begin{equation}
r = \frac{R \sinh\alpha}{\cosh\alpha - \cos\beta},\qquad\qquad z = \frac{R \sin\beta}{\cosh\alpha - \cos\beta},
\end{equation}
and the azimuthal angle $\phi$ has the same meaning as in cylindrical coordinates.  Solution to the Laplace equation in toroidal coordinates involves the Legendre functions of fractional degree and was derived in a book by Lebedev~\cite{lebedev}.  The expression for the electrostatic potential or vapor density in toroidal coordinates obtained in that book is independent of the azimuthal angle $\phi$ and reads
$$ n(\alpha,\beta) = n_\infty + (n_s - n_\infty) \sqrt{2(\cosh\alpha - \cos\beta)} \times$$
\begin{equation}
\times \int_0^\infty \frac{\cosh\theta\tau \cosh(2\pi - \beta)\tau}{\cosh\pi\tau \cosh(\pi - \theta)\tau} P_{-1/2 + i\tau}(\cosh\alpha) \, d\tau.
\end{equation}
Here $n_s$ is the density of the saturated vapor just above the liquid-air interface (or the potential of the conductor), $n_\infty$ is the ambient vapor density (or the value of the potential at infinity), and $P_{-1/2 + i\tau}(x)$ are the Legendre functions of the first kind (despite the presence of $i\tau$ in the index, these functions are real valued).  The surface of the lens is described by the two coordinate surfaces $\beta_1 = \pi - \theta$ and $\beta_2 = \pi + \theta$, and the $\beta$ derivative is normal to the surface.  Evaporation rate from the surface of the drop is therefore given by
\begin{equation}
J(\alpha) = D \frac{1}{h_\beta} \left.\partial_\beta n(\alpha,\beta)\right|_{\beta = 2\pi + \beta_1} = D \frac{\cosh\alpha - \cos\beta}R \left.\partial_\beta n(\alpha,\beta)\right|_{\beta = 3\pi - \theta},
\end{equation}
where $D$ is the diffusion constant and $h_\beta = R/(\cosh\alpha - \cos\beta)$ is the metric coefficient in coordinate $\beta$.  [Note that an incorrect expression for $J$ with a plus sign in the metric coefficient was used in Eq.~(A2) of Ref.~\cite{deegan1}.]  Carrying out the differentiation, one can obtain an exact analytical expression for the absolute value of the evaporation rate from the surface of the drop as a function of the polar coordinate $r$:
$$J(r) = \frac{D(n_s - n_\infty)}R \left[\frac{1}2 \sin\theta + \sqrt{2} \left(\cosh\alpha + \cos\theta\right)^{3/2} \times \right.$$
\begin{equation}
\left. \times \int_0^\infty \frac{\cosh\theta\tau}{\cosh\pi\tau} \tanh\left[(\pi - \theta)\tau\right] P_{-1/2 + i\tau}(\cosh\alpha) \, \tau d\tau\right],
\label{j-circular}\end{equation}
where the toroidal coordinate $\alpha$ and the polar coordinate $r$ are uniquely related on the surface of the drop:
\begin{equation}
r = \frac{R \sinh\alpha}{\cosh\alpha + \cos\theta}.
\end{equation}
Thus, expression~(\ref{j-circular}) provides the exact analytical expression for the evaporation rate $J$ as a function of distance $r$ to the center of the drop for an arbitrary contact angle $\theta$.  This expression also corrects an earlier expression of Ref.~\cite{hu} [Eq.~(28)] where a factor of $\sqrt{2}$ in the second term inside the square bracket is missing.

The expression for the evaporation rate is not operable analytically in most cases, as it represents an integral of a non-trivial special function (which, in its turn, is an integral of some simpler elementary functions).  In most cases, we will need to recourse to an asymptotic expansion for small contact angles $\theta$ in order to obtain any meaningful analytical expressions.  However, there is one exception to this general statement (not reported in the literature previously).  An important quantity is the total rate of water mass loss by evaporation $dM/dt$, which sets the time scale for all the processes.  This total rate can be expressed as an integral of the evaporation rate (defined as evaporative mass loss per unit surface area per unit time) over the surface of the drop:
\begin{equation}
\frac{dM}{dt} = - \int_A J(r,\phi) \sqrt{1+(\nabla h)^2} \, r dr d\phi = - \int_0^R J(r) \sqrt{1 + (\partial_r h)^2} \, 2 \pi r dr,
\label{dmdt-circular}\end{equation}
where the first integration is over the substrate area $A$ occupied by the drop.  This expression actually involves triple integration: one in the expression above as an integral of $J(r)$, another in expression for $J(r)$ as an integral of the Legedre function of the first kind, and the third as an integral representation of the Legendre function in terms of elementary functions.  However, one can significantly simplify the above expression and reduce the number of integrations from three to one.  Substituting $\sqrt{1 + (\partial_r h)^2} = 1/\sqrt{1 - (r \sin\theta / R)^2}$ and evaporation rate~(\ref{j-circular}) into Eq.~(\ref{dmdt-circular}) and changing variables and integration order a few times, one can obtain a substantially simpler result:
$$\frac{dM}{dt} = - \pi R D (n_s - n_\infty) \left[\frac{\sin\theta}{1+\cos\theta} + \right.$$
\begin{equation}
\left. + 4 \int_0^\infty \frac{1 + \cosh 2\theta\tau}{\sinh 2\pi\tau} \tanh\left[(\pi - \theta)\tau\right] \, d\tau \right].
\label{dmdtexact}\end{equation}
[Note: Eq.~(2.17.1.10) of Ref.~\cite{prudnikov} was used in this calculation.]  This result together with the time derivative of expression~(\ref{watermass-circular}) for total mass
\begin{equation}
\frac{dM}{dt} = \pi \rho R^3 \bar V'(\theta) \frac{d\theta}{dt} = \frac{\pi \rho R^3}{(1+\cos\theta)^2} \frac{d\theta}{dt}
\end{equation}
provides a direct method for finding the time dependence of $\theta$ for an {\em arbitrary\/} value of the contact angle.  Combining the last two equations, one can obtain a single differential equation for $\theta$ as a function of time $t$:
$$\frac{d\theta}{dt} = - \frac{D (n_s - n_\infty)}{\rho R^2} (1+\cos\theta)^2 \left[\frac{\sin\theta}{1+\cos\theta} + \right.$$
\begin{equation}
\left. + 4 \int_0^\infty \frac{1 + \cosh 2\theta\tau}{\sinh 2\pi\tau} \tanh\left[(\pi - \theta)\tau\right] \, d\tau \right].
\label{dthetadtexact}\end{equation}
Having determined the dependence $\theta(t)$ from this equation, one can obtain the time dependence of any other quantities dependent on the contact angle, for instance, the time dependence of mass from relation~(\ref{watermass-circular}), or any other geometrical quantities of the preceding section.

In practice, however, we will always use the limit of small contact angles in all the analytical calculations of this and the subsequent chapters.  This is the limit of most practical importance, and the usage of this limit will help us keep the physical essence of the problem clear from the unnecessary mathematical elaborations.  While for numerical purposes our exact expressions are absolutely adequate, the analytical calculations in a closed form cannot be conducted any further for an arbitrary contact angle.

Expanding the right-hand side of Eq.~(\ref{dthetadtexact}) in small $\theta$, we immediately obtain that the contact angle decreases {\em linearly\/} with time in the main order of this expansion:
\begin{equation}
\theta = \theta_i \left( 1 - \frac{t}{t_f} \right),
\label{theta-circular}\end{equation}
where we introduced the total drying time $t_f$ defined in terms of the initial contact angle $\theta_i = \theta(0)$:
\begin{equation}
t_f = \frac{\pi \rho R^2 \theta_i}{16 D (n_s - n_\infty)}.
\label{dryingtime}\end{equation}
In the main order, the total rate of water mass loss is constant and water mass also decreases linearly with time:
\begin{equation}
M = \frac{\pi \rho R^3 \theta_i}4 \left( 1 - \frac{t}{t_f} \right).
\label{massofwater}\end{equation}
This linear time dependence during the vast majority of the drying process was directly confirmed in the experiments~\cite{deegan3, deegan4}; see Fig.~\ref{thetatimeeps}.  The dependence of the evaporation rate on radius (linearity in $R$) was also confirmed experimentally and is known to hold true for the case of the diffusion-limited evaporation~\cite{davies}.

\begin{figure}
\begin{center}
\includegraphics{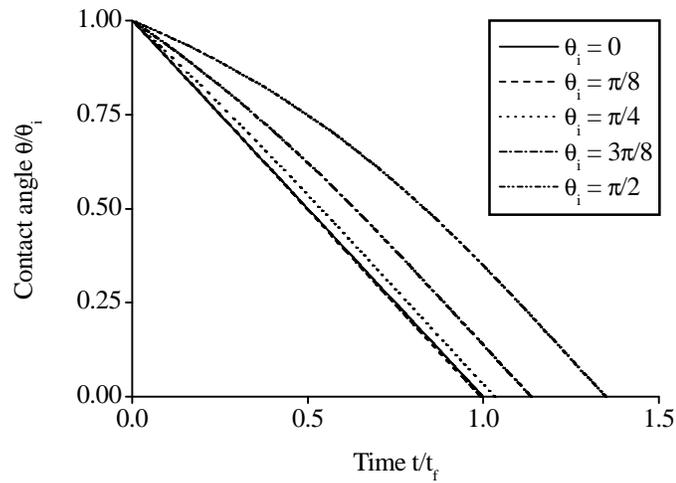}
\caption{Numerical results: Dependence of contact angle $\theta$ on time $t$.  Different curves correspond to different initial contact angles; values of parameter $\theta_i$ are shown at each curve.  The analytical result [Eq.~(\ref{theta-circular})] in limit $\theta_i \to 0$ is also provided (the solid curve).}
\label{theta-teps}
\end{center}
\end{figure}

\begin{figure}
\begin{center}
\includegraphics{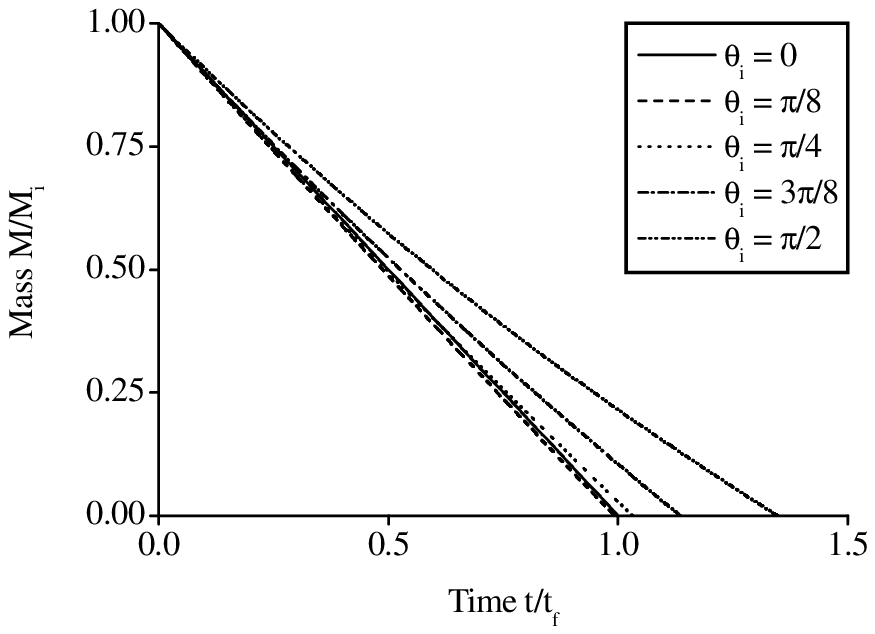}
\caption{Numerical results: Dependence of water mass $M$ on time $t$.  Different curves correspond to different initial contact angles; values of parameter $\theta_i$ are shown at each curve.  The analytical result [Eq.~(\ref{massofwater})] in limit $\theta_i \to 0$ is also provided (the solid curve).}
\label{m-teps}
\end{center}
\end{figure}

\begin{figure}
\begin{center}
\includegraphics{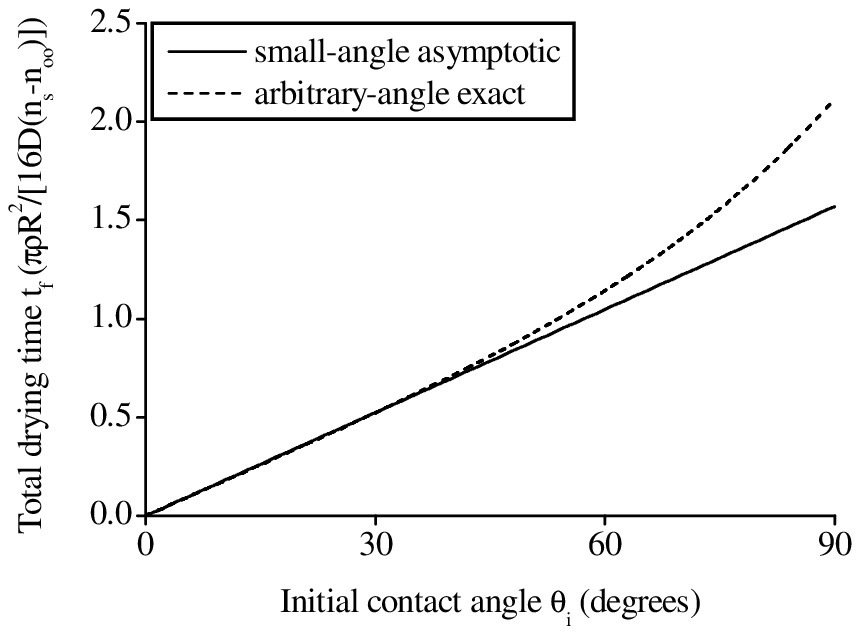}
\caption{Dependence of the total drying time on the initial contact angle.  The solid curve demonstrates the analytical result for small contact angles [Eq.~(\ref{dryingtime})]; the dashed curve represents the exact numerical result [inferred from Eq.~(\ref{dthetadtexact})].}
\label{tf-thetaieps}
\end{center}
\end{figure}

Before we entirely switch to the case of the small contact angles for the rest of this chapter, it is instructive to compare the small-angle analytical asymptotics of the preceding paragraph to the exact arbitrary-angle numerical results based on Eqs.~(\ref{dmdtexact}) and (\ref{dthetadtexact}).  In Figs.~\ref{theta-teps} and \ref{m-teps}, we plot the numerical solutions for $\theta(t)$ and $M(t)$, respectively, for several values of the initial contact angle and compare them to the small-angle asymptotics of Eqs.~(\ref{theta-circular}) and (\ref{massofwater}).  In these figures, $\theta_i$ and $M_i$ are the initial contact angle and the initial mass of water in the drop, respectively.  Note that in these graphs $t_f$ is {\em not\/} the total drying time for each $\theta_i$; instead, it is just the combination of the problem parameters defined in Eq.~(\ref{dryingtime}), which coincides with the total drying time only when $\theta_i \to 0$.  As is clear from these graphs, the total drying time {\em increases\/} with the increasing initial contact angle.  However, it does not increase linearly as prescribed by the asymptotic expression~(\ref{dryingtime}); instead, it grows faster (Fig.~\ref{tf-thetaieps}).  Fig.~\ref{tf-thetaieps} demonstrates that the small-angle approximation works amazingly well up to the angles as large as 45~degrees (this can also be seen in Figs.~\ref{theta-teps} and \ref{m-teps}).  Therefore, working in the limit of small contact angles for the rest of this chapter, we will not loose any precision or generality for the typical experimental values of this parameter.  Lastly, we note that the large-angle corrections may be responsible for the observed non-linearity of the experimentally measured dependence $M(t)$, as is clear from the comparison of Fig.~\ref{m-teps} (theory) and Fig.~\ref{thetatimeeps} (experiment).

Expression for the evaporation rate~(\ref{j-circular}) becomes particularly simple in the limit of small contact angles.  Employing one of the integral representations of the Legendre function in terms of elementary functions [Eq.~(7.4.7) of Ref.~\cite{lebedev}], taking limit $\theta \to 0$ and conducting a number of integrations, it is relatively straightforward to obtain the following result:
\begin{equation}
J(r) = \frac{D (n_s - n_\infty)}R \frac{2}\pi \cosh\frac{\alpha}2\qquad\qquad(\theta \to 0),
\end{equation} 
which, upon identification $\cosh\alpha = (R^2 + r^2)/(R^2 - r^2)$ for $\theta = 0$, can be further reduced to
\begin{equation}
J(r) = \frac{2}\pi \frac{D (n_s - n_\infty)}{\sqrt{R^2 - r^2}}\qquad\qquad(\theta \to 0).
\label{evaprate-circular}\end{equation}
Thus, for thin drops the expression for the evaporation rate reduces to an extremely simple result featuring the square-root divergence near the edge of the drop.  The same result could have been obtained directly if we solved an equivalent electrostatic problem for an infinitely thin disk instead of the double-convex lens.  It is particularly rewarding that after all the laborious calculations the asymptotic of our result is in exact agreement with elementary textbook's predictions [see Ref.~\cite{jackson} for the derivation of the one-over-the-square-root divergence of the electric field near the edge of a conducting plane in the three-dimensional space].  Eq.~(\ref{evaprate-circular}) will be widely used below for all thin drops of circular geometry.

For the sake of completeness, it is also interesting to note the opposite limit of the expression~(\ref{j-circular}), when the surface of the drop is a hemisphere ($\theta = \pi / 2$).  In this limit, a similar calculation can be conducted [also employing Eq.~(7.4.7) of Ref.~\cite{lebedev} and a couple of integrations], and the following simple result can be obtained:
\begin{equation}
J(r) = \frac{D (n_s - n_\infty)}R\qquad\qquad(\theta \to \pi/2).
\end{equation}
Thus, a uniform evaporation rate is recovered.  Again, this result is in perfect agreement with the expectations; the same result could have been obtained if we directly solved the Laplace equation for a sphere (the hemispherical drop and its reflection in the substrate).  The uniform evaporation rate is a result of the full spherical symmetry of such a system.  Similar exact results can also be obtained for a few other discrete values of the contact angle ({\em e.g.}\ for $\theta = \pi/4$).

\section{Flow field}

With $h$ and $J$ in hand, we are in position to find the flow inside the drop according to prescription~(\ref{psi})-(\ref{vpsi}).  In circular geometry this task is particularly easy, since, due to the symmetry, there is no $\phi$ component of the velocity ({\em i.e.}\ the flow is directed radially outwards) and $v_r$ is independent of $\phi$.  Equations~(\ref{psi}) and (\ref{vpsi}) can be combined to yield
\begin{equation}
v_r(r,t) = - \frac{1}{r h} \int _0^r \left( \frac{J}\rho + \partial_t h \right) \, r dr
\label{vdef-circular}\end{equation}
for the radial component of the velocity.  Plugging Eqs.~(\ref{h-circular}), (\ref{theta-circular}), and (\ref{evaprate-circular}) into the above equation, one can obtain the following expression for the flow inside the drop under assumption of small contact angle:
\begin{equation}
v_r(r,t) = \frac{R^2}{4(t_f - t)r} \left( \frac{1}{\sqrt{1 - \left( \frac{r}R \right)^2}} - \left[ 1 - \left( \frac{r}R \right)^2 \right] \right).
\label{v-circular}\end{equation}
This is the final expression for the velocity that we were looking for.

The velocity diverges near the edge of the drop with a one-over-the-square-root singularity in $(R - r)$ at the contact line.  This could have been deduced directly from the conservation of mass~(\ref{consmass}), where the divergent evaporation rate must be compensated by the divergent velocity (since the free-surface height is a regular function of coordinates and, moreover, vanishes near the contact line).  Physically, change of volume near the edge becomes increasingly smaller as the contact line is approached and hence the outgoing vapor flux must be matched by an equally strong incoming flow of liquid.  

In addition, the velocity diverges at the end of the drying time.  Since the amount of liquid removed by evaporation from the immediate vicinity of the contact line of a thin drop remains approximately constant, the amount of incoming fluid must also stay approximately constant over the drying time.  This flux to the region in the vicinity of the contact line is proportional to $v_r h$, and hence velocity must scale as $1/h$ in terms of its time dependence.  Therefore, since the height decreases linearly with time (in the main order in $\theta$), time dependence $1/(t_f - t)$ is to be expected for the velocity.  The result above displays this expected behavior.

\section{Solute transfer and deposit growth}

Shape of the streamlines in the highly symmetric case of circular drops can be predicted without any calculations:  these are the straight lines from the center of the drop to the contact line along the radius.  The parameter characterizing each of these streamlines is simply the polar angle $\phi$ of the corresponding radial direction, and the variable along the streamline is $r$.  The time it takes an element of fluid to reach the contact line can be inferred from the differential equation $v_r(r,t) dt = dr$.  Since dependences on distance and time in $v_r(r,t)$ of Eq.~(\ref{v-circular}) are separable (factorize), the integration of one side of this differential equation from $r_i$ (the initial location of an element of fluid) to $R$ in distance and the integration of the other side from 0 to $t$ in time are straightforward to carry out.  The resulting dependence of the time it takes an element of fluid to reach the contact line on the initial position of this element of fluid can be written as
\begin{equation}
\left( 1 - \frac{t}{t_f} \right)^{3/4} + \left[ 1 - \left(\frac{r_i}{R}\right)^2 \right]^{3/2} = 1.
\label{time-circular}\end{equation}
Clearly, $t = 0$ when $r_i = R$, and $t = t_f$ when $r_i = 0$, as expected from the physical intuition; however, the intermediate behavior is quite non-trivial, and no intuition would be of much help.

The mass of the deposit accumulated at the contact line by time $t$ is given by Eq.~(\ref{mass-def}), which for the case of azimuthal symmetry can be rewritten as
\begin{equation}
m = c \int_{r_i}^R h(r,0) \, 2\pi r dr,
\label{mass-aux}\end{equation}
where $h(r,0)$ refers to the expression for the equilibrium surface shape~(\ref{h-circular}) at the beginning of the drying process (with $\theta = \theta_i$ at $t = 0$).  Simple integration yields
\begin{equation}
m = m_0 \left[ 1 - \left(\frac{r_i}{R}\right)^2 \right]^2,
\label{depositmass-circular}\end{equation}
where $m_0 = \pi c R^3 \theta_i / 4$ is the total mass of the solute present initially in the drop.  When an element of fluid starts from a vicinity of the contact line ($r_i \approx R$), the accumulated deposit mass is still virtually zero when it arrives at the contact line.  When an element of fluid starts from the center of the drop ($r_i \ll R$), virtually all the solute ($m \approx m_0$) is already at the contact line by the time the element of fluid reaches it.

Eliminating $r_i$ from expressions~(\ref{time-circular}) and (\ref{depositmass-circular}), one can finally obtain the dependence of the deposit accumulated at the contact line on drying time $t$:
\begin{equation}
m = m_0 \left[ 1 - \left( 1 - \frac{t}{t_f} \right)^{3/4} \right]^{4/3},
\end{equation}
which can be rewritten in a more symmetric form
\begin{equation}
\left( \frac{m}{m_0} \right)^{3/4} + \left( 1 - \frac{t}{t_f} \right)^{3/4} = 1.
\end{equation}
(We do not have a simple and intuitive explanation for this symmetry.)  No solute ($m = 0$) is at the contact line at the beginning of the drying process ($t = 0$), and all the solute ($m = m_0$) is accumulated at the contact line when the drop has completely dried ($t = t_f$).  At early times, the deposit mass scales with the drying time as a power law with exponent $4/3$:
\begin{equation}
m \approx m_0 \left(\frac{3 t}{4 t_f}\right)^{4/3}\qquad\qquad(t \ll t_f).
\end{equation}
As we will show in the next chapter, this scaling with time is universal for early drying times and should be observed in any geometry of the drop (as long as the contact line is locally smooth).  At the end of the drying process, the rate of the deposit accumulation diverges as $(t_f - t)^{- 1/4}$:
\begin{equation}
m \approx m_0 \left[ 1 - \frac{4}3 \left( 1 - \frac{t}{t_f} \right)^{3/4} \right]\qquad\qquad(|t_f - t| \ll t_f).
\end{equation}
It is this final divergence that is responsible for the experimentally observed 100\% transfer of the solute to the edge.

Thus, in the case of circular geometry, it is relatively simple to obtain the desired scaling of the deposit mass at the contact line with time.  The situation is far more complex in the case of angular geometry.  This case is considered in the following chapter.

\chapter{Deposit growth for zero-volume particles in pointed evaporating drops}

\begin{figure}
\begin{center}
\includegraphics{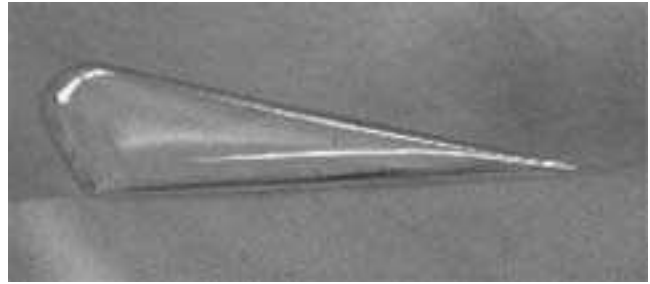}

(a)

\includegraphics{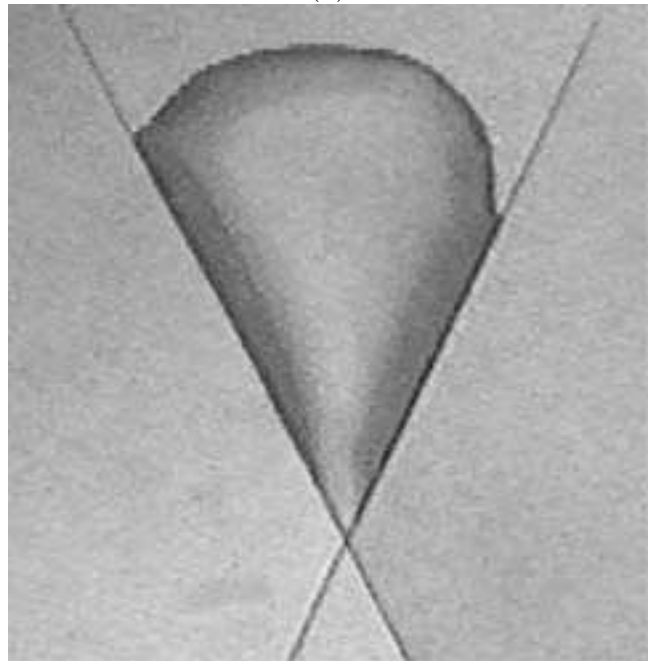}

(b)

\caption{(a) A water droplet with a sector-shaped boundary on the plane substrate (side view).  (b) The same droplet pictured from another point (top view).  Black lines are the grooves on the substrate necessary to ``pin'' the contact line.  (Courtesy Itai Cohen.)}
\label{cohen}
\end{center}
\end{figure}

\begin{figure}
\begin{center}
\includegraphics{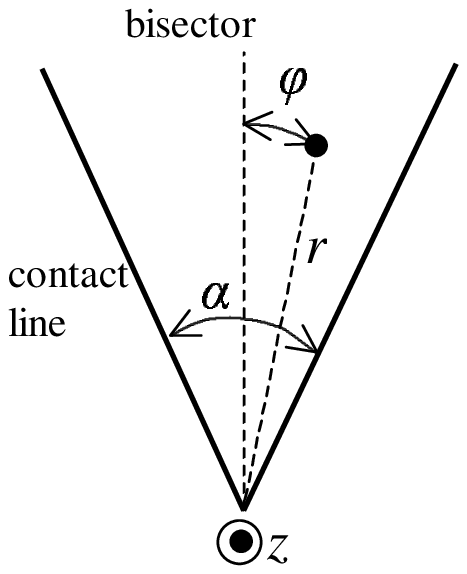}
\caption{Geometry of the problem.  The plane of the figure coincides with the substrate.}
\label{geometry}
\end{center}
\end{figure}

In this chapter we will provide the asymptotic solution to the problem of solute transfer in the case of angular drops.  In this geometry, the droplet is bounded by an angle $\alpha$ in the plane of the substrate (Fig.~\ref{cohen}).  We choose the origin of the cylindrical coordinates at the vertex of the angle, so that the angle occupied by the drop on the substrate is $0<r<\infty$ and $-\alpha/2<\phi<\alpha/2$ (Fig.~\ref{geometry}).  Given the complexity of the geometry, we seek an approximate solution that captures the essential physical features and correct at least asymptotically.  Here, in the geometry of an angular sector, the only possible locations of singularities and divergences (which normally govern properties of the solutions to differential equations) are at the vertex of the angle ({\em i.e.}\ at $r = 0$) and at its sides ({\em i.e.}\ at $\phi = \pm \alpha/2$).  Therefore, the most important physical features will be correctly reflected if asymptotic results as $r \to 0$ and as $\phi \to \pm \alpha/2$ are found.  Thus, we limit our task to determining analytically only the asymptotic power-law scaling for most quantities, and this task proves to be sufficiently challenging by itself.  We will also provide some numerical results that do not rely on these assumptions.

Most of the results presented in this chapter have been previously published in two our earlier papers~\cite{popov1, popov2}, and hence we will often refer to those works for further details.

\section{Surface shape}

\begin{figure}
\begin{center}
\includegraphics{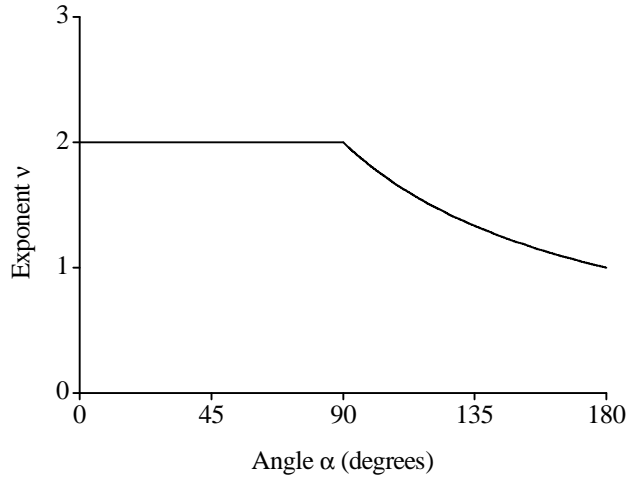}
\caption{Dependence of exponent $\nu$ in the power law $h(r)$ [Eq.~(\ref{surfshap})] on opening angle $\alpha$, after Ref.~\cite{popov1}.}
\label{nueps}
\end{center}
\end{figure}

The boundary problem for the equilibrium surface shape consists of the differential equation~(\ref{laplace}) and the boundary conditions at the vertex and at the sides of the angle:
\begin{equation} h(0,\phi,t) = h(r,-\alpha/2,t) = h(r,\alpha/2,t) = 0.
\label{boundary} \end{equation}
Equation~(\ref{laplace}) represents the fact that the local mean curvature is spatially uniform, but changes with time as the right-hand side ($\Delta p$) changes during the drying process.  The asymptotic solution to the boundary problem~(\ref{laplace}), (\ref{boundary}) was found in our earlier paper~\cite{popov1}.  The result turned out to have two qualitatively different regimes in opening angle $\alpha$ (acute and obtuse angles) and can be written as
\begin{equation}
h(r,\phi,t) = \frac{r^\nu \tilde h(\phi)}{R_0^{\nu-1}}.
\label{surfshap}\end{equation}
Here $R_0(t) = \sigma/\Delta p$ and is the only function of time in this expression; exponent $\nu$ has a discontinuous derivative at $\alpha = \pi/2$ and is shown in Fig.~\ref{nueps}; and
\begin{equation}
\tilde h(\phi) = \left\{\begin{array}{lll} \frac 14 \left(\frac{\cos 2\phi}{\cos\alpha}-1\right) &\qquad\qquad\mbox{if}\quad 0\le\alpha<\frac \pi 2 &\qquad(\nu=2),\\ \\ C \cos\frac{\pi\phi}{\alpha} &\qquad\qquad\mbox{if}\quad \frac \pi 2 <\alpha\le\pi &\qquad(\nu=\pi/\alpha). \end{array}\right.
\label{phi}\end{equation}
The constant $C$ cannot be determined without imposing boundary conditions on $h$ at some curve on the side of the drop furthest from the vertex of the angle.  It is restricted by neither the equation nor the side boundary conditions, and thus, is not a universal feature of the solution near the vertex of the angle.  The constant $C$ can (and does) depend on the opening angle $\alpha$.  As we showed in the earlier paper, this constant must have the following diverging form near $\alpha = \pi/2$:
\begin{equation}
C = \frac 1{4\alpha - 2\pi} + C_0 + O(\alpha - \pi/2)
\label{c-choice}\end{equation}
where $C_0$ is independent of $\alpha$.  We will adopt this form of $C$ (with $C_0$ set to unity) for all numerical estimates for obtuse opening angles.

Two different values of $\nu$ corresponding to the acute and obtuse angles give rise to the two qualitatively different regimes for surface shape.  This difference can best be seen from the fact that the principal curvatures of the surface stay finite as $r \to 0$ for acute angles and diverge as a power of $r$ for obtuse angles.  This qualitative difference can be observed in a simple experimental demonstration, which we provided in our earlier work~\cite{popov1}.  We refer to that earlier work for further details and discussion.  We only note here that the asymptotic $r \to 0$ at the vertex of the angle actually means $r \ll R_0$ (which is typically of the order of a few millimeters for water under normal conditions), and that $\nabla h$ is indeed small for $r \ll R_0$ and can again be safely neglected with respect to unity ({\em i.e.}\ the free surface is nearly horizontal in the vicinity of the tip of the angle) as was asserted earlier.

\section{Evaporation rate}

As we explained in Chapter~2, in order to determine the evaporation rate one needs to solve an equivalent electrostatic problem for a conductor of the shape of the drop and its reflection in the plane of the substrate kept at constant potential.  In geometry of the angular sector this shape resembles a dagger blade, and one has to tackle the problem of finding the electric field around the tip of a dagger blade at constant potential in infinite space.

If one decides to account for the thickness of the blade [given by doubled $h(r,\phi,t)$ of Eq.~(\ref{surfshap})] accurately, it becomes apparent that there is no hope for any analytical solution in this complex geometry.  This is suggested by both the bulkiness of the expressions for a round drop with a non-zero contact angle in the preceding chapter and direct attempts to find the solution.  However, taking into account that near the tip $\nabla h$ is very small and hence the thickness of the blade itself is very small, we can approximate our thick dagger blade with a dagger blade of thickness zero and the same opening angle ({\em i.e.}\ with a flat angular sector).  In the limit $r \to 0$ the contact angle $\theta$ scales with $r$ as $(r/R_0)^{\nu-1}$ and hence goes to zero.  Thus, only the flat blade can be considered up to the main order in $r$.  This approximation would not be adequate for determining the surface shape or the flow field, but it is perfectly adequate for finding the evaporation rate.  We will discuss possible corrections to this result later in this section.

\begin{figure}
\begin{center}
\includegraphics{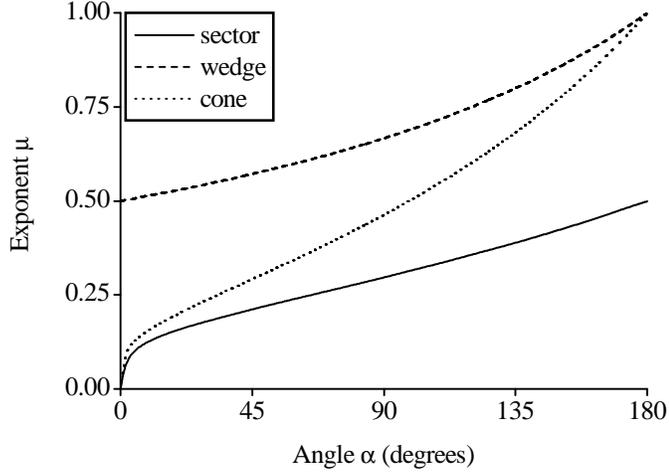}
\caption{Solid line: Dependence of exponent $\mu$ in the power law $J(r)$ [Eqs.~(\ref{j-def}) and (\ref{evaprate})] on opening angle $\alpha$ for an angular sector, after Refs.~\cite{blume1, blume2, desmedt}.  The same dependencies for a wedge (dashed line) and a cone (dotted line) are also shown.}
\label{mueps}
\end{center}
\end{figure}

The problem of finding the electric field and the potential for an infinitely thin angular sector in the three-dimensional space requires introduction of the so-called conical coordinates (the orthogonal coordinates of the elliptic cone) and heavily involves various special functions.  Luckily, it was studied extensively in the past~\cite{kraus, blume1, blume2, desmedt}, although the results cannot be expressed in a closed form.  An important conclusion from these studies is that the $r$ and $\phi$ dependences separate and that the electric field near the vertex of the sector scales with $r$ as a power law with an exponent depending on the opening angle $\alpha$:
\begin{equation}
J(r,\phi) \propto r^{\mu - 1} \tilde J(\phi).
\label{j-def}\end{equation}
Here
\begin{equation}
\tilde J(\phi) \propto \frac 1{|\cos\phi^*|} \left.\frac{\partial Y_\mu(\theta^*,\phi^*)}{\partial\theta^*}\right|_{\theta^* = \pi},
\label{theta-def}\end{equation}
and $\mu$ and $Y_\mu(\theta^*,\phi^*)$ are the eigenvalue and the eigenfunction, respectively, of the eigenvalue problem
\begin{equation}
- {\bf L}^2 Y_\mu(\theta^*,\phi^*) = \mu (\mu + 1) Y_\mu(\theta^*,\phi^*)
\label{eigenvalue}\end{equation}
with Dirichlet boundary conditions on the surface of an elliptic cone (degenerating to an angular sector as $\theta^* \to \pi$).  In the last relation ${\bf L}^2$ is the angular part of the Laplacian in conical coordinates ($r$, $\theta^*$, $\phi^*$).  On the surface of the sector ({\em i.e.}\ at $\theta^* = \pi$) the relation between the conical coordinate $\phi^*$ and the usual polar coordinate $\phi$ is $\sin\phi = \sin(\alpha/2) \sin\phi^*$.  We refer to work~\cite{blume2} for further details.  Here we notice only that neither $\mu$ nor $Y_\mu(\theta^*,\phi^*)$ can be expressed in a closed analytic form; however, the exponent $\mu$ can be computed numerically and is shown in Fig.~\ref{mueps} as a function of $\alpha$.  Note that this exponent is {\em lower\/} than similar exponents for corresponding angles for both a wedge (a two-dimensional corner with an infinite third dimension) and a circular cone, which are also shown in the figure.  Both these shapes (wedge and cone) allow simple analytical solutions~\cite{jackson} but none of them would be appropriate for the zero-thickness sector.

Despite the unavailability of an explicit analytical expression for $\tilde J(\phi)$, its analytic properties at $\phi = 0$ and at $\phi = \pm \alpha/2$ are quite straightforward to infer.  Indeed, $\tilde J(\phi)$ is an even function of $\phi$; therefore, $\tilde J'(0) = 0$ (as well as any other odd derivative on the bisector) and
\begin{equation}
\tilde J(\phi) = \tilde J(0) + O(\phi^2)
\label{theta-bisector}\end{equation}
for small $\phi$.  Obviously, $\tilde J(0)$ is positive.  On the other hand, at $\phi = \pm \alpha/2$ the leading asymptotic of the evaporation rate (or the electric field) is known to be $(\Delta\phi)^{-1/2}$ with exponent $-1/2$ corresponding to the edge of an infinitely thin half plane in the three-dimensional space~\cite{jackson}.  (We have introduced notation $\Delta\phi = \alpha/2 - |\phi|$ in the previous line.)  If one were to correct this asymptotic in order to reflect the non-zero contact angle at the edge of the sector, the asymptotic form at $\phi = \pm \alpha/2$ would have to be written as
\begin{equation}
\tilde J(\phi) \approx J^* (\Delta\phi)^{-\lambda}
\label{theta-edge}\end{equation}
where $J^*$ is a positive constant and
\begin{equation}
\lambda = \frac{\pi-2\theta}{2\pi-2\theta}.
\label{lambda}\end{equation}
This result corresponds to the divergence of the electric field along the edge of a wedge of opening angle $2\theta$ [both the drop and its reflection contribute to the opening angle, hence a factor of 2]~\cite{jackson}.  However, accounting for the non-zero $\theta$ is a first-order correction to the main-order result $\lambda = 1/2$.  This can be seen from the expression for the contact angle:
\begin{equation}
\theta = \arctan\left[\left(\frac{r}{R_0}\right)^{\nu-1} \left|\tilde h'\left(\alpha/2\right)\right|\right] \propto \left(\frac{r}{R_0}\right)^{\nu-1}.
\label{contactangle}\end{equation}
For all opening angles $\nu > 1$ (except $\alpha = \pi$ where $\nu = 1$).  Thus, the correction due to the non-zero contact angle can indeed be neglected in the main-order results, and $\lambda$ should indeed be set to $1/2$.  Nevertheless, we will keep the generic notation $\lambda$ for this exponent in order to keep track of the origin of different parts of the final result and in order to account properly for the case $\alpha = \pi$ in addition to the range of opening angles below $\pi$.  The numerical value of $\lambda$ will be assumed to be $1/2$ in all estimates.  The analytical results below will employ the asymptotics of this paragraph.

Thus, we will use the following expression for the evaporation rate $J$:
\begin{equation}
J(r,\phi) = J_0 \left(\frac{r}{\sqrt{A}}\right)^{\mu-1} \tilde J(\phi),
\label{evaprate}\end{equation}
where function $\tilde J(\phi)$ is defined in Eq.~(\ref{theta-def}) with asymptotics~(\ref{theta-bisector}) and (\ref{theta-edge}).  Here we broke down the constant prefactor into two pieces:  a distance scale $\sqrt{A}$ (where $A$ is the substrate area occupied by the drop) and all the rest $J_0$ (which is of the dimensionality of the evaporation rate).  Trivially, $J_0$ is directly proportional to the product of the diffusion constant and the difference of the saturated and the ambient vapor densities $D (n_s - n_\infty)$, as was the case in the round-drop geometry.

The evaporation rate above does not depend on time and the same form of $J$ applies during the entire drying process, since the diffusion process is steady and the variation of the contact angle with time does not influence the main order result~(\ref{evaprate}).  The same is true for the total rate of mass loss $dM/dt$ since
\begin{equation}
\frac{dM}{dt} = - \int_A J \sqrt{1+(\nabla h)^2} \, r dr d\phi \approx - \int_A J \, r dr d\phi \propto - J_0 A,
\label{dmdt}\end{equation}
where the integrations are over the substrate area occupied by the drop.  We saw the same behavior in circular geometry.  The constancy of this rate during most of the drying process was also confirmed experimentally~\cite{deegan3}.  This fact can be used to determine the time dependence of the length scale $R_0$ of Eq.~(\ref{surfshap}) [and hence of the pressure $p_0$ of Eq.~(\ref{laplace})] explicitly, as the mass $M$ of a sufficiently thin drop is inversely proportional to the mean radius of curvature $R_0$:
\begin{equation}
M \propto \frac{\rho A^2}{R_0},
\label{totalmass}\end{equation}
where we retained only the dimensional quantities and suppressed all the numerical prefactors sensitive to the details of the drop shape.  From the last two equations one can conclude that
\begin{equation}
\frac{d}{dt}\left(\frac 1{R_0}\right) \propto - \frac{J_0}{\rho A}
\end{equation}
and remains constant during most of the drying process.  Hence,
\begin{equation}
R_0(t) = \frac{R_{0i}}{1-t/t_f},
\label{r-t}\end{equation}
where $R_{0i}$ is the initial mean radius of curvature ($R_{0i} = R_0(0)$) and $t_f$ is the total drying time:
\begin{equation}
t_f \propto \frac{\rho A}{J_0 R_{0i}}.
\label{tf}\end{equation}
Thus, at early drying stages ($t \ll t_f$) scale $R_0$ grows linearly with time; this time dependence will be implicitly present in the results below.  However, it is very weak at sufficiently early times and will be occasionally ignored (by setting $R_0 \approx R_{0i}$) when only the main-order results are of interest.

\section{Flow field}

As in the case of circular geometry, with $h$ and $J$ in hand, we proceed by solving Eq.~(\ref{psi}) for the reduced pressure $\psi$.  Assuming power-law divergence of $\psi$ as $r \to 0$ and leaving only the main asymptotic (which effectively means that we neglect the regular term $\partial_t h$ with respect to the divergent one $J/\rho$), we arrive at the following asymptotically-correct expression for $\psi$:
\begin{equation}
\psi(r,\phi,t) = \frac{J_0}\rho \frac{r^{\mu - 3\nu + 1}}{\sqrt{A}^{\mu - 1} R_0^{- 3\nu + 3}} \tilde\psi(\phi),
\label{psi-result}\end{equation}
where time-dependence is implicitly present via $R_0$ and the function $\tilde\psi(\phi)$ is a solution to the following differential equation:
\begin{equation}
\tilde\psi^{\prime\prime}(\phi) + 3 \frac{\tilde h'(\phi)}{\tilde h(\phi)} \tilde\psi'(\phi) - (\mu + 1)(3\nu - \mu - 1) \tilde\psi(\phi) = - \frac{\tilde J(\phi)}{\tilde h^3(\phi)}
\label{psi-equation}\end{equation}
(the combination $(\mu + 1)(3\nu - \mu - 1)$ is positive for all opening angles).  Computing $\bfv$ according to prescription~(\ref{vpsi}), we obtain the depth-averaged flow field
\begin{equation}
\bfv = v_r {\bf \hat r} + v_\phi {\bf \hat\phi}
\label{vresult}\end{equation}
with components
\begin{equation}
v_r(r,\phi,t) = - (3\nu - \mu - 1) \frac{J_0}\rho \frac{r^{\mu - \nu}}{\sqrt{A}^{\mu - 1} R_0^{- \nu + 1}} \tilde h^2(\phi) \tilde\psi(\phi)
\label{vr}\end{equation}
and
\begin{equation}
v_\phi(r,\phi,t) = \frac{J_0}\rho \frac{r^{\mu - \nu}}{\sqrt{A}^{\mu - 1} R_0^{- \nu + 1}} \tilde h^2(\phi) \tilde\psi'(\phi).
\label{vphi}\end{equation}
Thus, one needs to solve Eq.~(\ref{psi-equation}) with respect to $\tilde\psi(\phi)$ in order to know the flow velocity.

We were not able to find an exact analytical solution to this equation; however, we succeeded in finding approximate solutions on the bisector ($|\phi| \ll \alpha/2$) and near the contact line ($\Delta\phi \ll \alpha/2$), which represent the two opposite limits of the range of $\phi$.  (Again, we define $\Delta\phi = \alpha/2 - |\phi|$.)  We describe these two solutions in great detail in our earlier work~\cite{popov2}, and since the solutions themselves are not needed in this abbreviated account, we refer to this earlier work for further details.  Here we will only mention the asymptotics of both solutions (which can actually be inferred directly from the differential equation, without even solving it).  Near the contact line, in limit $\Delta\phi \to 0$, the asymptotic of the solution is
\begin{equation}
\tilde\psi(\phi) \propto \frac{J^* (\Delta\phi)^{-\lambda-1}}{(1 - \lambda^2) \left|\tilde h'\left(\alpha/2\right)\right|^3}.
\label{psiedge}\end{equation}
On the bisector, in the opposite limit $\phi \to 0$, the asymptotic is
\begin{equation}
\tilde\psi(\phi) \propto \tilde\psi(0) + \frac 12 \left[(\mu + 1)(3\nu - \mu - 1)\tilde\psi(0) - \frac{\tilde J(0)}{\tilde h^3(0)}\right] \phi^2.
\label{psibisector}\end{equation}
The reduced pressure on the bisector $\tilde\psi(0)$ is positive.  The value of $\tilde\psi(0)$ cannot be determined from the original differential equation; one needs to employ an integral condition resulting from the equality of the total influx into a sector of radius $r$ by flow from the outer regions of the drop and the total outflux from this sector by evaporation:
\begin{equation}
\rho \int_{-\alpha/2}^{\alpha/2} |v_r| h \, r d\phi = \int_0^r \int_{-\alpha/2}^{\alpha/2} J \, r dr d\phi.
\end{equation}
This condition is similar to an analogous condition for the circular drop (which was written for the entire drop instead of only a certain part of it, since, unlike the infinite sector, the round drop was finite).  Upon simplification this condition reduces to the following equation defining the constant prefactor $\tilde\psi(0)$:
\begin{equation}
\int_0^{\alpha/2} \left[(\mu + 1)(3\nu - \mu - 1) \tilde h^3(\phi) \tilde\psi(\phi) - \tilde J(\phi)\right] \, d\phi = 0.
\label{psi0}\end{equation}
Obviously, $\tilde\psi(0)$ is proportional to $\tilde J(0) \tilde h^{-3}(0)$.

In order to compensate for the unavailability of the exact analytical solution to Eq.~(\ref{psi-equation}) we also approached this problem numerically.  We will not describe the numerical procedure in full here as it was described in great detail in our earlier paper~\cite{popov2}.  Here, we will only mention that we use two different trial forms of $\tilde J(\phi)$ that are simple and at the same time satisfy proper asymptotics~(\ref{theta-bisector}) and (\ref{theta-edge}).  These model forms are neither exact nor the only ones satisfying asymptotics, but they allow one to avoid the numerical solution of the eigenvalue problem~(\ref{eigenvalue}) and thus not to repeat the elaborate treatment of works~\cite{kraus, blume1, blume2, desmedt}.  We use the results based on these trial forms only for illustrative purposes in order to picture general behavior of the solution for arbitrary values of the argument (not only in the limiting cases).  As we discussed in work~\cite{popov2}, the difference between the numerical results based on the two model forms of $\tilde J(\phi)$ did not exceed 10--15\% in most cases, and we have all reasons to believe that at least the orders of magnitude obtained by this approximation are correct.  We would like to emphasize that only the {\em numerical\/} graphs based on the choice of $\tilde J(\phi)$ are affected by these simplified forms; all the {\em analytical\/} results below do not rely on a particular form of $\tilde J(\phi)$ and use only the analytical asymptotics of the preceding section.

\begin{figure}
\begin{center}

\includegraphics{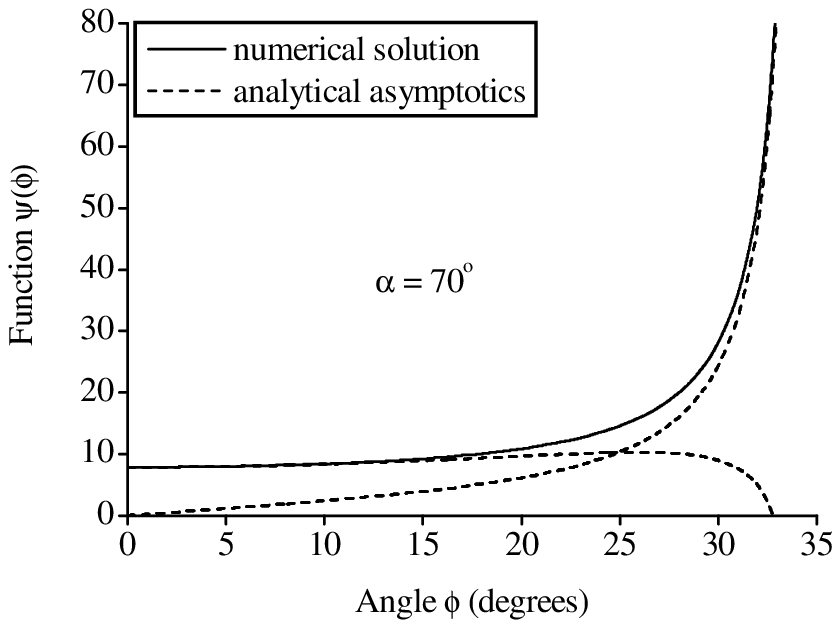}

\includegraphics{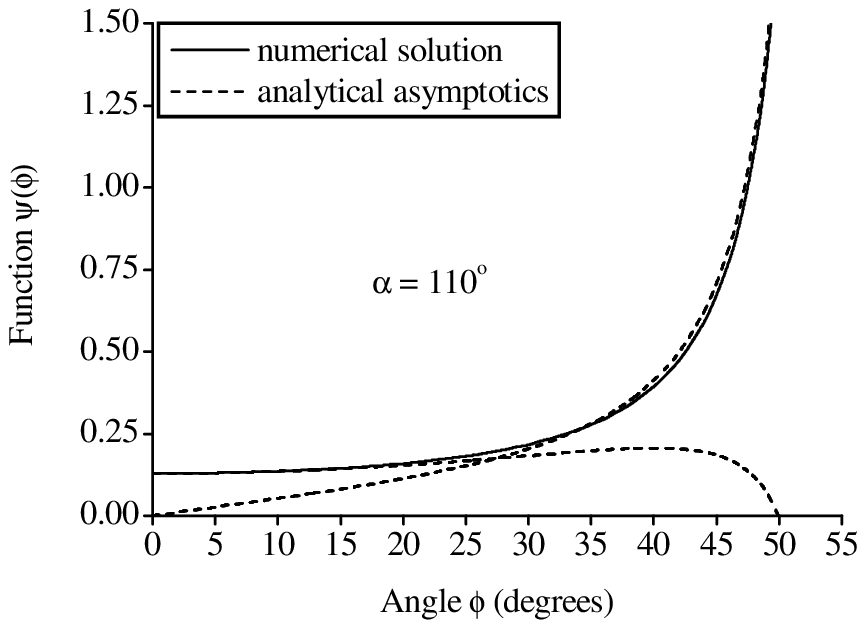}

\caption{Typical behavior of the numerical solution and the analytical asymptotics of function $\tilde\psi(\phi)$ for two values of opening angle.}
\label{psieps}
\end{center}
\end{figure}

The numerical solution to Eq.~(\ref{psi-equation}) satisfying conditions~(\ref{psi0}) for $\tilde\psi(0)$ and $\tilde\psi'(0) = 0$ for $\tilde\psi'(0)$ was found for the two model forms of $\tilde J(\phi)$ and for approximately 20 different values of the opening angle.  In all cases perfect agreement between the numerical solution and the analytical asymptotics of this section was observed.  Two examples of the numerical solution together with the analytical asymptotics are provided in Fig.~\ref{psieps} for opening angles $70^{\circ}$ and $110^{\circ}$.

\begin{figure}
\begin{center}
\includegraphics{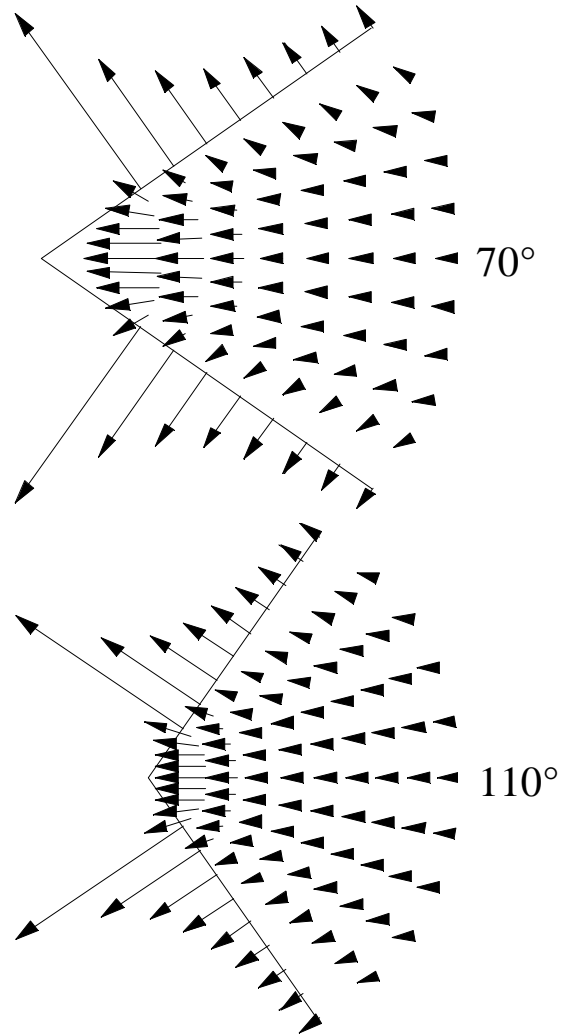}
\caption{Characteristic behavior of flow field for two values of opening angle.  Each arrow represents the absolute value and the direction of velocity $\bfv$ at the point of arrow origin.}
\label{flowfield}
\end{center}
\end{figure}

Characteristic behavior of the velocity field~(\ref{vresult}) is shown in Fig.~\ref{flowfield} for $\alpha = 70^{\circ}$ and $\alpha = 110^{\circ}$.  Note that despite the fact that the exponent $(\mu-\nu)$ of the power law in $r$ is not a smooth function of $\alpha$, the qualitative behavior of the flow field does not visibly change as the opening angle increases past the right angle.

The velocity diverges near the edge of the drop.  At the sides of the angle, only the $\phi$ component diverges, as can be seen from Fig.~\ref{flowfield} and expression~(\ref{vphi}).  This divergence near the sides is of exactly the same one-over-the-square-root dependence on distance as the divergence in the circular drop and has exactly the same origin.  In addition, there is a divergence of both components of the velocity as $r^{\mu - \nu}$ when $r \to 0$, as is apparent from expressions~(\ref{vr}) and (\ref{vphi}).  This divergence is entirely new and is due to the presence of the vertex.  The exponent of this divergence depends on the opening angle as both $\mu$ and $\nu$ do.  This is the first in a set of indices for the pointed drop that are universal but depend on the geometry.  Similar indices will also be encountered for other physical quantities throughout this chapter.

\section{Streamlines}

One must know the shape of the streamlines in order to be able to predict the scaling of the deposit growth, and now, with velocity field in hand, we have everything needed to compute it.  Integration of the velocity field~(\ref{vr})--(\ref{vphi}) according to Eq.~(\ref{vel-ratio}) yields the streamline equation, {\em i.e.}\ the trajectory of each particle as it moves with the fluid:
\begin{equation}
r(\phi) = r_0 \exp\left[(3\nu - \mu - 1) \int_\phi^{\alpha/2} \frac{\tilde\psi(\xi) \, d\xi}{\tilde\psi'(\xi)}\right],
\label{streamline}\end{equation}
where we assume that $\phi$ is positive here and everywhere below (the generalization to the case of negative $\phi$ is obvious as all functions of $\phi$ are even).  Thus, $r = r_0$ when $\phi = \alpha/2$, so that $r_0$ is the distance from the terminal endpoint of the trajectory to the vertex.  In limit $\Delta\phi \to 0$ the integral in the exponent goes to zero (quadratically in $\Delta\phi$) and the streamline equation reduces to
\begin{equation}
r \approx r_0
\label{streamline-edge}\end{equation}
[see Ref.~\cite{popov2} for details of all calculations in this section].  The streamlines are perpendicular to the contact line (up to the quadratic terms in $\Delta\phi$).  This is in good agreement with what one would expect near the edge of the drop, since the azimuthal component of the fluid velocity diverges at the side contact line while the radial component goes to zero.  In limit $\phi \to 0$ the integral in the exponent diverges as $\ln[(\alpha/2)/\phi]$ and hence the result reads
\begin{equation}
r \approx r_0 \left(\frac{\alpha/2}\phi\right)^\epsilon.
\label{streamline-bisector}\end{equation}
Here we introduced exponent $\epsilon$:
\begin{equation}
\epsilon = \frac 1{(\mu + 1)(1 - I)},
\label{epsilon}\end{equation}
where
\begin{equation}
I = \frac{\int_0^{\alpha/2} \left(\frac{\tilde h(\phi)}{\tilde h(0)}\right)^3 \frac{\tilde\psi(\phi)}{\tilde\psi(0)} \, d\phi}{\int_0^{\alpha/2} \frac{\tilde J(\phi)}{\tilde J(0)} \, d\phi}.
\label{i-def}\end{equation}
As we showed in Ref.~\cite{popov2}, $0 < I < 1$ for all $\alpha$.  The positiveness of exponent $\epsilon$ follows both from this fact and from the fact that the trajectory $r(\phi)$ necessarily has to diverge as $\phi \to 0$ (as solute comes from the {\em outer\/} regions of the drop).

Note that the last expression for parameter $I$ is independent of the prefactors of each function of $\phi$, particularly, it is independent of the evaporation intensity $J_0$ and the constant non-universal prefactor $C$ of the surface shape, reinforcing the general conclusion of Chapter~2 that the shape of the streamlines is universal for each geometry.  The same conclusion can also be reached directly from Eq.~(\ref{streamline}), which is independent of $J_0$ and $C$ and thus is universally correct regardless of the overall intensity of evaporation and the non-universal features of the drop boundary.
 
\begin{figure}
\begin{center}
\includegraphics{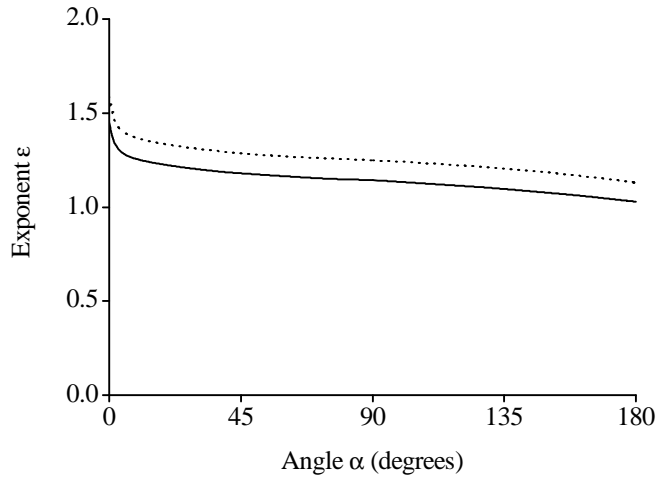}
\caption{Dependence of exponent $\epsilon$ in the power law $r(\phi)$ [Eq.~(\ref{streamline-bisector})] on opening angle $\alpha$.  The two curves correspond to the two model forms for function $\tilde J(\phi)$.}
\label{epsiloneps}
\end{center}
\end{figure}

We cannot compute $I$ and $\epsilon$ explicitly, since we do not have an analytical expression for $\tilde\psi(\phi)$.  However, we can gain some idea of the behavior of these parameters by using approximate forms of $\tilde J(\phi)$ as we did in the preceding section.  Fig.~\ref{epsiloneps} demonstrates the characteristic behavior of exponent $\epsilon$ as a function of the opening angle, obtained numerically on the basis of the two model forms for function $\tilde J(\phi)$.  In order to obtain this plot, equation~(\ref{psi-equation}) was solved with respect to $\tilde\psi(\phi)$ numerically for each $\alpha$, and then expression~(\ref{i-def}) for $I$ was employed.  As can be observed in this graph, the two model forms of $\tilde J(\phi)$ lead to the plots of very similar shape, but shifted by no more than 10\% in the whole range of the opening angles.  Thus, we conclude that Fig.~\ref{epsiloneps} provides a good estimate for the qualitative behavior and the order of magnitude of exponent $\epsilon(\alpha)$.

\begin{figure}
\begin{center}
\includegraphics{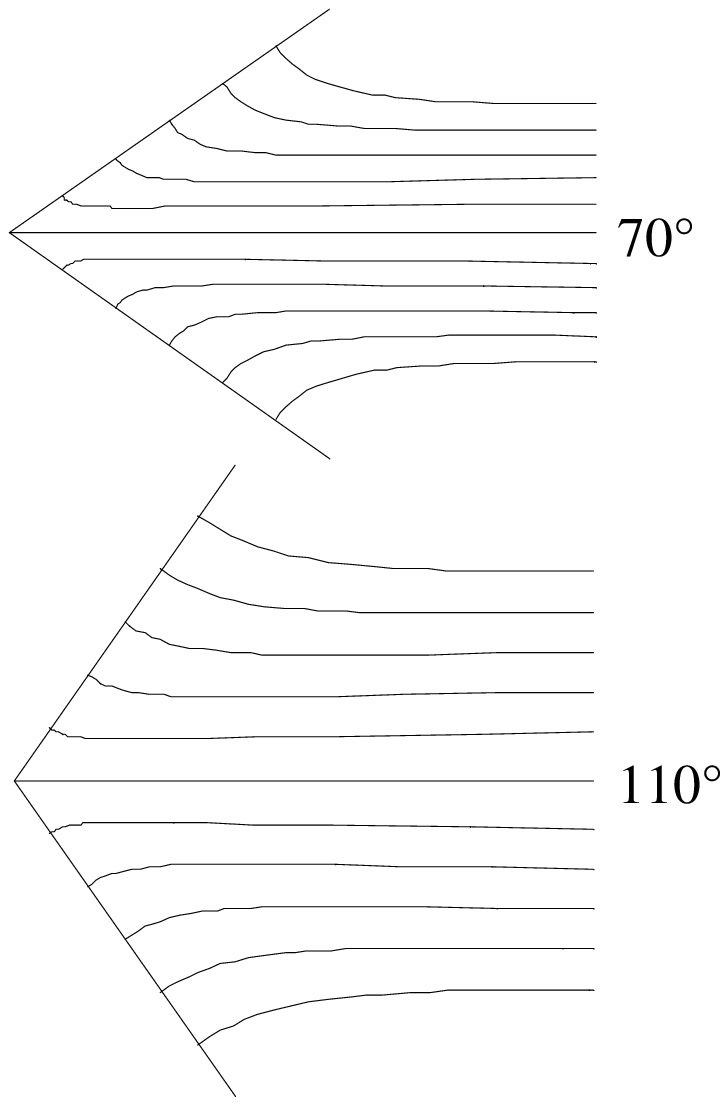}
\caption{Typical shape of the streamlines for two values of opening angle (for the same values as in Figs.~\ref{psieps} and \ref{flowfield}).}
\label{streamlines}
\end{center}
\end{figure}

Typical shape of the streamlines is shown in Fig.~\ref{streamlines} for $\alpha = 70^{\circ}$ and $\alpha = 110^{\circ}$.  It was based on one of the model forms for function $\tilde J(\phi)$, and involved the corresponding numerical solutions for function $\tilde\psi(\phi)$ (Fig.~\ref{psieps}) employed in Eq.~(\ref{streamline}).  This shape is practically insensitive to the model form of $\tilde J(\phi)$, and almost an identical copy of this graph was obtained for the other model form.

The distance from a point on a streamline to the bisector scales with $\phi$ as $\phi \, r(\phi) \propto \phi^{1-\epsilon}$ when $\phi \to 0$.  Since $\epsilon > 1$ (Fig.~\ref{epsiloneps}), this distance increases when $\phi$ decreases.  Thus, the streamlines diverge away from the bisector when $\phi \to 0$, and hence they do {\em not\/} originate on the bisector.  An incoming element of fluid initially located close to the bisector moves towards this bisector, reaches a minimum distance and then veers away towards the contact line.  One can also arrive at the same conclusion having started from the ratio of velocity components.  As can be shown [see Ref.~\cite{popov2}], for small $\phi$ (in limit $\phi \to 0$) the ratio of the velocity components is $v_\phi / v_r = - \phi / \epsilon$.  This ratio represents the angle between a streamline and a coordinate line $\phi = \phi_0$ at any point $(\phi_0, r(\phi_0))$ on that streamline.  Since $\epsilon > 1$, the absolute value of this angle is less than $|\phi|$, and therefore, despite the opposite sign of this angle, the streamline diverges away from the bisector for small $\phi$.  This tendency can also be observed directly in Fig.~\ref{streamlines}.

Another feature apparent from Fig.~\ref{streamlines} is the self-similarity of all the streamlines.  As is clear from equation~(\ref{streamline}), the only scaling parameter of the family of streamlines is $r_0$, and therefore all the streamlines can be obtained from a single streamline (say, the one with $r_0 = 1$) by multiplying its $r$-coordinate by different values of $r_0$.

\section{Solute transfer and deposit growth}

\begin{figure}
\begin{center}
\includegraphics{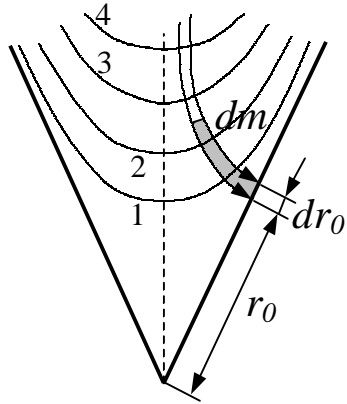}
\caption{Qualitative sketch: mutual location of streamlines (the two lines with arrows) and isochrones (the four numbered lines).  Solute moves along the streamlines towards the contact line (the bold line).  Shaded area is swept by an infinitesimal element of fluid between the two infinitesimally close streamlines as that element moves towards the contact line.  The isochrones are the geometric locations, starting from which the solute reaches the contact line at the same time.  Solute from isochrone 1 reaches the contact line first; solute from isochrone 4 reaches the contact line last.}
\label{soltrans}
\end{center}
\end{figure}

The time it takes an element of fluid to reach the contact line can be inferred from either $v_r(r,\phi,t) dt = dr$ or $ v_\phi(r,\phi,t) dt = r \, d\phi$ on a streamline ({\em i.e.}\ with relation~(\ref{streamline}) between $r$ and $\phi$).  The only term with time dependence in expressions~(\ref{vr}) and (\ref{vphi}) is $R_0^{\nu - 1}$.  As we mentioned earlier, the time dependence of this term is weak and can be safely ignored for times $t \ll t_f$, where $t_f$ is the total drying time defined in Eq.~(\ref{tf}).  It will be seen later in this section that these times are the only times of interest in the geometry of the pointed drop since everything that happens closer to the end of the drying process is not universal and depends on the details of the boundary shape outside the sector of interest ({\em i.e.}\ on the shape of the contact line on the side of the drop furthest from the vertex).  Therefore, restricting our attention to only the universal features of the solution, we can consider only the times $t \ll t_f$ and ignore the dependence of $R_0$ (and hence $v_r$ and $v_\phi$) on time.  Then the time and the coordinate dependences in the differential equations above separate trivially and the time it takes an element of fluid to reach the contact line can be calculated as
\begin{equation}
t = \int_{\phi_i}^{\alpha/2} \frac{r\, d\phi}{v_\phi} = \int_{r_i}^{r_0} \frac{dr}{v_r},
\end{equation}
where $(r_i,\phi_i)$ are the initial coordinates of that element of fluid.  Taking into account expressions~(\ref{vr}) and (\ref{vphi}) for the velocity components and relation~(\ref{streamline}) between $r$ and $\phi$ on a streamline, one can reduce both integrals to
\begin{equation}
t = t_0 \, \int_{\phi_i}^{\alpha/2} \frac{\exp\left[(\nu - \mu + 1)(3\nu - \mu - 1) \int_\phi^{\alpha/2} \frac{\tilde\psi(\xi) \, d\xi}{\tilde\psi'(\xi)}\right]}{\tilde h^2(\phi) \tilde\psi'(\phi)} \, d\phi,
\label{time}\end{equation}
where $t_0$ is a combination of system parameters with the dimensionality of time:
\begin{equation}
t_0 = \frac\rho{J_0} \sqrt{A}^{\mu - 1} R_{0i}^{- \nu + 1} r_0^{\nu - \mu + 1}
\label{t0}\end{equation}
and $R_{0i} = R_0(0)$.  Within this time {\em all\/} the solute that lays on the way of this element of fluid as it moves toward the contact line becomes part of the deposit (highlighted area in Fig.~\ref{soltrans}).  The mass $dm$ of this deposit (accumulated on the contact line between $r_0$ and $r_0 + dr_0$) can be found by integrating $h(r,\phi,0)$ of Eq.~(\ref{surfshap}) over area $dA$ swept by this infinitesimal volume and multiplying the result by the initial concentration $c$ of the solute [see Eq.~(\ref{mass-def})].  Employing relation~(\ref{streamline}) once again, we obtain:
\begin{equation}
dm = c \, \frac{r_0^{\nu + 1} \, dr_0}{R_{0i}^{\nu - 1}} \int_{\phi_i}^{\alpha/2} \tilde h(\phi) \exp\left[(\nu + 2)(3\nu - \mu - 1) \int_\phi^{\alpha/2} \frac{\tilde\psi(\xi) \, d\xi}{\tilde\psi'(\xi)}\right] \, d\phi.
\label{mass}\end{equation}
Dependence $dm(t)$ can now be found by eliminating $\phi_i$ from results~(\ref{time}) and (\ref{mass}).

Exact analytical calculation of the dependence $m(t)$ is not possible for an arbitrary starting point $(r_i,\phi_i)$ on a streamline since no analytical expression for $\tilde\psi(\phi)$ is available for arbitrary $\phi$ and since integrals in Eqs.~(\ref{time}) and (\ref{mass}) cannot be computed analytically for arbitrary $\phi$ even if $\tilde\psi(\phi)$ were known.  However, there are two important cases that {\em can\/} be tackled analytically: (a) {\em early times,} when the initial point is close to the contact line ({\em i.e.}\ when $\Delta\phi_i \ll \alpha/2$ and $r_i \approx r_0$) and only the solute between that initial point and the contact line is swept into the edge deposit (the starting point is on isochrone 1 of Fig.~\ref{soltrans} or closer to the contact line), and (b) {\em intermediate times,} when the initial point is close to the bisector ({\em i.e.}\ when $|\phi_i| \ll \alpha/2$ and $r_i \gg r_0$) and virtually all the solute between the bisector and the contact line is swept into the edge deposit (the starting point is on isochrone 4 of Fig.~\ref{soltrans} or further from the vertex).  Situations between these two limiting cases (highlighted area in Fig.~\ref{soltrans} demonstrates one of them, starting points on isochrones 2 and 3 would correspond to some other) can be extrapolated on the basis of continuity of the results.  Since our region is indefinitely smaller than the drop as a whole, we may treat regions (a) and (b) assuming that a negligible fraction of the drop has evaporated.  At some later stage that we call the {\em late-time\/} regime, an appreciable fraction of the drop has evaporated, and the fluid trajectories have reached back into the bulk of the drop.  In this late regime our asymptotic treatments are clearly not adequate to describe the flow as we did not specify any details of the drop geometry far from the vertex.  Thus we cannot treat this regime by our methods, and only the properties of drying process at early and intermediate stages can be found from information in hand.

Apart from the definitions based on the trajectories, the three regimes can be equivalently defined in terms of time $t$: (a) early times: $t \ll t_0$; (b) intermediate times: $t_0 \ll t \ll t_f$; and (c) late times: $t \approx t_f$.  Here $t_0$ is the characteristic time defined in Eq.~(\ref{t0}) (this characteristic time depends on $r_0$) and $t_f$ is the total drying time defined in Eq.~(\ref{tf}).  The equivalence of the definitions in terms of the initial position on a trajectory and in terms of time can be seen from equation~(\ref{time}).  As we explained in the preceding paragraph, only the early and the intermediate times are of our interest as being independent on the rest of the boundary shape, and therefore the assumption $t \ll t_f$ employed at the beginning of this section is coherent with our intention to determine only the universal features of the results.

As is clear from the definition of the intermediate-time regime, the necessary condition for its existence is $t_0 \ll t_f$, which can be reduced to $(r_0/\sqrt{A})^{3 - \mu} (r_0/R_{0i})^{\nu - 2} \ll 1$ by combining Eqs.~(\ref{t0}) and (\ref{tf}).  For acute opening angles the latter condition is always true~\cite{popov2}, and hence the intermediate regime is well defined for acute opening angles.  For obtuse opening angles the condition above can be transformed into $(r_0/\sqrt{A})^{\nu - \mu + 1} \ll \theta_i^{2 - \nu}$, where $\theta_i$ is the typical initial value of the contact angle in the bulk of the drop, {\em i.e.}\ far away from the vertex.  This condition should be expected to be satisfied for not too small initial values of the bulk contact angle $\theta_i$.  The closer to the vertex the trajectory endpoint is, the better this condition is obeyed.  On the other hand, this condition is obeyed worse for larger opening angles.  At exactly $\alpha = \pi$ the intermediate-time regime is indistinguishable from the early-time regime, and hence should not exist~\cite{popov2}.

\begin{figure}
\begin{center}
\includegraphics{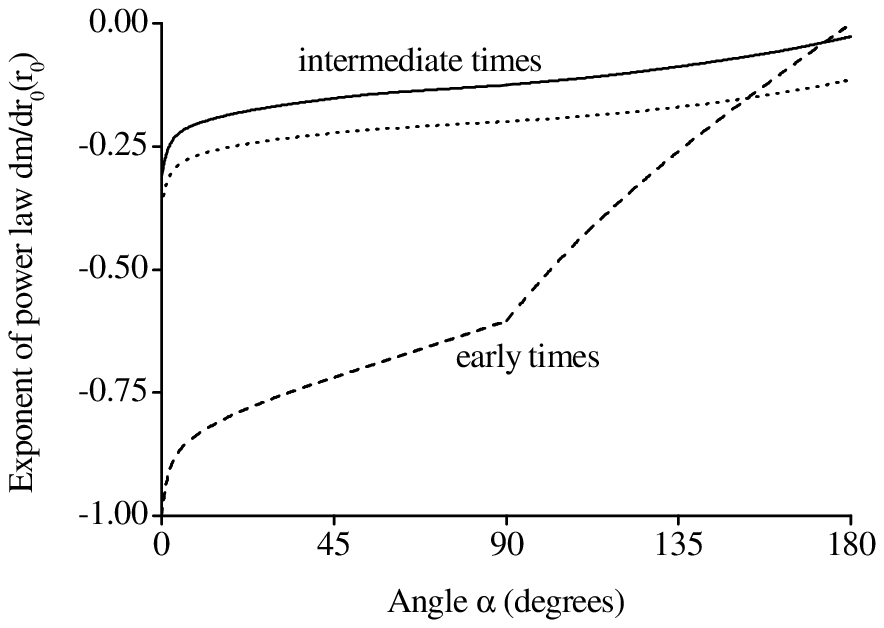}
\caption{Exponent of distance $r_0$ in the power law $dm/dr_0(r_0)$ [Eqs.~(\ref{exponents-early}) and (\ref{exponents-interm})] as a function of the opening angle for the two time regimes.  The early-time curve corresponds to the exponent $\beta$ of Eq.~(\ref{beta}); the intermediate-time curves correspond to the exponent $\gamma$ of Eq.~(\ref{gamma}).  The two curves for the intermediate-time exponent correspond to the two model forms for function $\tilde J(\phi)$.}
\label{massdist}
\end{center}
\end{figure}

At early times, the deposit growth is entirely due to the transfer of particles originally located near the contact line.  Evaluating the integrals in expressions~(\ref{time}) and (\ref{mass}) in limit $\Delta\phi_i \to 0$, expressing $\Delta\phi_i$ in terms of time and then substituting the result into the expression for mass, we obtain the mass of the deposit as a function of time:
\begin{equation}
\frac{dm}{dr_0}(t,r_0) \approx c \, \frac{r_0^{\nu + 1}}{R_{0i}^{\nu - 1}} \frac{\left|\tilde h'\left(\alpha/2\right)\right|}2 \left(\frac{1 + \lambda}{1 - \lambda} \frac{J^*}{\left|\tilde h'\left(\alpha/2\right)\right|} \frac{t}{t_0}\right)^{\frac{2}{1 + \lambda}}.
\label{early-result}\end{equation}
Note that $t_0$ also depends on $r_0$.  Thus, at early times the deposit grows in time as a power law
\begin{equation}
\frac{dm}{dr_0}(t,r_0) \propto t^{2/(1 + \lambda)} r_0^\beta,
\label{exponents-early}\end{equation}
where the $r_0^\beta$ arises from the $r_0^{\nu + 1}$ prefactor and from the $r_0$-dependence of $t_0$.  Using Eq.~(\ref{t0}), we find
\begin{equation}
\beta = - \frac{(1-\lambda)(1+\nu)-2\mu}{1+\lambda}
\label{beta}\end{equation}
and plot it in Fig.~\ref{massdist} as a function of opening angle (the early-time curve).  Note that this result is independent of function $\tilde\psi(\phi)$ and hence precise as long as the factor $J^*$ in the asymptotic of the evaporation rate~(\ref{theta-edge}) is known.

There are two important conclusions to be drawn from this result.  One is that the power-law exponent of time $2/(1+\lambda) = 4/3$ is exactly the same as in the case of a round drop considered in the preceding chapter.  This should be of no surprise since close to the side of the angle (as well as close to the circumference of a round drop) the contact line looks locally like a straight line, and the solute ``does not know'' about the vertex of the angle or the curvature of the circumference.  This exponent is determined entirely by the local properties of an infinitesimal segment of the contact line of length $dr_0$ and is independent of larger geometrical features of the system.

\begin{figure}
\begin{center}
\includegraphics{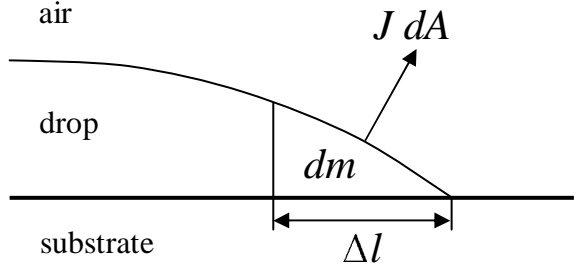}
\caption{An illustration of the derivation of the four-thirds law at a straight segment of the contact line.  The contact line is normal to the plane of the figure.  Length $dr_0$ is along the contact line and hence not shown.  The flow is in the plane of the figure from left to right.}
\label{fourthirds}
\end{center}
\end{figure}

The value $2/(1+\lambda) = 4/3$ of the exponent of time can be obtained from a very simple argument, relying only on the assumptions that (a) the contact line is straight, (b) the streamlines are perpendicular to the contact line, and (c) the distribution of the solute is uniform.  Indeed, the mass of both the water and the solute is proportional to the volume of an element of fluid near the contact line (Fig.~\ref{fourthirds}): $dm \propto (\Delta l)^2 dr_0$.  All this mass should be evaporated from the surface of this volume element in some time $t$.  The evaporation rate (per unit area) scales as $J \propto (\Delta l)^{-\lambda}$ and therefore the full rate of mass loss is $J dA \propto (\Delta l)^{-\lambda+1} dr_0$.  The time it takes this volume to evaporate can now be found as the ratio of its mass to the rate of mass loss: $t = dm/(J dA) \propto (\Delta l)^{1+\lambda}$.  Thus, $(\Delta l) \propto t^{1/(1 + \lambda)}$ and hence $dm/dr_0 \propto (\Delta l)^2 \propto t^{2/(1 + \lambda)}$ as asserted.

The other observation is the dependence on $r_0$.  First of all, the dependence on distance to the vertex itself is entirely new compared to the round-drop case (for obvious reasons).  Second, since exponent $\beta$ is always between $-1$ and 0, the singularity in $r_0$ is always integrable at $r_0 = 0$.  Physically, this corresponds to the statement that the vertex of the sector does {\em not\/} dominate the sides and that the deposit accumulation at the vertex is {\em not\/} qualitatively different from the deposit accumulation on the sides.

Similar calculation can also be conducted analytically for the intermediate times as well.  The starting point of a streamline in the intermediate-time regime lies near the bisector, and the appropriate limit is $\phi_i \to 0$.  Calculating the time it takes an element of fluid to reach the contact line [Eq.~(\ref{time})] and the mass accumulated at the contact line between $r_0$ and $r_0 + dr_0$ for this time [Eq.~(\ref{mass})] in limit $\phi_i \to 0$, and then eliminating $\phi_i$ from the two results, we arrive at the dependence of mass on time:
$$\frac{dm}{dr_0}(t,r_0) \approx c \, \frac{r_0^{\nu + 1}}{R_{0i}^{\nu - 1}} \frac{(\mu + 1)(1 - I)}{(\nu - \mu + 1) + (\mu + 1)I} \,\tilde h(0)\, \frac{\alpha}2 \times$$
\begin{equation}
\times\left((\nu - \mu + 1)(3\nu - \mu - 1)\tilde h^2(0)\tilde\psi(0) \frac{t}{t_0}\right)^{1 + (\mu + 1)I/(\nu - \mu + 1)}.
\label{interm-result}\end{equation}
Taking into account that $t_0$ also depends on $r_0$, we finally conclude that the deposit mass grows as a power law
\begin{equation}
\frac{dm}{dr_0}(t,r_0) \propto t^\delta r_0^\gamma,
\label{exponents-interm}\end{equation}
where we introduced notations for the exponent of time
\begin{equation}
\delta = 1 + \frac{(\mu + 1)I}{(\nu - \mu + 1)}
\label{delta}\end{equation}
and for the exponent of $r_0$, originating from both the prefactor $r_0^{\nu + 1}$ and the $r_0$-dependence of $t_0$ [Eq.~(\ref{t0})]:
\begin{equation}
\gamma = \mu - (\mu + 1)I.
\label{gamma}\end{equation}
Unlike the early-time limit, this result depends on the entire functions $\tilde J(\phi)$ and $\tilde\psi(\phi)$, not only their asymptotics, via parameter $I$ defined in Eq.~(\ref{i-def}).  This is quite natural because an element of fluid passes through the entire drop when it moves from the bisector to the contact line, and hence knowledge of all quantities in the intermediate points is required.  In the early regime, only the asymptotic properties near the edge are of importance.

The exponent of time stays greater than one in the intermediate-time regime.  Thus, the rate of mass accumulation $dm/dt$ continues to grow with time in this regime, and the deposit mass grows faster and faster.  This result has a simple explanation for both the early- and the intermediate-time regimes.  Since the initial distribution of the solute is uniform, and since the solvent evaporates, the solute concentration at any given volume {\em increases\/} with time.  Thus, even though the fluid and the particles move along the same streamlines in practically constant velocity field (assuming that $R_0(t) \approx R_{0i}$ at sufficiently early stages), the rate of mass accumulation also {\em increases\/} with time, since portions of solution arriving at the contact line at approximately constant rate have higher and higher solute concentration.  Note that this mechanism and this general conclusion are in good agreement with the exact analytical result for the circular geometry that the rate of mass accumulation must diverge at the end of the drying process (as $t \to t_f$) and that {\em all\/} the deposit must accumulate at the contact line by $t = t_f$.

Another observation is related to the exponent of $r_0$.  Since $I < 1$, then $\gamma > -1$.  Therefore, the mass is integrable at $r_0 = 0$, and the statement of the early-time regime that the deposit accumulation at the vertex is not qualitatively different from the deposit accumulation on the sides continues to hold in the intermediate-time regime as well.  Trivially, $\gamma < \mu$, as $I > 0$.

The exponent of $r_0$ must be identically zero at {\em any\/} time for the opening angle of exactly $\alpha = \pi$.  Indeed, at $\alpha = \pi$ the contact line is just a straight line ({\em i.e.}\ there is no angle at all), and therefore there is a full translational symmetry with respect to which point of this line should be called ``vertex.''  Thus, the choice of $r_0 = 0$ is absolutely arbitrary, and there can be no dependence on $r_0$ whatsoever.

\begin{figure}
\begin{center}
\includegraphics{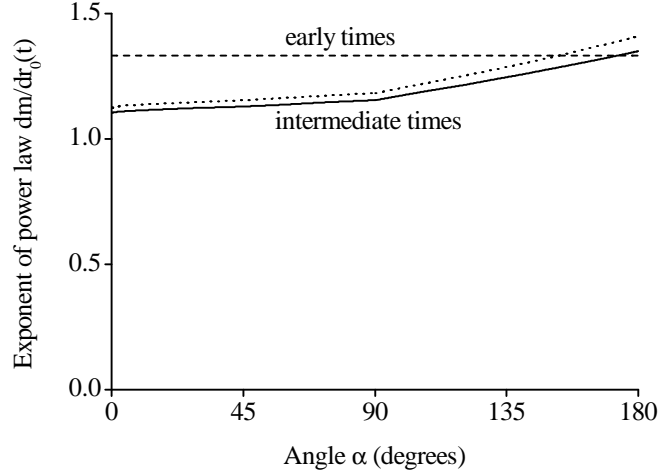}
\caption{Exponent of time $t$ in the power law $dm/dr_0(t)$ [Eqs.~(\ref{exponents-early}) and (\ref{exponents-interm})] as a function of the opening angle for the two time regimes.  The early-time curve corresponds to the exponent $2/(1+\lambda) = 4/3$; the intermediate-time curves correspond to the exponent $\delta$ of Eq.~(\ref{delta}).  The two curves for the intermediate-time exponent correspond to the two model forms for function $\tilde J(\phi)$.}
\label{masstime}
\end{center}
\end{figure}

Indices $\gamma$ and $\delta$ are plotted in Figs.~\ref{massdist} and \ref{masstime}, respectively, as functions of the opening angle (the intermediate-time curves).  The two intermediate-time curves on each graph correspond to the same two model forms for function $\tilde J(\phi)$ as we used in Fig.~\ref{epsiloneps}.  To facilitate the comparison of the results, we plot the exponents for the early- and the intermediate-time regimes in Figs.~\ref{massdist} and \ref{masstime} together.

The intersection of the exponents near $\alpha = \pi$ on both graphs can be attributed to a couple of reasons.  First, it should be kept in mind that the plotted results for $\gamma$ and $\delta$ are based on a relatively arbitrary choice of model forms for function $\tilde J(\phi)$, which, as we suspect, become increasingly inaccurate for large $\alpha$.  Second, as was explained just after Eq.~(\ref{contactangle}), at exactly $\alpha = \pi$ the contact angle $\theta$ is not small even for $r \ll R_0$, and the correction to the exponent $\lambda$ due to this contact angle [see Eq.~(\ref{lambda})] is comparable to the value $1/2$ assumed in all numerical estimates.  All in all, we believe that this intersection of the early- and intermediate-time exponents is an artifact of our formalism and should not be observed in reality, since the results for the two time regimes should be identical at exactly $\alpha = \pi$.  At exactly $\alpha = \pi$ the exponent of $r_0$ must be equal to zero at any time and the exponent of time must be equal to $2/(1+\lambda) = 4/3$ at any time.

\begin{figure}
\begin{center}
\includegraphics{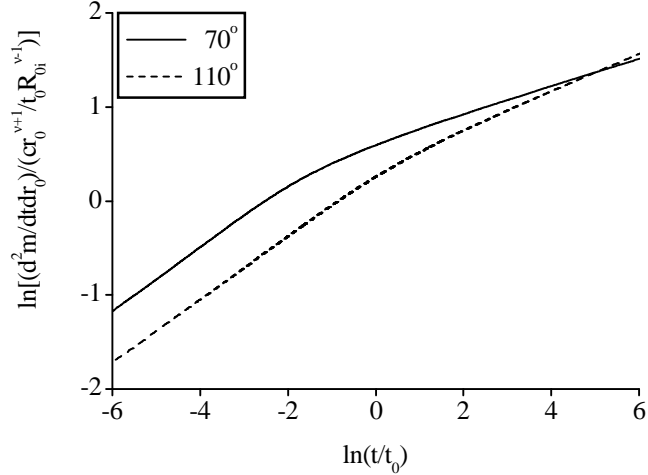}
\caption{Log-log plot of the numerical solution $d^2 m/dtdr_0(t)$ for two values of opening angle (for the same values as in Figs.~\ref{psieps}, \ref{flowfield} and \ref{streamlines}).}
\label{lnmasslntime}
\end{center}
\end{figure}

In addition to the early- and the intermediate-time analytical asymptotics above, we also find the numerical solution for $d^2 m/dtdr_0(t)$.  We find the time derivative of $dm/dr_0$ instead of $dm/dr_0$ itself in order to demonstrate the amount of mass arriving at the contact line at time $t$ rather than the total mass accumulated by the time $t$.  We employ the chain rule to obtain $d^2 m/dtdr_0$ on the basis of Eq.~(\ref{time}) for $t(\phi_i)$ and Eq.~(\ref{mass}) for $dm/dr_0(\phi_i)$:
$$ \frac{d}{dt}\left(\frac{dm}{dr_0}\right) = \frac{\frac{d}{d\phi_i}\left(\frac{dm}{dr_0}\right)}{\frac{dt}{d\phi_i}} = $$ \begin{equation}
= \frac{c}{t_0} \, \frac{r_0^{\nu + 1}}{R_{0i}^{\nu - 1}} \, \tilde h^3(\phi_i) \tilde\psi'(\phi_i) \exp\left[(\mu + 1)(3\nu - \mu - 1) \int_{\phi_i}^{\alpha/2} \frac{\tilde\psi(\xi) \, d\xi}{\tilde\psi'(\xi)}\right],
\label{dmassdtime}\end{equation}
then use the numerical result for $\tilde\psi(\phi)$ (Fig.~\ref{psieps}) in order to find $t(\phi_i)$ [Eq.~(\ref{time})] and $d^2 m/dtdr_0(\phi_i)$ [Eq.~(\ref{dmassdtime})] numerically, and finally create a log-log parametric plot $d^2 m/dtdr_0$ {\em vs.}\ $t$, as shown in Fig.~\ref{lnmasslntime}.  The two curves in Fig.~\ref{lnmasslntime} correspond to the two values of $\alpha$ we used earlier ($70^{\circ}$ and $110^{\circ}$).  As all numerical results of this section are, the plot is based on one of the model forms for function $\tilde J(\phi)$, but very insensitive to the particular form of this function.  This plot clearly demonstrates two different slopes (and hence two different time regimes) of each curve.  The crossover between the two regimes (slopes) occurs around time $t \approx t_0$ ({\em i.e.}\ near $\ln(t/t_0) \approx 0$), and the early-time slopes are equal for both values of the opening angle (and equal to $2/(1+\lambda) - 1 = 1/3$ as to be expected from our early time results).  All these numerical results are in excellent agreement with our analytical predictions, and the numerical values of time exponents compare very well with those of Fig.~\ref{masstime} (which should be corrected by $-1$ due to the differentiation with respect to time in Fig.~\ref{lnmasslntime}).

\begin{figure}
\begin{center}
\includegraphics{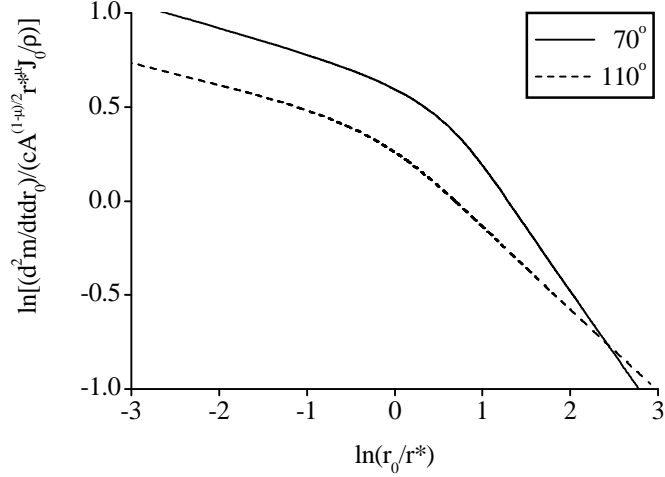}
\caption{Log-log plot of the numerical solution $d^2 m/dtdr_0(r_0)$ for two values of opening angle (for the same values as in Figs.~\ref{psieps}, \ref{flowfield}, \ref{streamlines} and \ref{lnmasslntime}).  Parameter $r^*$ is defined by Eq.~(\ref{r-star}).}
\label{lnmasslndist}
\end{center}
\end{figure}

In a similar fashion we obtain a log-log plot for $d^2 m/dtdr_0$ as a function of $r_0$.  We fix $t$, then express $r_0$ in terms of $\phi_i$ by combining Eqs.~(\ref{time}) and (\ref{t0}), and finally determine $r_0(\phi_i)$ and $d^2 m/dtdr_0(\phi_i)$ [Eq.~(\ref{dmassdtime})] numerically on the basis of the numerical result for $\tilde\psi(\phi)$ of Fig.~\ref{psieps}.  The resulting log-log parametric plot $d^2 m/dtdr_0(r_0)$ is shown in Fig.~\ref{lnmasslndist} for the two values of the opening angle ($70^{\circ}$ and $110^{\circ}$).  The purpose of this graph is to provide a snapshot of the deposit growth at any given moment of time $t$.  For small $r_0$ the accumulation of the solute at the contact line is already in the intermediate-time regime, while for large $r_0$ the growth is still in the early-time regime.  The threshold between the two regimes is defined by $t = t_0$.  This condition can be reversed by solving Eq.~(\ref{t0}) with respect to $r_0$.  The resulting value
\begin{equation}
r^* = \left( \frac{J_0}\rho \sqrt{A}^{1-\mu} R_{0i}^{\nu-1} t \right)^{\frac 1{\nu-\mu+1}}
\label{r-star}\end{equation}
defines the threshold in terms of $r_0$ (at any moment of time $t$):  the early regime corresponds to $r_0 \gg r^*$, and the intermediate regime corresponds to $r_0 \ll r^*$.  As can be seen from the numeric plot, the regimes indeed switch at $r_0 \approx r^*$ ({\em i.e.}\ near $\ln(r_0/r^*) \approx 0$).  The intermediate-time slopes are almost equal for both graphs since the intermediate-time exponent $\gamma$ (the upper curves in Fig.~\ref{massdist}) varies very weakly with $\alpha$ ($\gamma \approx -0.135$ for $\alpha = 70^{\circ}$, and $\gamma \approx -0.111$ for $\alpha = 110^{\circ}$).  Again, the numerical results are in excellent agreement with the analytical asymptotics, and the numerical values of exponents compare very well with those of Fig.~\ref{massdist}.

Figure~\ref{lnmasslndist} suggests that the rate of increase of $dm/dr_0$ has a sharp change of behavior as a function of $r_0$ for any given time.  For small $r_0$ (intermediate times) function $d^2 m/dtdr_0(r_0)$ varies weakly, while for large $r_0$ (early times) it falls off more dramatically with increasing $r_0$.  The crossover point $r^*$ [Eq.~(\ref{r-star})] moves outwards as a power of time $t$ with exponent $1/(\nu-\mu+1)$.  (Note that this exponent involves only the accurately known functions of $\alpha$.)  This crossover point and its outward motion provide a clear-cut signature of our mechanism, and this signature should be the strongest for small opening angles (as Fig.~\ref{massdist} suggests).  (In order to avoid possible non-universal effects from late times, one needs to measure the system before the late-time regime.)

Probably, the most exciting feature of the angular-sector solution is its dependence on the opening angle.  Unlike the round-drop case, there is an extra free parameter of the problem --- the opening angle of the sector.  All the results, including the exponents of the power laws, depend on this opening angle.  Note that these exponents are {\em universal, i.e.}\ they do not depend on any other parameters of the system, except for the opening angle.  At the same time, the only parameter they depend on is extremely easy to control --- preparing an evaporating drop one can adjust the opening angle of the contact line at his will without any technical elaborations.  Thus, for example, by suitably choosing the opening angle (and the time regime), one can create a predetermined power-law distribution of the solute along the contact line with virtually any exponent of distance between $-1$ and 0 (Fig.~\ref{massdist}).  In principle, this feature may have significant practical applications for all the processes mentioned in the Introduction. 

\chapter{Deposit growth for finite-volume particles in circular evaporating drops}

In the preceding parts of this work we considered how mass of the contact-line deposit grows with time and how it depends on such geometrical characteristics of the drop as its radius (circular drops) or its opening angle and distance from the vertex (pointed drops).  However, we never attempted to describe the geometrical characteristics of the contact-line deposit itself, {\em e.g.}\ the width and the height of the deposit ring in the case of circular geometry.  At the same time, there is solid experimental data~\cite{deegan3} on various geometrical characteristics of the ring and their dependence on time, the initial concentration of the solute, and the geometry of the drop.  In this chapter we will provide a simple model that addresses this lack of theoretical understanding of the geometrical properties of the contact-line deposit and provides analytical results that compare favorably to the experimental data.  This model is as universal (in its range of validity) as all the consideration above and provides the description that depends only on the geometry of the problem.  This model is based on the assumption that particles have finite volume and they simply cannot pack denser than some volume fraction, {\em e.g.}\ fraction of close packing.  It turns out that this model is sufficient to explain most of the observed phenomena.

The notion that the profile of the deposit could be found by the simple assumption that the solute becomes immobilized when the volume fraction reaches a threshold was originally suggested by Todd Dupont~\cite{dupont}.  First efforts to create a model were conducted by Robert Deegan~\cite{deegan3, deegan4} who formulated most of the physical assumptions and wrote them down mathematically.  Here we present the entire problem, including its full formulation and its analytical and numerical solutions (not reported in the literature previously).

\section{Model, assumptions, and geometry}

In this chapter we will restrict our attention to the circular geometry of the drop only, as this geometry is the easiest to deal with mathematically and the most important practically.  We will continue to use cylindrical coordinates $(r,\phi,z)$ with the origin in the center of the drop.  This will allow us to employ some results obtained for the circular drops earlier in this work.

As we already saw, it is necessary to assume that the contact angle is small in order to be able to obtain any analytical results in a closed form.  This case is also most important practically, as virtually always $\theta \ll 1$ in experimental realizations, including experiments of Ref.~\cite{deegan3}.  Thus, we do not lose any generality assuming that contact angle is small.  This will be the only small parameter in this chapter.  It is clear that a drop with a small contact angle is necessarily thin, {\em i.e.}\ its maximal height is much smaller than its radius.  Therefore, it makes good sense to continue considering vertically averaged quantities, like the vertically averaged velocity $\bfv$ of Eq.~(\ref{defv}) that we used in the three preceding chapters.

Our model pictures the drop as a two-component system (the components are ``fluid'' and ``solute''), which has two ``phases'': the ``liquid phase'' in the middle of the drop and the ``deposit phase'' near the contact line.  Both components are present in both phases, and the difference between the phases lies in the concentration of the solute.  In the deposit phase, volume fraction of solute $p$ is high ({\em e.g.}\ comparable to close packing fraction or 1) and {\em fixed\/} in both space and time.  Thus, $p$ is just a constant number, one can think of it as close packing fraction or 1.  (The case of $p = 1$ may seem to be special as there is no water in the deposit phase; however, as will be seen below, for small initial concentrations of the solute this case leads to exactly the same main order results.)  In the liquid phase, volume fraction of solute $\chi$ varies in space and changes with time, and it is relatively small compared to $p$.  The initial volume fraction $\chi_i = \chi(0)$ is constant throughout the drop; at later moments the solute gets redistributed due to the hydrodynamic flows, and the concentration becomes different in different parts of the liquid phase.  Volume fraction of fluid is then $(1-p)$ in the deposit phase and $(1-\chi)$ in the liquid phase.  Note that we do not require $\chi_i \ll p$ in this section, although we do assume $\chi_i < p$.  We should also emphasize that we do not presume there is any real ``phase difference'' between the so-called phases: one phase is just defined as having the maximal reachable solute fraction $p$ (solute cannot move in this phase) while the other phase is characterized by lower solute fraction $\chi$ (in this phase solute can move and hence concentration can change in time and space).  Besides this difference, phases are essentially identical.  The idea that the solute loses its mobility when its concentration exceeds some threshold was suggested by Todd Dupont~\cite{dupont}.

\begin{figure}
\begin{center}
\includegraphics{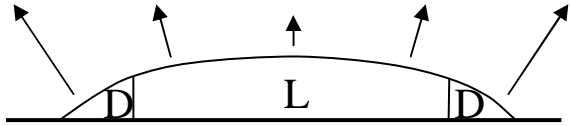}
\caption{Mutual location of two ``phases'' in the drying drop:  L is the liquid phase, and D is the deposit phase.}
\label{phaseseps}
\end{center}
\end{figure}

\begin{figure}
\begin{center}
\includegraphics{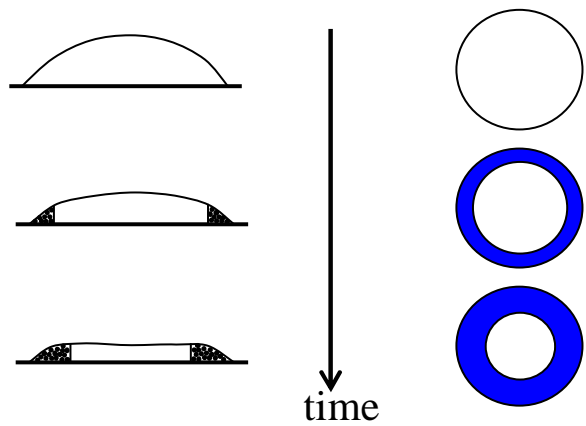}
\caption{Time evolution of the deposit phase growth: side view (left) and top view (right).  Only the deposit phase is shown.  Thickness of the ring is exaggerated compared to the typical experimental results.}
\label{evolutioneps}
\end{center}
\end{figure}

Since the drop is thin and since we employ vertically averaged velocity ({\em i.e.}\ the problem is two-dimensional in certain sense), it is natural to assume the boundary between the phases is vertical.  Thus, the particles get stacked uniformly at all heights when being brought to the phase boundary by the hydrodynamic flow (which does not depend on vertical coordinate $z$).  This boundary can be pictured as a vertical wall at some radius $R(t)$ from the center of the drop, and this wall propagates from the contact line (located at $R_i = R(0)$) to the center of the drop.  Fig.~\ref{phaseseps} illustrates the mutual location of the phases, and Fig.~\ref{evolutioneps} schematically shows the time evolution of the drying process and growth of the deposit phase.

\begin{figure}
\begin{center}
\includegraphics{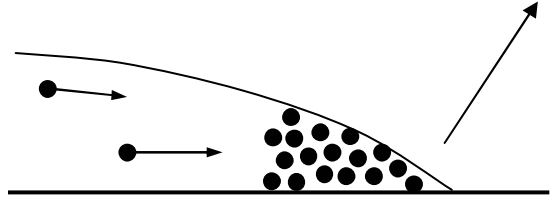}
\caption{Presence of particles in the deposit does not obstruct fluid evaporation at the edge of the drop.  All the necessary fluid is supplied, and it is this motion of the fluid that brings the particles to the deposit phase.  Also shown schematically is the fact that the boundary between the phases is vertical and the particles get stacked at full height between the substrate and the free surface of the drop.}
\label{suckseps}
\end{center}
\end{figure}

The evaporation rate depends only on the overall shape of the drop, and evaporation occurs in the same fashion from both phases.  We assume that evaporation is not influenced by any motion of the solute inside the drop, and the necessary amount of fluid can always be supplied to the regions of highest evaporation near the contact line.  Physically, high evaporation near the edge is what brings the solute to the contact line, and we assume that presence of the deposit does not obstruct the motion of fluid (Fig.~\ref{suckseps}).  Since the drop is thin and the contact angle is small, we can use expression~(\ref{evaprate-circular}) for the evaporation rate, which we used for the case of zero-volume particles, but with a slight modification --- now, the radius of the drop is $R_i$ (while $R(t)$ stands for the radius of the inner boundary of the deposit phase, so that $R_i = R(0)$):
\begin{equation}
J(r) = \frac{2}\pi \frac{D (n_s - n_\infty)}{\sqrt{R_i^2 - r^2}}.
\label{evaprate-finite}\end{equation}
Here $D$ the diffusion constant, and $n_s$ and $n_\infty$ are the saturated and ambient vapor densities, respectively.  The real situation may be different from the assumed above when $p$ is large or comparable to 1, and the edge of the area where evaporation occurs may be located near the boundary of the phases instead of the contact line.  However, for small initial concentrations of the solute, the main order result will be insensitive to the exact location of the singularity of the evaporation rate: whether it is located at the contact line or near the boundary of the phases.  We will further comment on this case of ``dry deposit'' when we obtain the full system of equations.

\begin{figure}
\begin{center}
\includegraphics{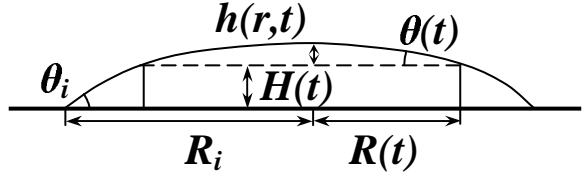}
\caption{Geometry of the problem.  Vertical scale is exaggerated in order to see the details, typically $H \ll R_i$ and $h \ll R_i$.}
\label{geometry-ring}
\end{center}
\end{figure}

The geometry of the problem is shown in Fig.~\ref{geometry-ring}.  The radius of the drop is $R_i$, the radius of the phase boundary is $R(t)$, and $R(0) = R_i$.  The height of the phase boundary is $H(t)$, and the initial condition is $H(0) = 0$.  We conveniently split the height of the free surface of the liquid phase into the sum of $H(t)$ and $h(r,t)$.  Since $H$ is independent of $r$, function $h(r)$ satisfies the Young-Laplace equation~(\ref{laplace}).  As is easy to see, the solution to this equation remains the same as in Chapter~2, and hence the shape of the upper part (above the dashed line in Fig.~\ref{geometry-ring}) is just a spherical cap.  Thus, we can use expression~(\ref{sphericalcap}) obtained for $h(r,t)$ earlier, which reduces to Eq.~(\ref{h-circular}) in limit of small contact angles:
\begin{equation}
h(r,t) = \frac{R^2(t) - r^2}{2R(t)} \theta(t).
\label{h-finite}\end{equation}
Here $\theta(t)$ is the angle between the liquid-air interface and the substrate at phase boundary (an equivalent of the contact angle).  Note that we do not assume that $\theta(t)$ and $h(r,t)$ are necessarily positive at all times: both can be negative at later drying stages, and the shape of the liquid-air interface may be concave.  Both convex and concave solutions for $h(r,t)$ are consistent with Eq.~(\ref{laplace}); the right hand side of this equation can have either sign.  By definition, both $\theta(t)$ and $h(r,t)$ are positive when the surface is convex (and therefore they are positive at the beginning of the drying process) and negative when the surface is concave.  The initial value of $\theta(t)$ coincides with the initial contact angle $\theta_i = \theta(0)$.

Clearly, there are three unknown functions of time in this geometry: $\theta(t)$, $R(t)$ and $H(t)$.  These three time dependences are to be determined in the subsequent sections of this chapter.  However, these quantities are not independent of each other.  If we assume that solute particles fill up the entire space between the substrate and the liquid-air interface when being brought to the phase boundary, it is easy to see that the three geometrical functions are related by the constraint
\begin{equation}
\frac{dH}{dt} = - \theta \frac{dR}{dt}.
\label{constraint}\end{equation}
Physically, this assumption means that the angles between the liquid-air interface and the substrate are identical on both sides of the phase boundary ($\theta = |dH/dR|$), and hence $h(r)$ and its first derivative are continuous past the phase boundary.  This condition was first introduced by Robert Deegan~\cite{deegan3, deegan4}.  Thus, there are actually only two {\em independent\/} functions of time, $\theta(t)$ and $R(t)$.  In the case of zero-volume particles we had only one independent function of time --- contact angle $\theta(t)$.

The geometrical definitions above allow us to find volume of each of the two phases.  Volume of the liquid phase is simply
\begin{equation}
V_L = 2\pi \int_0^{R(t)} \left( h(r,t) + H(t) \right) \, r dr = 2\pi \left( \frac{R^3 \theta}8 + \frac{R^2 H}2 \right).
\end{equation}
Taking into account relation~(\ref{constraint}), an infinitesimal variation of this volume can be related to the infinitesimal variations of $\theta$ and $R$:
\begin{equation}
dV_L = \frac{\pi R^4}4 d\left( \frac{\theta}R \right) + 2\pi H R dR.
\label{dvl}\end{equation}
The first term is responsible for the motion of the liquid-air interface, and the second term corresponds to the shift of the phase boundary.  It is also straightforward to obtain an expression for the differential of the volume of the deposit phase, which has only the term related to the inward motion of the phase boundary:
\begin{equation}
dV_D = - 2\pi H R dR.
\label{dvd}\end{equation}
We will use the last two expressions in the following section.  We will also adopt the notation that subscripts L and D refer to the liquid and deposit phases, respectively.

Having formulated assumptions and having stated the model and the geometry, we are now in position to write the governing equations.

\section{Main equations: conservation of mass}

If one were asked to express the essence of the entire theory in one sentence, this sentence would be: ``It is all about conservation of mass.''  Indeed, as we will see by the end of this section, all three governing equations obtained here represent conservation of mass (or volume) in one form or another.

We start from the global conservation of mass in the drop.  Since the amount of solute within the drop does not change, the change of the entire drop volume is equal to the change of the amount of fluid only.  This fluid gets evaporated from the surface of the drop, and the total amount evaporated per unit of time from the entire surface of the drop equals to the total change of fluid volume in the drop.  Hence the total change of drop volume is equal to volume of fluid evaporated from the surface:
\begin{equation}
\left. dV \right|_{tot} = \left. dV^F \right|_{surf}.
\label{aux1}\end{equation}
By convention, superscripts F and S refer to the fluid and the solute components, respectively (while subscripts L and D continue to denote phases).  Now, the total change of drop volume is the sum of volume changes of each phase:
\begin{equation}
\left. dV \right|_{tot} = dV_L + dV_D = \frac{\pi R^4}4 d\left( \frac{\theta}R \right),
\label{aux2}\end{equation}
where $dV_L$ and $dV_D$ were found in the preceding section [Eqs.~(\ref{dvl}) and (\ref{dvd})].  On the other hand, the volume of fluid evaporated from the surface of the drop during time $dt$ can be determined from
\begin{equation}
\left. \frac{dV^F}{dt} \right|_{surf} = - \int_0^{R_i} \frac{J(r)}\rho \sqrt{1 + (\partial_r h)^2} \, 2\pi r dr \approx - 2\pi \int_0^{R_i} \frac{J(r)}\rho \, r dr,
\end{equation}
where, as in the rest of this work, $\rho$ is the density of the fluid and we neglected the gradient of $h(r)$ with respect to unity (which is always legitimate for thin drops).  Simple integration with $J(r)$ of Eq.~(\ref{evaprate-finite}) yields
\begin{equation}
\left. dV^F \right|_{surf} = - \frac{4 D (n_s - n_\infty) R_i}\rho \, dt.
\label{aux3}\end{equation}
Combining Eqs.~(\ref{aux1}), (\ref{aux2}), and (\ref{aux3}), we finally obtain the first main differential equation of this chapter:
\begin{equation}
R^4 \frac{d}{dt}\left(\frac\theta{R}\right) = - \frac{16 D (n_s - n_\infty) R_i}{\pi\rho}.
\label{main1}\end{equation}
This equation represents the global conservation of mass in the drop and relates time dependencies of $\theta(t)$ and $R(t)$.

Our next equation represents the {\em local\/} conservation of mass and is analogous to Eq.~(\ref{consmass}).  Now, one has to take into account that there are two components in the liquid phase and write a separate equation for each component.  Since solute particles are carried along by the flow [a free particle of an appropriate size reaches the speed of the flow in about 50~ns in water under normal conditions~\cite{popov3}], the velocities of each component should be identical at each point within the liquid phase.  Conservation of fluid can be written in exact analogy with Eq.~(\ref{consmass}):
\begin{equation}
\bfnabla\cdot\left[(1 - \chi)(h + H)\bfv\right] + \frac{J}{\rho} + \partial_t \left[(1 - \chi)(h + H)\right] = 0,
\label{cons-fluid}\end{equation}
where we took into account that the surface height is now $(h + H)$ and neglected its gradient with respect to unity in the second term as we did everywhere in this work.  Similar equation can be written for the conservation of solute, but without the evaporation term:
\begin{equation}
\bfnabla\cdot\left[\chi(h + H)\bfv\right] + \partial_t \left[\chi(h + H)\right] = 0.
\label{cons-solute}\end{equation}
In the last two equations $\chi$ is the volume fraction of the solute at a given point within the liquid phase.  Adding the two equations and employing the linearity of differential operations, we obtain a direct analog of Eq.~(\ref{consmass}):
\begin{equation}
\bfnabla\cdot[(h + H)\bfv] + \frac{J}{\rho} + \partial_t (h + H) = 0.
\label{cons-vol}\end{equation}
This equation could have been obtained if we considered only one component with volume fraction 1 in the liquid phase, and this equivalence should not be of any surprise: we explicitly assumed that the solute particles move in exactly the same fashion as the fluid does, and hence any differentiation between the two (from the point of view of the conservation of volume) is completely lost.  Note that if evaporation were too intensive, this equivalence would not hold, as there might be an insufficient amount of fluid coming into a volume element and the solution could get completely dry ({\em i.e.}\ only the solute component would be left).  We implicitly assume this is not the case for our liquid phase where the solute fraction is relatively small and the evaporation is not too strong.

In circular geometry Eq.~(\ref{cons-vol}) can now be resolved with respect to the flow velocity, as was done in Eq.~(\ref{vdef-circular}):
\begin{equation}
v_r(r,t) = - \frac{1}{r (h + H)} \int _0^r \left( \frac{J}\rho + \partial_t h + \partial_t H \right) \, r dr.
\end{equation}
Straightforward integration with $J(r)$ of Eq.~(\ref{evaprate-finite}), $h(r,t)$ of Eq.~(\ref{h-finite}), and $dH/dt$ of Eq.~(\ref{constraint}) and employment of Eq.~(\ref{main1}) for $d(\theta/R)/dt$ yield
\begin{equation}
v_r(r,t) = \frac{2 D (n_s - n_\infty)}{\pi\rho} \frac{R_i}r \frac{\sqrt{1 - \left( \frac{r}{R_i} \right)^2} - \left[ 1 - \left( \frac{r}R \right)^2 \right]^2}{\frac{R \theta}2 \left[ 1 - \left( \frac{r}R \right)^2 \right] + H}.
\label{vr-finite}\end{equation}
This is an analog to Eq.~(\ref{v-circular}); however, in this expression the velocity does not diverge at $r = R$, but rather reaches a finite value (which is quite natural as the evaporation rate does {\em not\/} diverge at $r = R$).  This equation is a direct consequence of the local conservation of mass.

As we did in the preceding chapters, we label $r_i(t)$ the initial location of the solute particles that reach the phase boundary (and become part of the deposit ring) at time $t$.  This function is monotonically decreasing, as the solute particles from the outer areas of the drop reach the deposit phase sooner than the particles from the inner areas do.  The derivative of this function is related to the velocity found in the preceding paragraph in a very simple fashion:
\begin{equation}
\frac{dr_i}{dt} = - v_r(r_i,t).
\end{equation}
We explained the origin of this relation in Chapter~2.  Thus, this equation and the result for the flow velocity of the preceding paragraph yield the second principal equation of this section:
\begin{equation}
\frac{dr_i}{dt} = - \frac{2 D (n_s - n_\infty)}{\pi\rho} \frac{R_i}{r_i} \frac{\sqrt{1 - \left( \frac{r_i}{R_i} \right)^2} - \left[ 1 - \left( \frac{r_i}R \right)^2 \right]^2}{\frac{R \theta}2 \left[ 1 - \left( \frac{r_i}R \right)^2 \right] + H}.
\label{main2}\end{equation}
This equation relates $r_i(t)$ to time dependencies of the geometrical parameters of the drop [$\theta(t)$, $R(t)$, and $H(t)$].

The volume of the solute in the deposit phase $V_D^S$ at time $t$ is equal to the volume of the solute located outside the circle of radius $r_i(t)$ at time 0 (since all the solute between $r_i(t)$ and $R_i$ becomes part of the deposit by time $t$).  The latter volume can be found similarly to Eqs.~(\ref{mass-def}) and (\ref{mass-aux}) (which are written for the mass instead of the volume) and hence
\begin{equation}
V_D^S = \chi_i \int_{r_i}^{R_i} h(r,0) \, 2\pi r dr = V^S \left[ 1 - \left( \frac{r_i}{R_i} \right)^2 \right]^2,
\label{vds}\end{equation}
where $V^S = \pi \chi_i R_i^3 \theta_i / 4$ is the total volume of the solute in the drop.  On the other hand, the volume of the solute in the deposit phase is just the constant fraction $p$ of the volume of the entire deposit phase:
\begin{equation}
V_D^S = p V_D.
\end{equation}
Equating the right-hand sides of these two equations, taking the time derivatives of both sides, and making use of the already determined $dV_D$ of Eq.~(\ref{dvd}), we obtain the third principal equation of this section:
\begin{equation}
\chi_i R_i^3 \theta_i \left[ 1 - \left( \frac{r_i}{R_i} \right)^2 \right] \frac{d}{dt} \left[ 1 - \left( \frac{r_i}{R_i} \right)^2 \right] = - 4 p H R \frac{dR}{dt}.
\label{main3}\end{equation}
As the other two principal equations~(\ref{main1}) and (\ref{main2}) are, this equation is a consequence of the conservation of mass.  While Eq.~(\ref{main1}) comes from the global conservation of the fluid, this equation represents the global conservation of the solute.

Thus, we have four unknown functions of time: $\theta(t)$, $R(t)$, $H(t)$, and $r_i(t)$, and four independent differential equations for these functions: Eqs.~(\ref{constraint}), (\ref{main1}), (\ref{main2}), and (\ref{main3}).  In reality we need only three of these functions: $\theta(t)$, $R(t)$, and $H(t)$; however, there is no simple way to eliminate $r_i(t)$ from the full system of equations and reduce the number of equations.  Having solved this system of the four differential equations, we will be able to fully characterize the dimensions of the deposit phase and describe the evolution of the deposit ring.  The following sections are devoted to the details of this solution.

Here we will only comment on how this system of equations would change if we were to consider the case of the completely dry solute $p = 1$.  In this case there would be no evaporation from the surface of the deposit phase, and the effective edge of the evaporating area would be somewhere in the vicinity of the phase boundary.  Assuming the same one-over-the-square-root divergence of the evaporation rate at $r = R$ instead of $r = R_i$ [which mathematically means substitution of $R$ in place of $R_i$ in Eq.~(\ref{evaprate-finite})] and conducting a derivation along the lines of this section, one can obtain a very similar system of four differential equations.  These equations would be different from Eqs.~(\ref{constraint}), (\ref{main1}), (\ref{main2}), and (\ref{main3}) in only two minor details.  First, Eqs.~(\ref{main1}) and (\ref{main2}) would lose all indices $i$ at all occasions of $R_i$ ({\em i.e.}\ one should substitute $R$ for all $R_i$ in both equations).  Second, $p$ should be set to 1 in Eq.~(\ref{main3}).  Apart from these details, the two systems would be identical.  As we will see in the following section, this difference between the two systems is not important in the main order in a small parameter introduced below, and thus this ``dry-solute'' case does not require any special treatment contrary to the intuitive prudence.

\section{Analytical results in the limit of small initial concentrations of the solute}

So far we have not introduced any small parameters in this problem other than the initial contact angle $\theta_i \ll 1$.  In particular, equations~(\ref{constraint}), (\ref{main1}), (\ref{main2}), and (\ref{main3}) were obtained without assuming any relation between $p$ and $\chi_i$ other than the non-restrictive condition $\chi_i < p$.  In order to find the analytical solution to this system, we will have to assume that $\chi_i \ll p$.  Then, we will solve the same system of differential equations numerically for an arbitrary relation between $\chi_i$ and $p$.

Assumption $\chi_i \ll p$ physically means that the solute concentration in the liquid phase is small --- it is much smaller than the concentration of close packing or any other comparable number of the order of 1.  This is the case for most practical realizations of the ring deposits in experiments and observations: the solute concentration rarely exceeds 10\% of volume, and in most cases it is far lower.  If the volume fraction of the solute in the drop is small, then the solute volume is also small compared to the volume of the entire drop.  Hence, the deposit phase, which consists mostly of the solute and whose volume is of the order of magnitude of the volume of the entire solute in the drop, must have small volume compared to the volume of the entire drop.  Thus, if the initial volume fraction of the solute in the drop is small, then the dimensions of the deposit ring must be small compared to the corresponding dimensions of the entire drop ({\em i.e.}\ the ring width must be much smaller than the drop radius and the ring height must be much smaller than the drop height).

Let us now introduce parameter $\epsilon$ that is small when $\chi_i/p$ is small.  However, we do not fix its functional dependence on $\chi_i/p$ for the moment:
\begin{equation}
\epsilon = f\left(\frac{\chi_i}p\right) \ll 1,
\label{epsilon-def}\end{equation}
where $f$ is an arbitrary increasing function of its argument.  Then we postulate that the ring width is proportional to this parameter:
\begin{equation}
R(t) = R_i \left[ 1 - \epsilon \tilde W(t) \right],
\label{tilde-w-def}\end{equation}
where $\tilde W(t)$ is an arbitrary dimensionless function and we explicitly introduced the dimensionality via $R_i$.  Obviously, $\tilde W(0) = 0$.  So far we have not done anything but writing mathematically that the ring width is small whenever the initial volume fraction of the solute is small ({\em i.e.}\ the statement of the preceding paragraph).  Next, we introduce a dimensionless variable for the angle $\theta(t)$:
\begin{equation}
\theta(t) = \theta_i \tilde\theta(t),
\label{tilde-theta-def}\end{equation}
where both $\theta(t)$ and $\theta_i$ are small everywhere in this chapter, while the newly introduced function $\tilde\theta(t)$ is arbitrary (in particular, $\tilde\theta(0) = 1$).  From the definitions above and from the geometrical constraint~(\ref{constraint}), it is straightforward to conclude that for small initial concentrations of the solute the height of the ring $H(t)$ must be linear in small parameters $\epsilon$ and $\theta_i$ and directly proportional to the only dimensional scale $R_i$:
\begin{equation}
H(t) = \epsilon \theta_i R_i \tilde H(t),
\label{tilde-h-def}\end{equation}
where $\tilde H(t)$ is a dimensionless function of time ($\tilde H(0) = 0$).  Equation~(\ref{constraint}) fixes the relation between this function and the previously introduced functions $\tilde W(t)$ and $\tilde\theta(t)$:
\begin{equation}
\frac{d\tilde H}{dt} = \tilde\theta \frac{d\tilde W}{dt}.
\label{cons-dim-less}\end{equation}
Finally, we introduce the last dimensionless variable in place of the fourth unknown function $r_i(t)$:
\begin{equation}
\tilde V(t) = 1 - \left( \frac{r_i(t)}{R_i} \right)^2,
\label{tilde-v-def}\end{equation}
with the initial condition $\tilde V(0) = 0$.  So far, we simply introduced four new dimensionless variables in place of the four original unknown functions of time and explicitly separated dependence of these unknown functions on small parameters $\epsilon$ and $\theta_i$.  As a final step of this procedure, we introduce the dimensionless time $\tau$:
\begin{equation}
\tau = \frac{t}{t_f},
\label{tau}\end{equation}
where $t_f$ is a combination of system parameters with the dimensionality of time:
\begin{equation}
t_f = \frac{\pi \rho R_i^2 \theta_i}{16 D (n_s - n_\infty)}.
\label{t-f}\end{equation}
In the limit $\chi_i/p \to 0$ this combination represents the time at which all the solute reached the deposit phase; for finite $\chi_i/p$ it does not have so simple interpretation.

Substitution of all the definitions for the dimensionless variables of the preceding paragraph into the original system of equations~(\ref{constraint}), (\ref{main1}), (\ref{main2}), and (\ref{main3}) and retention of only the leading and the first correctional terms in $\epsilon$ lead to the following simplified system of equations:
\begin{equation}
\frac{d\tilde H}{d\tau} = \tilde\theta \frac{d\tilde W}{d\tau},
\label{dimless1}\end{equation}
\begin{equation}
\frac{d\tilde\theta}{d\tau} + \epsilon \tilde\theta \frac{d\tilde W}{d\tau} - 3 \epsilon \tilde W \frac{d\tilde\theta}{d\tau} = -1,
\label{dimless2}\end{equation}
\begin{equation}
\frac{d\tilde V}{d\tau} = \frac{\sqrt{\tilde V} - \tilde V^2 \left[ 1 - 4 \epsilon \tilde W \left(\tilde V^{-1} - 1\right) \right]}{2 \tilde\theta \tilde V \left[ 1 - \epsilon \tilde W \left(2\tilde V^{-1} - 1\right) \right] + 4 \epsilon \tilde H},
\label{dimless3}\end{equation}
\begin{equation}
\frac{\chi_i}p \tilde V \frac{d\tilde V}{d\tau} = 4 \epsilon^2 \tilde H \frac{d\tilde W}{d\tau} \left(1 - \epsilon \tilde W\right).
\label{dimless4}\end{equation}
These differential equations are still coupled, but the coupling is simpler than in the original system.  As is apparent from the last equation, parameter $\epsilon^2$ must be proportional to $\chi_i/p$.  Since the separation of the ring width into $\epsilon$ and $\tilde W$ in Eq.~(\ref{tilde-w-def}) is absolutely arbitrary, parameter $\epsilon$ is defined up to a constant multiplicative factor.  Therefore, we {\em set\/} this factor in such a way that $\epsilon^2$ is {\em equal\/} to $\chi_i/p$:
\begin{equation}
\epsilon = \sqrt{\frac{\chi_i}p}.
\end{equation}
This fixes the function $f$ from the original definition~(\ref{epsilon-def}).

Next, we expand all four unknown functions of $\tau$ in small parameter $\epsilon$:
\begin{equation}
\tilde\theta = \tilde\theta_0 + \epsilon \tilde\theta_1 + \cdots,
\end{equation}
\begin{equation}
\tilde V = \tilde V_0 + \epsilon \tilde V_1 + \cdots,
\end{equation}
\begin{equation}
\tilde H = \tilde H_0 + \epsilon \tilde H_1 + \cdots,
\label{height-expansion}\end{equation}
\begin{equation}
\tilde W = \tilde W_0 + \epsilon \tilde W_1 + \cdots,
\label{width-expansion}\end{equation}
and keep only the main-order terms after substitution into system~(\ref{dimless1})--(\ref{dimless4}).  The resulting main-order system of differential equations becomes very simple:
\begin{equation}
\frac{d\tilde H_0}{d\tau} = \tilde\theta_0 \frac{d\tilde W_0}{d\tau},
\label{main-order-1}\end{equation}
\begin{equation}
\frac{d\tilde\theta_0}{d\tau} = -1,
\label{main-order-2}\end{equation}
\begin{equation}
\frac{d\tilde V_0}{d\tau} = \frac{\sqrt{\tilde V_0} - \tilde V_0^2}{2 \tilde\theta_0 \tilde V_0},
\label{main-order-3}\end{equation}
\begin{equation}
\tilde V_0 \frac{d\tilde V_0}{d\tau} = 4 \tilde H_0 \frac{d\tilde W_0}{d\tau}.
\label{main-order-4}\end{equation}
Clearly, equations {\em decouple\/}: the second equation can be solved with respect to $\tilde\theta_0(\tau)$ independently of all the others, then the third equation can be solved with respect to $\tilde V_0(\tau)$ independently of the first and the fourth, and finally the first and the fourth equations can be solved together as well (since $d\tilde W_0 / d\tau$ can be eliminated from the two and the result can be solved with respect to $\tilde H_0(\tau)$).  Thus, it is a matter of technical effort and time to obtain the following solution to the system of equations above with the appropriate initial conditions:
\begin{equation}
\tilde\theta_0(\tau) = 1 - \tau,
\label{tilde-theta-res}\end{equation}
\begin{equation}
\tilde V_0(\tau) = \left[1 - (1 - \tau)^{3/4}\right]^{2/3},
\label{tilde-v-res}\end{equation}
\begin{equation}
\tilde H_0(\tau) = \sqrt{\frac{1}3 \left[ {\rm B}\left(\frac{7}3,\frac{4}3\right) - {\rm B}_{(1 - \tau)^{3/4}}\left(\frac{7}3,\frac{4}3\right) \right]},
\label{tilde-h-res}\end{equation}
\begin{equation}
\tilde W_0(\tau) = \int_0^\tau \frac{1}{8 \tilde H_0(\tau')} \frac{\left[ 1 - (1 - \tau')^{3/4} \right]^{1/3}}{(1 - \tau')^{1/4}} \, d\tau'.
\label{tilde-w-res}\end{equation}
Here ${\rm B}(a,b) = \int_0^1 x^{a-1} (1-x)^{b-1} \, dx$ is the complete beta-function, ${\rm B}_z(a,b) = \int_0^z x^{a-1} (1-x)^{b-1} \, dx$ is the incomplete beta-function ($a > 0$, $b > 0$, and $0 \le z \le 1$), and the integral in the last equation cannot be expressed in terms of the standard elementary or special functions.  In a similar fashion, systems of equations of the first and higher orders in $\epsilon$ can be written, however, they cannot be resolved analytically as easily as the main-order system above.
{\sloppy

}
A system of equations similar to our system~(\ref{dimless1})--(\ref{dimless4}) was presented by Robert Deegan in works~\cite{deegan3, deegan4}.  Small concentrations {\em are\/} assumed in those works, but no general system of equations similar to our Eqs.~(\ref{constraint}), (\ref{main1}), (\ref{main2}), and (\ref{main3}) is written {\em before\/} introducing the small parameter and expanding in it.  The equations in that system mix terms of the main and the correctional order in concentration arbitrarily.  Upon examination, that system does include all the main-order terms of our system; however, the first-order corrections are incomplete, and most corrections are missing.  Moreover, no analysis quantifying how small the neglected and the retained ``small corrections'' are is made in Refs.~\cite{deegan3, deegan4}, and therefore the importance of different terms is difficult to infer from that system.  Equations are not written as a single system of four differential equations for four unknown functions; instead, equations for two variables are solved before even writing the other two equations, thereby undermining the equality of all the four functions of time that must be determined {\em simultaneously\/} and making it impossible to compare the neglected terms.  Finally, no analytical solution to the system of equations is provided in Refs.~\cite{deegan3, deegan4}; the author tackles the system numerically.  The numerical results are not presented explicitly either; they are only used to convert the data between the experimental graphs.  All these deficiencies are rectified in our approach above.  The analytical solution~(\ref{tilde-theta-res})--(\ref{tilde-w-res}) is reported here for the first time.  Higher-order terms can also be constructed (with sufficient labor invested), and the procedure can be conducted up to an arbitrary order.

\begin{figure}
\begin{center}
\includegraphics{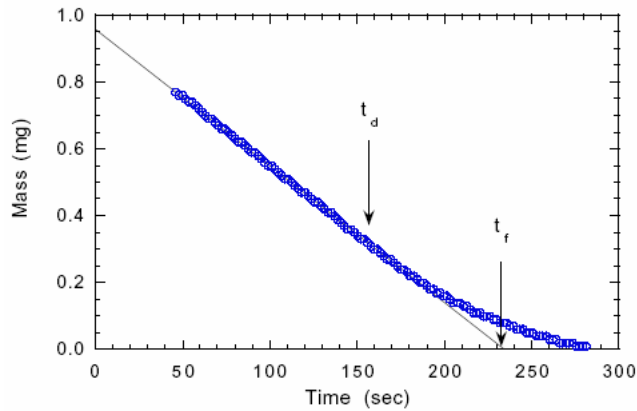}
\caption{Mass of a drying drop as a function of time.  Experimental results, after Refs.~\cite{deegan3, deegan4}.  The line running through the data is a linear fit.  (Courtesy Robert Deegan.)}
\label{thetatimeeps}
\end{center}
\end{figure}

How do our results~(\ref{tilde-theta-res})--(\ref{tilde-w-res}) translate into the original variables?  The first two of them [Eqs.~(\ref{tilde-theta-res}) and (\ref{tilde-v-res})] reproduce the earlier results obtained for the zero-volume deposit.  In terms of the original (dimensional) variables Eq.~(\ref{tilde-theta-res}) represents the linear decrease of the angle between the liquid-air interface and the substrate with time:
\begin{equation}
\theta(t) = \theta_i \left(1 - \frac{t}{t_f}\right).
\label{theta-result}\end{equation}
Taking into account the definition of $t_f$ [Eq.~(\ref{t-f})], one can see that this is a direct analog of Eq.~(\ref{theta-circular}) for the contact angle in the zero-volume-particle case.  So, angle $\theta$ in the finite-volume case depends on time in exactly the same fashion as the contact angle in the zero-volume case does.  This expression also provides an interpretation of $t_f$: it is the time at which the free surface of the liquid phase becomes flat.  Before time $t_f$ this surface is convex, after $t_f$ it becomes concave and bows inward (until it touches the substrate).  Thus, $t_f$ is {\em not\/} the total drying time, but rather the time at which $\theta$ becomes zero.  In the limit $\chi_i/p \to 0$ or equivalently $\epsilon \to 0$ the height of the deposit is going to zero and the two times are the same.  For finite values of both parameters the total drying time is longer than the time at which the liquid-air interface becomes flat.  Eq.~(\ref{theta-result}) has been verified in experiments~\cite{deegan3, deegan4} where the mass of the drop was measured as a function of time (Fig.~\ref{thetatimeeps}).  Since the mass of the drop is directly proportional to $\theta$ these results confirm the linearity of $\theta(t)$ during most of the drying process.

\begin{figure}
\begin{center}
\includegraphics{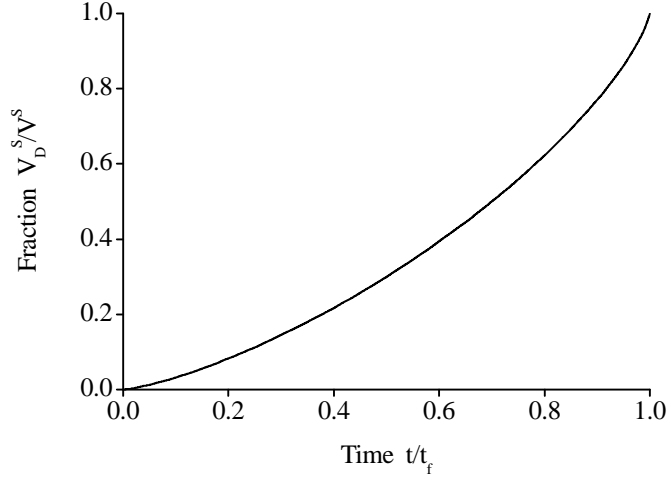}
\caption{Fraction of the solute in the deposit phase $V_D^S / V^S$ as a function of time [Eq.~(\ref{fraction-result})].}
\label{fractiontimeeps}
\end{center}
\end{figure}

The second equation~(\ref{tilde-v-res}) also has a direct analog in the zero-volume case.  In the original variables it can be rewritten as
\begin{equation}
\left( 1 - \frac{t}{t_f} \right)^{3/4} + \left[ 1 - \left(\frac{r_i(t)}{R_i}\right)^2 \right]^{3/2} = 1,
\label{ri-result}\end{equation}
which is identical to Eq.~(\ref{time-circular}).  Clearly, $r_i = 0$ when $t = t_f$.  Thus, $t_f$ can also be interpreted as the time at which all the solute particles become part of the deposit.  The same interpretation can be obtained differently: according to Eqs.~(\ref{vds}), (\ref{tilde-v-def}), and (\ref{tilde-v-res}), the fraction of the solute particles in the deposit phase $V_D^S / V^S$ is simply
\begin{equation}
\frac{V_D^S}{V^S} = \tilde V_0^2 = \left[1 - \left( 1 - \frac{t}{t_f} \right)^{3/4} \right]^{4/3}
\label{fraction-result}\end{equation}
(plotted in Fig.~\ref{fractiontimeeps}).  This fraction becomes 1 when $t = t_f$, and thus all the solute particles reach the deposit by time $t_f$.  So far, the results of this finite-volume model coincide with the results of the zero-volume case considered earlier.

\begin{figure}
\begin{center}
\includegraphics{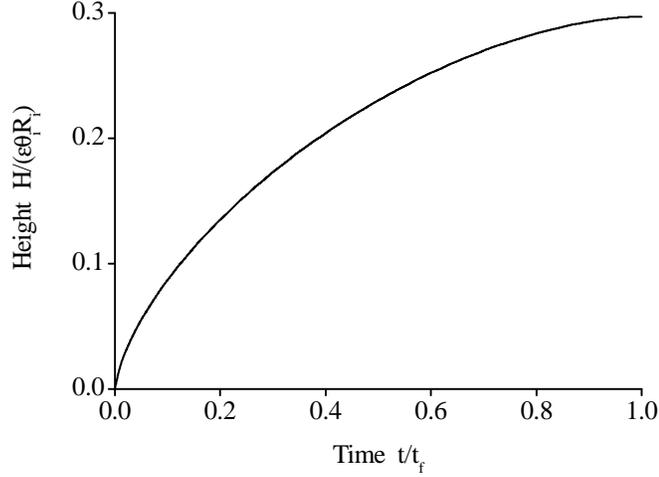}
\caption{Height of the phase boundary $H(t)$ in units of $\theta_i R_i \sqrt{\chi_i/p}$ as a function of time [Eq.~(\ref{height-result})].}
\label{heighttimeeps}
\end{center}
\end{figure}

\begin{figure}
\begin{center}
\includegraphics{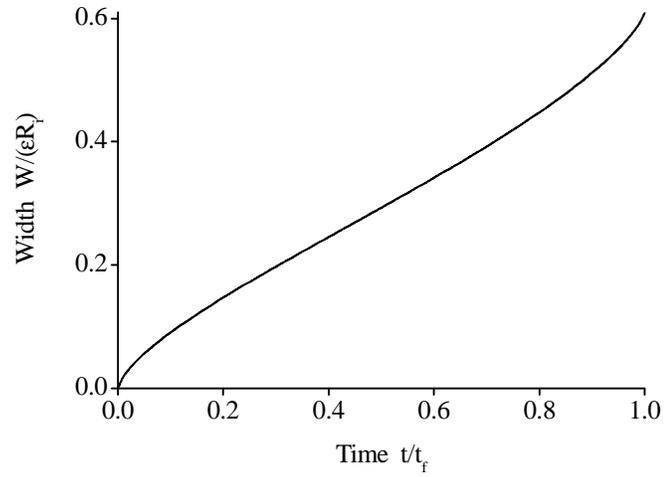}
\caption{Width of the deposit ring $W(t)$ in units of $R_i \sqrt{\chi_i/p}$ as a function of time [Eq.~(\ref{width-result})].}
\label{widthtimeeps}
\end{center}
\end{figure}

\begin{figure}
\begin{center}
\includegraphics{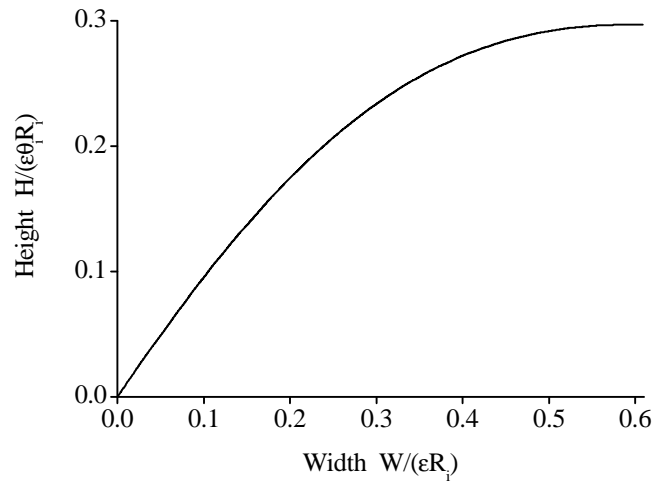}
\caption{Deposit ring profile: height {\em vs.}\ width.  The vertical scale is different from the horizontal scale by a factor of $\theta_i \ll 1$.}
\label{heightwidtheps}
\end{center}
\end{figure}

However, the third and the fourth equations [Eqs.~(\ref{tilde-h-res})--(\ref{tilde-w-res})] provide completely new results.  In the dimensional variables they yield the height of the phase boundary $H$ and the width of the deposit ring $W \equiv R_i - R$, respectively:
\begin{equation}
H(t) = \sqrt{\frac{\chi_i}p} \theta_i R_i \tilde H_0\left(\frac{t}{t_f}\right),
\label{height-result}\end{equation}
\begin{equation}
W(t) = \sqrt{\frac{\chi_i}p} R_i \tilde W_0\left(\frac{t}{t_f}\right),
\label{width-result}\end{equation}
where functions $\tilde H_0(\tau)$ and $\tilde W_0(\tau)$ are given by Eqs.~(\ref{tilde-h-res}) and (\ref{tilde-w-res}) and plotted in Figs.~\ref{heighttimeeps} and \ref{widthtimeeps}.  (Note that these expressions represent only the leading term in the expansions of the full functions for height and width in small parameter $\epsilon = \sqrt{\chi_i/p}$.)  These results provide the sought dependence of the geometrical characteristics of the deposit ring on all the physical parameters of interest: on the initial geometry of the drying drop ($R_i$ and $\theta_i$), on the initial concentration of the solute ($\chi_i$), and on the time elapsed since the beginning of the drying process ($t$).  If the time is considered as a parameter, they also provide all the necessary information to obtain the geometrical profile of the deposit ({\em i.e.}\ the dependence of height on width), which we plot in Fig.~\ref{heightwidtheps}.  Note that the vertical scale of this plot is highly expanded compared to the horizontal scale since there is an extra factor of $\theta_i \ll 1$ in the expression for height; in the actual scale the height is multiplied by $\theta_i$ (in addition to the same scaling parameters as in the width) and hence is much smaller than it appears in Fig.~\ref{heightwidtheps}.

The time dependence of $H$ and $W$ plotted in Figs.~\ref{heighttimeeps} and \ref{widthtimeeps} deserves a brief discussion. It is straightforward to obtain the asymptotics of $\tilde H_0(\tau)$ and $\tilde W_0(\tau)$ for small values of the argument and for values around 1 ({\em i.e.}\ for the early and the late drying stages).  At early times, both the height and the width scale with the drying time as a power law with exponent $2/3$:
\begin{equation}
H \approx \sqrt{\frac{\chi_i}p} \theta_i R_i \frac{(3 \tau)^{2/3}}{2^{7/3}} \left[ 1 + O(\tau) \right]\qquad\qquad\left(\tau = \frac{t}{t_f} \ll 1\right),
\end{equation}
\begin{equation}
W \approx \sqrt{\frac{\chi_i}p} R_i \frac{(3 \tau)^{2/3}}{2^{7/3}} \left[ 1 + O(\tau) \right]\qquad\qquad\left(\tau = \frac{t}{t_f} \ll 1\right).
\end{equation}
Thus, at early times $H \approx \theta_i W$, which can also be deduced directly from Eq.~(\ref{constraint}) without obtaining the complete solution above.  At the end of the drying process, the height and the width approach finite values (which, apart from the dimensional scales, are universal, {\em i.e.}\ simply numbers), and do so as power laws of $(t_f - t)$ with two different exponents:
\begin{equation}
H \approx \sqrt{\frac{\chi_i}p} \theta_i R_i \left[ \tilde H_0(1) - \frac{(1 - \tau)^{7/4}}{14 \tilde H_0(1)} + O(1 - \tau)^{5/2} \right]\qquad\quad\left(1 - \tau = 1 - \frac{t}{t_f} \ll 1\right),
\end{equation}
\begin{equation}
W \approx \sqrt{\frac{\chi_i}p} R_i \left[ \tilde W_0(1) - \frac{(1 - \tau)^{3/4}}{6 \tilde H_0(1)} + O(1 - \tau)^{3/2} \right]\qquad\quad\left(1 - \tau = 1 - \frac{t}{t_f} \ll 1\right),
\end{equation}
where $\tilde H_0(1)$ and $\tilde W_0(1)$ are just numbers:
\begin{equation}
\tilde H_0(1) = \sqrt{\frac{1}3 {\rm B}\left(\frac{7}3,\frac{4}3\right)} \approx 0.297,
\end{equation}
\begin{equation}
\tilde W_0(1) = \int_0^1 \frac{1}{8 \tilde H_0(\tau)} \frac{\left[ 1 - (1 - \tau)^{3/4} \right]^{1/3}}{(1 - \tau)^{1/4}} \, d\tau \approx 0.609.
\end{equation}
Clearly, $dH/dW = \theta_i (1 - \tau)$ and hence vanishes when $\tau \to 1$.  This fact can also be observed in flattening of the graph in Fig.~\ref{heightwidtheps} at late times.

\begin{figure}
\begin{center}
\includegraphics{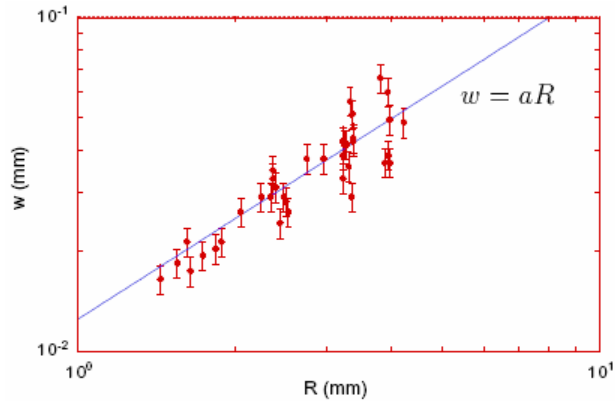}
\caption{Ring width {\em vs.}\ drop radius.  Experimental results, after Refs.~\cite{deegan3, deegan4}.  The line running through the data is a linear fit.  (Courtesy Robert Deegan.)}
\label{widthradiuseps}
\end{center}
\end{figure}

Dependence of the height and the width on the radius of the drop $R_i$, while intuitively obvious (since $R_i$ is the only scale in this problem with the dimensionality of the length), has been verified in experiments~\cite{deegan3, deegan4}.  A linear fit has been obtained for the dependence of the ring width on the radius (Fig.~\ref{widthradiuseps}), which agrees with our findings.

\begin{figure}
\begin{center}
\includegraphics{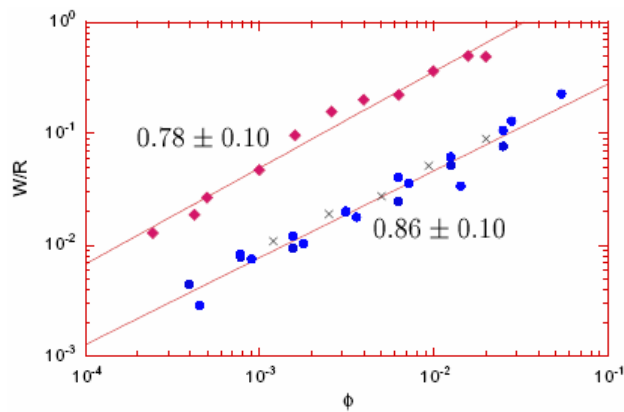}
\caption{Ring width normalized by the drop radius {\em vs.}\ initial concentration of the solute for two different particle sizes.  Experimental results, after Refs.~\cite{deegan3, deegan4}.  The two data sets are offset by a factor of 5 to avoid mixing of the data points related to the different particle sizes.  The lines running through the data are linear fits in the double-logarithmic scale, which upon conversion to the linear scale yield power laws with exponents $0.78 \pm 0.10$ and $0.86 \pm 0.10$.  (Courtesy Robert Deegan.)}
\label{widthconceps}
\end{center}
\end{figure}

\begin{figure}
\begin{center}
\includegraphics{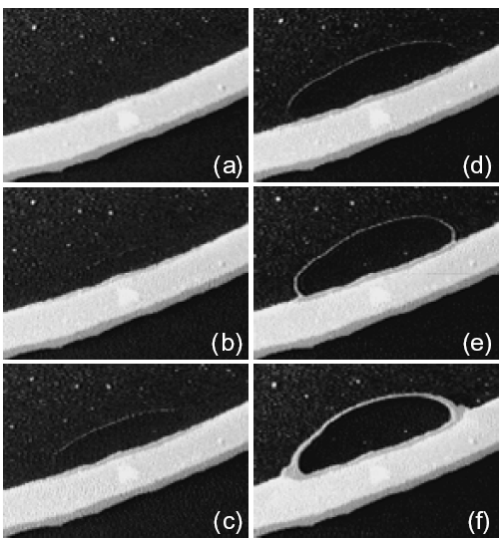}
\caption{A photographic sequence demonstrating a depinning event.  Experimental results, after Refs.~\cite{deegan3, deegan4}.  The view is from above, and the solid white band in the lower part of the frame is the ring; the rest of the drop is above the ring.  The time between the first and the last frames is approximately 6~s; the major axis of the hole is approximately 150~$\mu$m.  (Courtesy Robert Deegan.)}
\label{depinning}
\end{center}
\end{figure}

\begin{figure}
\begin{center}
\includegraphics{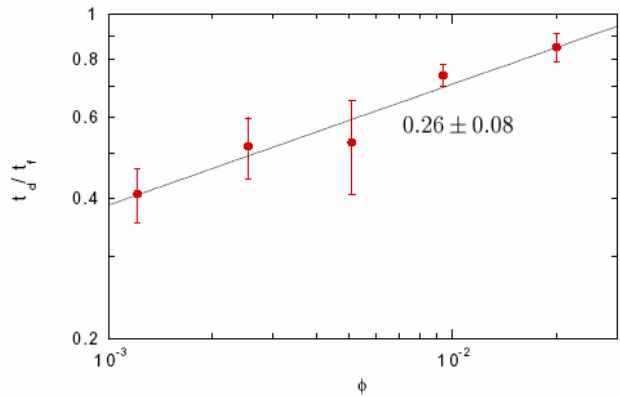}
\caption{Depinning time normalized by the extrapolated drying time {\em vs.}\ initial concentration of the solute.  Experimental results, after Refs.~\cite{deegan3, deegan4}.  The line running through the data is a linear fit in the double-logarithmic scale, which upon conversion to the linear scale yields a power law with exponent $0.26 \pm 0.08$.  (Courtesy Robert Deegan.)}
\label{timeconceps}
\end{center}
\end{figure}

Comparison to the experimental data for the dependence on the initial concentration of the solute is slightly less trivial.  Our results predict that both height $H$ and width $W$ scale with the initial concentration as $\chi_i^{1/2}$ (at least, in the leading order for small concentrations).  However, experimental results by Deegan~\cite{deegan3, deegan4} show a different exponent of $\chi_i$.  Exponent of $\chi_i$ was determined to be $0.78 \pm 0.10$ and $0.86 \pm 0.10$ for two different particles sizes (Fig.~\ref{widthconceps}).  Why is the difference?  The answer lies in the fact that the width measured in experiments~\cite{deegan3, deegan4} is not the full width of the ring at the end of the drying process, but rather the width of the ring at {\em depinning}.  Depinning is a process of detachment of the liquid phase from the deposit ring (Fig.~\ref{depinning}).  This detachment was observed experimentally in colloidal suspensions but has not been explained in full theoretically yet.\footnote{While the full explanation is yet to be developed, the naive reason for depinning seems relatively straightforward.  The pinning force depends only on the materials involved and is relatively insensitive to the value of the contact angle.  At the same time, the depinning force is simply the surface tension, which is directed along the liquid-air interface and which increases as the contact angle decreases (since only the horizontal component of this force is important).  Thus, the relatively constant pinning force cannot compensate for the increasing depinning force of the surface tension, and after the contact angle decreases past some threshold, the depinning force wins and causes detachment of the liquid phase from the deposit.}  An important observation, however, is that the depinning time ({\em i.e.}\ the time at which the detachment occurs and the ring stops growing) depends on the initial concentration of the solute.  This dependence was also measured by Deegan (Fig.~\ref{timeconceps}).  The resulting exponent was determined to be $0.26 \pm 0.08$.  Thus, the width of the ring at depinning $W_d$ scales with the initial concentration of the solute $\chi_i$ as
\begin{equation}
W_d \propto \chi_i^{1/2} \tilde W_0\left(\frac{t_d}{t_f}\right) \propto \chi_i^{0.5} \tilde W_0\left(\chi_i^{0.26 \pm 0.08}\right),
\end{equation}
where $t_d$ is the depinning time ($t_d / t_f \propto \chi_i^{0.26 \pm 0.08}$).  As is apparent from Fig.~\ref{timeconceps}, the typical values of the depinning time are of the order of 0.4--0.8~$t_f$.  In this time range, function $\tilde W_0$ is virtually linear (Fig.~\ref{widthtimeeps}).  Therefore, the dependence of $W_d$ on $\chi_i$ has the overall exponent of the order of $0.5 + (0.26 \pm 0.08) = 0.76 \pm 0.08$.  It is now clear that both experimental results $0.78 \pm 0.10$ and $0.86 \pm 0.10$ fall within the range of the experimental uncertainty of the approximate predicted value $0.76 \pm 0.08$.  All in all, it turns out the theoretical dependence of the ring width on the initial concentration agrees with the experimental results very well.

Note that Robert Deegan~\cite{deegan3, deegan4} did {\em not\/} report direct measurements of the height of the deposit.  He {\em calculated\/} the height from the data in hand, and thus direct comparison to the experimental data is not available for the height [and comparison between the plots of Refs.~\cite{deegan3, deegan4} and our predictions depends on the details of Deegan's calculation and its inherent assumptions].

The square-root dependence of the height and the width on the concentration is in good agreement with general physical expectations.  Indeed, the volume of the deposit ring is roughly proportional to the product of the height and the width.  On the other hand, the height is of the same order of magnitude as the width since the ratio of the two is of the order of $\theta_i$ (which is a constant).  Thus, both the height and the width scale approximately as a square root of the ring volume.  Finally, the volume of the deposit ring is proportional to the initial volume fraction of the solute: the more solute is present initially, the larger the volume of the deposit ring is at the end (since the ring is comprised of only the initial amount of the solute).  Therefore, both the height and the width must scale as a square root of the initial volume fraction.  It is rewarding that the results of our complex calculation match the results of this simple physical argument.
 
Thus, the complete analytical solution to our model is available in the limit $\chi_i / p \to 0$, and this solution compares quite well to the experimental results.  Since the main-order solution in $\chi_i / p$ is perfectly adequate, the difference between the original system of equations and the one for the ``completely dry'' case is not important: in the main order in $\chi_i / p$ the results are identical for both cases (since one case in different from the other only by presence of $R$ instead of $R_i$ in a few places in the main equations, and this difference is of the correctional order in $\chi_i / p$). 

\section{Numerical results for arbitrary initial concentrations of the solute}

Apart from approaching the original system of equations~(\ref{constraint}), (\ref{main1}), (\ref{main2}), and (\ref{main3}) analytically, we also solve the same system numerically.  During this numerical procedure we do not presume that $\chi_i / p$ is small, nor do we expand any quantities or equations in $\epsilon$ or any other small parameters.  Our main purpose is to reproduce the results of the preceding section and to determine the range of validity of our analytical asymptotics.

Typical values of $\chi_i / p$ in most experimental realizations are of the order of 0.001--0.01, and thus, only concentrations below approximately 0.1 are of practical interest.  Thus, we will mostly concentrate on this range of practical importance while describing the results despite the fact the numerical procedure can be (and have been) conducted for any ratio $\chi_i / p$.  It is to be noted that in the case of $\chi_i$ comparable to $p$ our model is not expected to produce any sensible results, as the entire separation of the drop into the two phases (the liquid phase and the deposit phase) is based on the assumption that the mobility of the solute is qualitatively different in the two regions.  When $\chi_i$ is comparable to $p$ the two phases are physically indistinguishable, while the model still assumes they are different.  Therefore, it should be of absolutely no surprise if any results for $\chi_i \approx p$ are unphysical or unrealistic.  When the assumptions of a theory are explicitly violated, its results are not to be taken seriously in that range of parameter values.  For completeness, we {\em do\/} provide results for $\chi_i = p$, however, in no way we attempt to claim these numerical results describe the real physical state of the system at this value of the parameter.  As we mentioned above, {\em only\/} the values of $\chi_i / p$ around 0.1 and below are presented as our final numerical results.  The case $\chi_i = p$ is provided only as an illustration of the general trend, without any attempt to draw any conclusions from this unphysical (within our model) value of the parameter.

We present our numerical results for the same quantities (and in the same order) as in our analytical results~(\ref{theta-result}), (\ref{fraction-result}), (\ref{height-result}), and (\ref{width-result}).  Since for arbitrary $\chi_i / p$ time $t_f$ is {\em not\/} exactly the total drying time, there is a question of where (at what time) to terminate the numerical curves.  By convention, we terminate all the curves (except $\chi_i / p = 1$) in all the graphs at value of $t / t_f$ when {\em all\/} the solute reaches the deposit phase.  In our model, it turns out that this time approximately coincides with the time the center-point of the liquid-air interface touches the substrate.  For all the initial concentrations (except $\chi_i = p$), the time the center-point touches the substrate was numerically found to be within 0.1\% of the time the very last solute particles reach the deposit phase.  Thus, within our model, the moment the center-point touches the substrate and the moment the last solute particles reach the deposit ring are about the same, and the curves are terminated at exactly this time.  Of course, in reality a small fraction of the solute stays in the liquid phase as long as the liquid phase exists, and so the moment the last solute particles reach the deposit phase must be {\em after\/} the moment the center-point touches the substrate; however, the amount of the solute remaining in the liquid phase at touchdown is insignificant, and thus practically all the deposit has already formed by that time (99.9\% of all the solute is already in the ring).

\begin{figure}
\begin{center}
\includegraphics{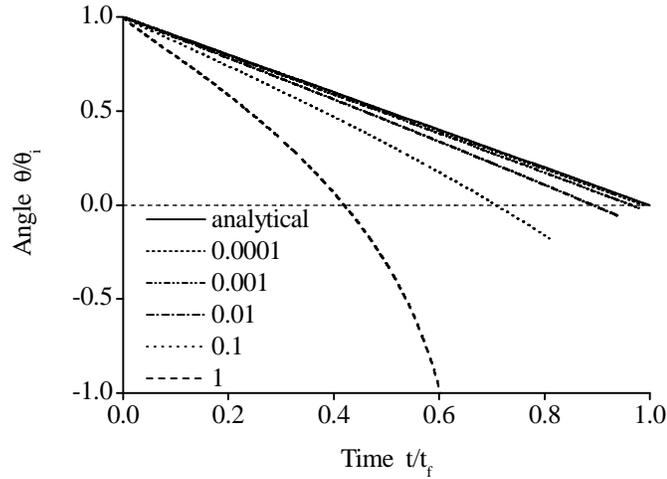}
\caption{Numerical results: Dependence of angle $\theta$ between the liquid-air interface and the substrate at the phase boundary on time $t$.  Different curves correspond to different initial concentrations of the solute; values of parameter $\chi_i/p$ are shown at each curve.  The analytical result in limit $\chi_i/p \to 0$ is also provided.}
\label{angleeps}
\end{center}
\end{figure}

Numerical results for angle $\theta$ as a function of time are shown in Fig.~\ref{angleeps}.  All curves (except $\chi_i/p = 1$) behave almost linearly (as expected), however, the slope increases with the concentration: formation of the ring in the drops with more solute finishes faster [in the relative scale of $t_f$ of Eq.~(\ref{t-f})].  The end of each curve demonstrates the value of the angle $\theta_t$ at the moment the liquid-air interface touches the substrate.  (The analytical expression for this angle is $\theta_t = - 2 H / R$ for a thin drop.)  The absolute value of this angle increases with concentration, which is quite natural since the height of the ring grows as a square root of the concentration while the radius of the liquid phase does not change substantially for small concentrations.  Clearly, the numerical results converge to the analytical curve when $\chi_i / p \to 0$.

\begin{figure}
\begin{center}
\includegraphics{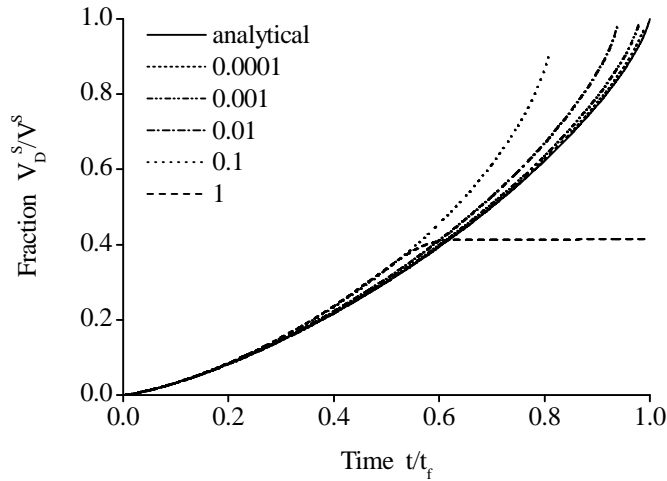}
\caption{Numerical results: Dependence of the volume fraction of the solute in the deposit phase $V_D^S/V^S$ on time $t$.  Different curves correspond to different initial concentrations of the solute; values of parameter $\chi_i/p$ are shown at each curve.  The analytical result in limit $\chi_i/p \to 0$ is also provided.}
\label{fractioneps}
\end{center}
\end{figure}

Growth of the volume fraction of the solute in the deposit phase with time is shown in Fig.~\ref{fractioneps} for various solute concentrations.  This graph reconfirms the observation of the preceding paragraph that the solute transfer happens faster (in units of the time scale $t_f$) for denser colloidal suspensions.  All curves are terminated when volume fraction $V_D^S/V^S$ becomes equal to 1.  The apparent termination of the curve for $\chi_i / p = 0.1$ earlier than that is an artifact of the plotting software, which discarded the very last point of this data set when creating a plot.  Presumably, this plot and the preceding one should hold true independently of the geometrical details of the solute accumulation in the ring (which cannot be expected from the following plots for the ring height and width).

\begin{figure}
\begin{center}
\includegraphics{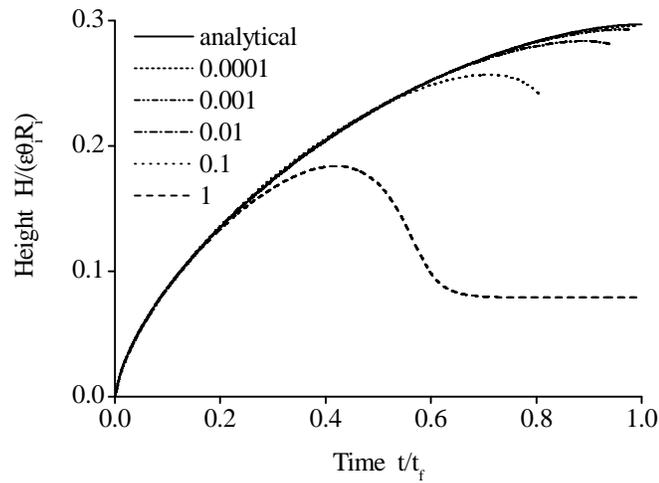}
\caption{Numerical results: Dependence of the height of the phase boundary $H$ on time $t$.  Different curves correspond to different initial concentrations of the solute; values of parameter $\chi_i/p$ are shown at each curve.  The analytical result in limit $\chi_i/p \to 0$ is also provided.}
\label{heighteps}
\end{center}
\end{figure}

\begin{figure}
\begin{center}
\includegraphics{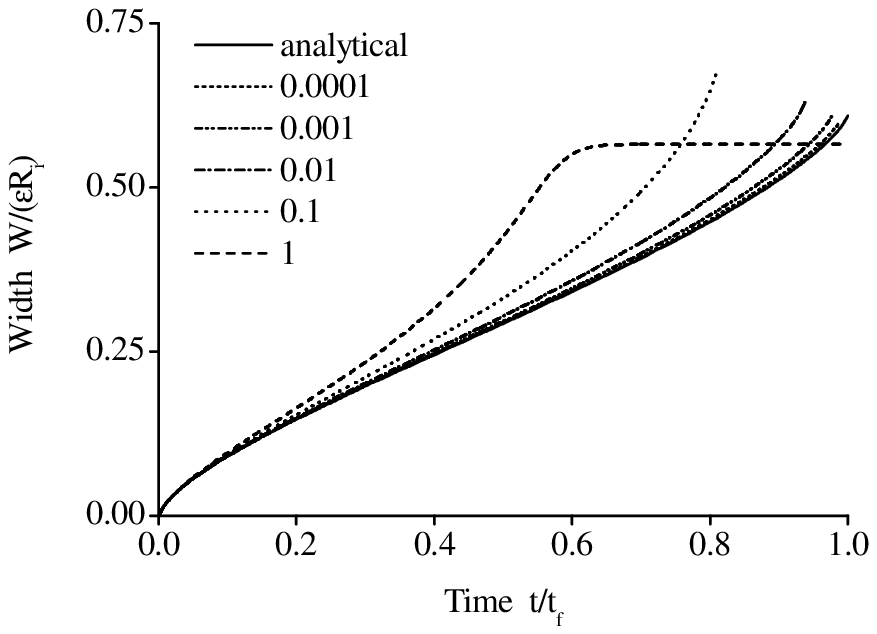}
\caption{Numerical results: Dependence of the width of the deposit ring $W$ on time $t$.  Different curves correspond to different initial concentrations of the solute; values of parameter $\chi_i/p$ are shown at each curve.  The analytical result in limit $\chi_i/p \to 0$ is also provided.}
\label{widtheps}
\end{center}
\end{figure}

\begin{figure}
\begin{center}
\includegraphics{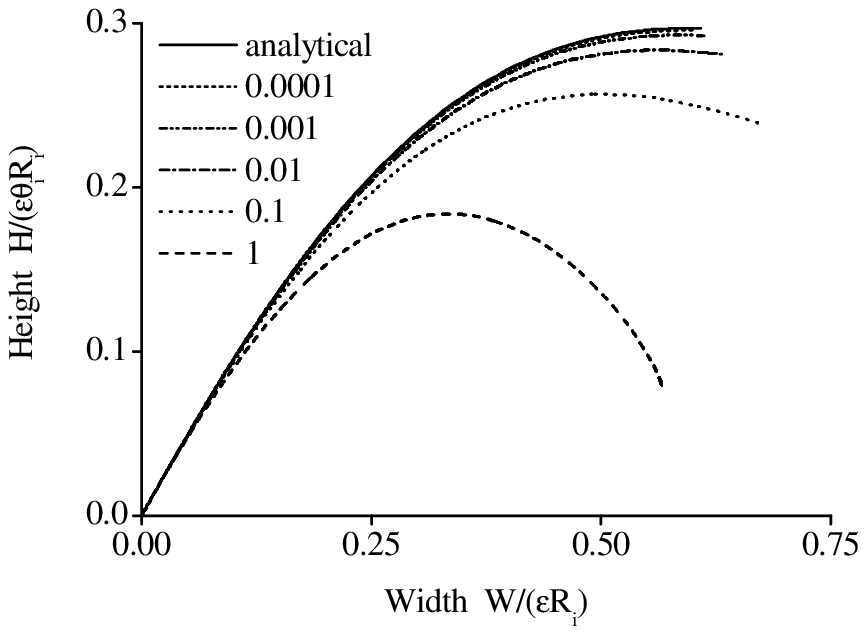}
\caption{Numerical results: Ring profile.  Dependence of the height of the phase boundary $H$ on the width of the deposit ring $W$.  Different curves correspond to different initial concentrations of the solute; values of parameter $\chi_i/p$ are shown at each curve.  The analytical result in limit $\chi_i/p \to 0$ is also provided.}
\label{profileeps}
\end{center}
\end{figure}

The next two graphs represent the numerical results for the geometrical characteristics of the deposit ring as functions of time: the height is shown in Fig.~\ref{heighteps}, while the width is in Fig.~\ref{widtheps}.  The ring profile, {\em i.e.}\ the dependence of height on width (obtained by elimination of time from the two results), is also provided in Fig.~\ref{profileeps}.  As the graphs depict, the ring becomes wider and lower at higher initial concentrations of the solute.  Since the volume of the ring is roughly proportional to the product of the height and the width, the decrease in height must be of the same percentage magnitude as the increase in width.  This can be qualitatively observed in the graphs.

\begin{figure}
\begin{center}
\includegraphics{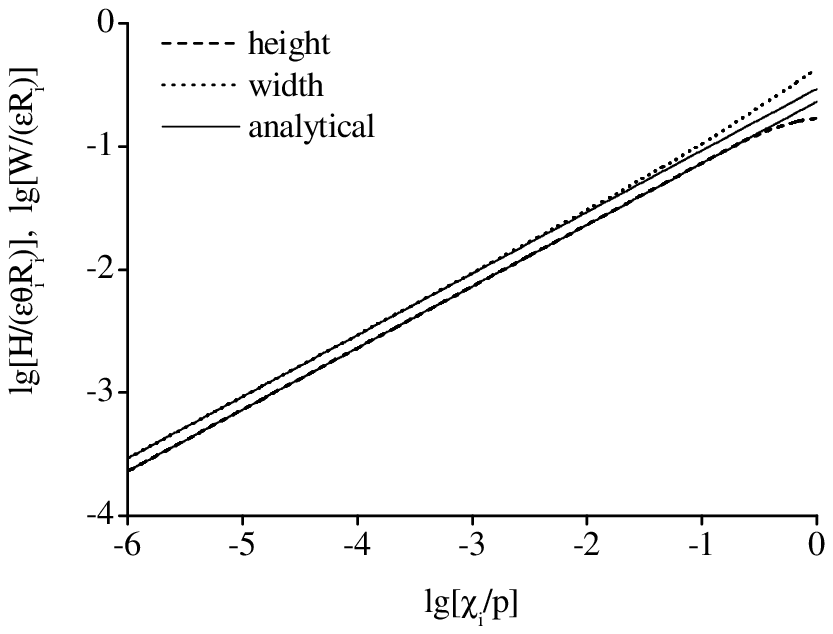}
\caption{Numerical results: Log-log plot of the dependence of the height of the phase boundary $H$ and the width of the deposit ring $W$ on the initial volume fraction of the solute.  ``Lg'' stands for the logarithm of base 10, and thus the numbers along the horizontal axis represent the order of magnitude of the ratio $\chi_i / p$.  The main-order analytical results $H \propto \sqrt{\chi_i/p}$ and $W \propto \sqrt{\chi_i/p}$ are also provided for comparison.}
\label{conceps}
\end{center}
\end{figure}

As a final piece of the numerical results, we create a double-logarithmic plot of the dependence of the height and the width on the initial concentration of the solute (Fig.~\ref{conceps}).  The predicted square-root dependence on the initial concentration is seen to hold true for volume fractions up to approximately $10^{-1/2} p$ for the height and up to approximately $10^{-3/2} p$ for the width.  Deviations for higher volume fractions are due to the increasing role of the correctional terms in $\epsilon$ compared to the main-order terms in expansions~(\ref{height-expansion}) and (\ref{width-expansion}).  In this graph, as in all the results of this section, it is clear that our main-order analytical results provide an adequate description of all the functional dependencies in the range of the initial concentrations of experimental importance (0.001--0.01).

In general, the numerical results of this section complement and reinforce the analytical results of the preceding section, and the good agreement between the results of the two independent solutions provides a crosscheck of both methods.

\section{Discussion}

The specific graphs of Figs.~\ref{angleeps}--\ref{profileeps}, both analytical and numerical, represent the original contribution of this chapter and may be determined experimentally giving validation to the model and our theory.  Measurements of the profiles in Fig.~\ref{profileeps} should be particularly easy to conduct (as there is no time dependence involved) and may confirm or refute the predicted robustness and universality of the deposition profiles.

One may notice that the curves in Fig.~\ref{profileeps} end at some positive (non-zero) height.  This indicates the solute is exhausted before the profile curves had a chance to return to the substrate, and the final shape of the deposit ring must have a vertical wall at its inner end.  We believe this is an artifact of our model that is inherently two-dimensional when flows inside the drop are concerned.  Thus, we used the depth-averaged velocity~(\ref{defv}) throughout this work, and we also explicitly assumed the phase boundary is vertical and the particles get stacked uniformly at all heights ({\em i.e.}\ the vertical distribution of the solute was assumed homogeneous).  This is equivalent to assuming that vertical mixing is complete.  This assumption is quite important, and the results are expected to get modified if the true velocity profile~(\ref{velocity-profile}) (or any other three-dimensional distribution) is used instead of the depth-averaged velocity.  We expect that if a three-dimensional model were built and the dependence on $z$ were taken into account for all quantities then the discontinuous wall of the phase boundary would get smoothened and the height would continuously return to zero.  A question remains whether such a model would be solvable analytically.

Our model relies on the assumption that the mobility of the solute is different in the so-called liquid and deposit phases.  Essentially, we assume that the mobility is 0 in the deposit phase and 1 in the liquid phase.  This assumption, while artificial in its nature, seems relatively reasonable when applied to this system.  Indeed, in the physical situations near the close packing, the loss of mobility typically occurs over a quite narrow range of the concentration values, and hence our assumption should work satisfactorily when the difference between $\chi_i$ and $p$ is in the orders of magnitude.  The higher the initial concentration is and the closer the two values are, the worse this assumption holds true and the more artificial the difference between the two phases is.  We observe this in our numerical data: the numerical results for $\chi_i = p$ are quite unphysical, while the low-concentration results seem to be consistent and coherent with expectations.  Thus, the validity of any model based on this separation of the mobility scales decreases for higher initial concentrations of the solute.

The model assumes that the free-surface slope between the liquid and the deposit phases is continuous.  In fact, assumption~(\ref{constraint}) expressing this continuity is one of the four basic equations of this chapter.  This assumption seems quite natural as well.  Indeed, if the liquid is present on both sides of the phase boundary, the change in the slope of its free surface would cost extra energy from the extra curvature at the phase boundary, since the liquid-air interface possesses effective elasticity.  Presence of this extra energy (or the extra pressure) at the location of the phase boundary is not justified by any physical reasons under the conditions of this chapter where processes are slow and the surface is in equilibrium.  In equilibrium the surface shape must have constant curvature past the phase boundary since the entire separation into the two phases is quite artificial as discussed above.  Presence of the particles below the liquid-air interface does not influence the surface tension, and thus the liquid surface (and its slope) should be continuous at the phase boundary.  On the other hand, if the density of the particles matches the density of the liquid (which was the case in the experiments), nothing prevents the particles from filling up the entire space between the substrate and the liquid-air interface, thus providing the growth of the upper edge of the deposit phase {\em along\/} the liquid-air interface.  This is particularly true for the thin drops discussed in this chapter ($\theta_i \ll 1$ was assumed throughout) where the vertical mixing is intensive, where the free surface is nearly horizontal, and where the problem is essentially two-dimensional.  However, the equality of the slopes on both sides of the phase boundary does {\em not\/} seem inevitable, and one may think of the situations when it does get violated.  One example might be the late drying times when the deposit growth is very fast (see Fig.~\ref{fractioneps}) and hence the deposition may occur in some non-regular manner inconsistent with the slow-process description of this chapter.  Other examples may be related to the gravity (slightly unequal densities of the particles and the liquid) or the convection.  In any case, this assumption can possibly be checked experimentally, and if condition~(\ref{constraint}) is found violated, an equivalent constraint dependent on the details of the deposit-growth mechanism must be constructed in place of Eq.~(\ref{constraint}) in order to relate functions $\theta(t)$, $R(t)$, and $H(t)$.

Another inherent assumption of our model is related to the evaporation rate $J(r)$.  As we discussed at the beginning of this chapter, we assumed that presence of the solute inside the drop does not affect the evaporation from its surface.  This is generally true when the evaporation is not too fast and the deposit phase is not too thick and not too concentrated.  When these conditions are not obeyed, presence of a thick or concentrated layer of the solute on the way of the liquid moving from the phase boundary to the contact-line divergence of the evaporation rate may create a strong viscous force.  This viscous force would prevent the necessary amount of the fluid from being supplied to the intensive-evaporation region near the contact line.  Generally, we assumed throughout this work that the viscous stresses are not important, and showed in Chapter~2 and the Appendix that this assumption is valid whenever $v \ll \sigma / 3 \eta$.  In the deposit phase, the velocity is large due to the proximity to the contact-line divergence of the evaporation rate, and the effective viscosity is large due to the high concentration of the solute.  Thus, this condition may get violated and the viscosity may become important in the deposit phase, slowing down the supply of the liquid and ultimately making the deposit dry.  Obviously, this affects the evaporation rate, and the functional form of the evaporation profile changes.  Simple assumption that the evaporation rate stays of the same functional form, but with the divergence at the phase boundary (at $R$) instead of the contact line (at $R_i$), was shown above not to affect our main-order results.  Thus, our results appear to be relatively insensitive to the exact location of this divergence within the (narrow) deposit phase.  (In reality the evaporation edge would be somewhere between the contact line and the phase boundary, {\em i.e.}\ the real situation is intermediate between the two considered.)  However, the deposit could modify the evaporation rate $J(r)$ in other ways.  When there is a dry deposit ring just outside the liquid phase, the entire functional form of $J$ may change, and the Laplace equation for an equivalent electrostatic problem must be solved anew with the additional boundary conditions responsible for the presence of the dry solute rim and the modified evaporation at the edge.  As we already saw in all the preceding chapters, this is the most complicated part of the problem, and the mathematics can become prohibitively complex.  Thus, finding the exact form of $J$ may be a formidable task.  One way around is in creating such evaporating conditions that the functional profile is simpler, for instance, $J$ is just a constant.  This would be more difficult to control experimentally, but would be much easier to treat analytically.  The unavailability of the exact analytical form for $J$ seems to be the biggest open question in this class of problems (the same difficulty was also encountered for the pointed-drop solution of Chapter~4).

The equilibrium surface shape of the liquid phase is a spherical cap~(\ref{h-finite}).  This is a rigorous result proved in the preceding chapters and valid during most of the drying process.  However, when $h(0,t)$ becomes negative and exceeds $H(t)$ in absolute value ({\em i.e.}\ when the center-point touches the substrate), the surface shape is no longer spherical.  Moreover, a new element of the contact line is introduced in the center of the drop in addition to the original contact line at the perimeter, and the entire evaporation profile gets modified in addition to the modified surface shape, thus influencing all the other quantities.  Our treatment does not account for the small fraction of the drying process occurring after this touchdown (which is a change in topology of the surface shape, and thus requires a separate treatment after it happened).  First of all, the amount of liquid remaining in the drop at this moment is of the order of $\epsilon$ compared to the original volume and hence would not modify our main-order analytical results.  Second, as our {\em numerical\/} calculations show, at touchdown practically all the solute is already in the deposit phase, and the remaining amount of the solute in the liquid phase is insignificant.  Thus, within our model, the remainder of the drying process cannot modify the deposit ring substantially, and hence this neglect of the late-time regime seems well justified.  Experimentally, the inner part of the deposit ring is different from our prediction (which is a vertical wall) and appears to have a spread shelf.  Presence of this tail in the deposit distribution can be caused by several features absent in our model.  Its inherent two-dimensionality may be one of these shortcomings (as discussed above); the account for the dynamical processes occurring after the deposit phase has already been formed ({\em e.g.}\ avalanches of the inner wall) may be another missing feature.  Absence of the account for the late-time regime after the center-point touches the substrate may be among these reasons influencing the final distribution of the deposit as well.  A more detailed account for the effects of this late-time regime might be required in the future.

\chapter{Conclusions}

The major results presented in this work are related to two problems.  One of them is the problem of the solute transfer and the deposit growth in angular evaporating drops, and the other is the problem of determining the geometrical characteristics of the deposition patterns in circular evaporating drops.

For the angular drops, we provide the full solution to the problem of the deposit accumulation at the contact line.  We determine all the necessary physical quantities (the surface shape, the evaporation rate, the flow field, and the mass of the deposit) and fully describe the deposition process near the vertex of the angle.  Not surprisingly, we find that the angular drops are rich in singularities that govern all these physical quantities and affect the deposition.  All the quantities scale as power laws of the distance to the vertex and the elapsed drying time with exponents dependent only on the opening angle of the drop.  These exponents are universal and do not depend on any free or fitting parameters, they depend only on the drop geometry.  This dependence on the geometry is useful for creating controlled deposition patterns and is a step towards complete understanding of the evaporative deposition for arbitrary contact-line geometries.

For the problem on the geometrical characteristics of the deposition patterns, we provide a model that accounts for the finite dimensions of the deposit on the basis of the assumption that the solute particles occupy finite volume and hence these dimensions are of the steric origin.  Within this model, we find the analytical solution for small initial concentrations of the solute and the numerical solution for arbitrary initial concentrations of the solute.  We demonstrate the agreement between our results and the experimental data and show that the observed dependence of the deposit dimensions on the experimental parameters (the initial concentration of the solute, the initial geometry of the drop, and the time elapsed from the beginning of the drying process) can be attributed mainly to the finite dimensions of the solute particles.  These results are also universal and important for understanding the deposition process and controlling the pattern formation.

The entire work provides all our results and derivations pertaining to the evaporative deposition and obtained to date in a single volume, which may serve as a reference guide to anyone interested in the theory of the evaporative deposition.  We hope this reference volume will simplify future theoretical and experimental efforts towards the complete understanding of these ubiquitous phenomena.

\appendix

\chapter{Dominance of the capillary forces over viscous stress, and legitimacy of employment of the equilibrium surface shape}

The purpose of this section is to demonstrate that for sufficiently slow flows one can employ the equilibrium surface shape for finding the pressure and the velocity fields instead of having to solve for all the dynamical variables simultaneously.  We will also quantify how slow ``sufficiently slow flows'' are.

We start from the equation of the mechanical equilibrium of the interface~(\ref{mechequil}), where we approximate the doubled mean curvature $2 K$ with $\nabla^2 h$.  This approximation holds true because the liquid-air interface is nearly horizontal for thin drops and small contact angles (this can be proved rigorously for both geometries of interest, see the corresponding sections in the main text), and the other terms of the functional $K[h]$ are unimportant.  Substitution of
\begin{equation}
p = - \sigma \nabla^2 h + p_{atm}
\label{p-h}\end{equation}
into the Darcy's law~(\ref{darcy}) yields
\begin{equation}
\bfv = v^* h^2 \bfnabla (\nabla^2 h),
\label{v-h}\end{equation}
where $v^* = \sigma/3\eta$.  Upon further substitution into the conservation of mass~(\ref{consmass}), we obtain
\begin{equation}
\bfnabla\cdot\left[v^* h^3 \bfnabla (\nabla^2 h)\right] + \frac{J}{\rho} + \partial_t h = 0,
\label{consmas1}\end{equation}
which, together with Eq.~(\ref{v-h}), constitutes the full system of equations for finding $h(r,\phi,t)$ and $\bfv(r,\phi,t)$.

Now, for water under normal conditions, $\eta = 1\mbox{ mPa}\cdot\mbox{s}$ and $\sigma = 72\mbox{ mN}/\mbox{m}$.  Hence, the velocity scale $v^*$ is of the order of
\begin{equation}
v^* = \frac\sigma{3\eta} \approx 24\mbox{ m}/\mbox{s}.
\end{equation}
Obviously, this is a huge value compared to the characteristic velocities encountered in usual drying process.  Therefore, one can develop a systematic series expansion in small parameter $\epsilon = \tilde v / v^*$ (where $\tilde v$ is some characteristic value of velocity, say, 1 or 10~$\mu$m/s):
\begin{equation}
h = h_0 + \epsilon h_1 + \cdots + \epsilon^n h_n + \cdots,
\end{equation}
\begin{equation}
\bfv = \bfv_0 + \epsilon \bfv_1 + \cdots + \epsilon^n \bfv_n + \cdots,
\end{equation}
and keep only the $h_0$ and $\bfv_0$ terms at the end in order to describe the process up to the main order in $\epsilon = \tilde v / v^*$.  A similar expansion can also be constructed for pressure:
\begin{equation}
p = p_0 + \epsilon p_1 + \cdots + \epsilon^n p_n + \cdots,
\end{equation}
where $p_i$ are related to $h_i$ by equation~(\ref{p-h}):
\begin{equation}
p_0 = - \sigma \nabla^2 h_0 + p_{atm},\qquad\qquad p_1 = - \sigma \nabla^2 h_1,\qquad\qquad\mbox{etc.}
\end{equation}
Physically, condition $\tilde v \ll v^*$ is equivalent to the statement that the viscous stress is negligible and that the capillary forces dominate.  Let us understand what $h_0$ and $\bfv_0$ physically correspond to.

Plugging the expansions for $h$ and $\bfv$ into the system~(\ref{v-h})--(\ref{consmas1}), one obtains a set of terms for each power of $\epsilon$, starting from $\epsilon^{-1}$ and up.  Equating terms of the main order in $\epsilon$ yields the following two equations
\begin{equation}
h_0^2 \bfnabla (\nabla^2 h_0) = \bfzero\qquad\qquad\mbox{and}\qquad\qquad\bfnabla\cdot\left[h_0^3 \bfnabla (\nabla^2 h_0)\right] = 0,
\end{equation}
which both can be satisfied if and only if $\nabla^2 h_0$ is a function of time only.  Writing it as
\begin{equation}
\nabla^2 h_0 = - \frac{p_0 - p_{atm}}\sigma = - \frac 1{R_0(t)},
\end{equation}
we immediately identify this equation with the statement of spatial constancy of the mean curvature of the interface, which describes the {\em equilibrium\/} surface shape at any given moment of time $t$ ({\em i.e.}\ we obtained equation~(\ref{laplace}) with the desired properties of $p_0$).  Thus, $h_0$ is indeed the equilibrium surface shape, and so is $h$ (up to the corrections of the order of $\tilde v / v^*$).

Repeating the same procedure for the terms of the next order in $\epsilon$, we arrive at another two equations:
\begin{equation}
\bfv_0 = \tilde v h_0^2 \bfnabla (\nabla^2 h_1),
\end{equation}
\begin{equation}
\tilde v \bfnabla\cdot\left[h_0^3 \bfnabla (\nabla^2 h_1)\right] + \frac{J}{\rho} + \partial_t h_0 = 0,
\end{equation}
which can be seen to be equivalent to the set of equations~(\ref{psi})--(\ref{vpsi}) upon identification $\psi = \tilde v \nabla^2 h_1 = - \epsilon p_1 / 3\eta$.  Knowing the equilibrium surface shape $h_0$, one can solve the second equation above with respect to the reduced pressure $\psi$, and then obtain velocity $\bfv_0$ by differentiating the result according to the first equation.  Thus, up to the corrections of the order of $\tilde v / v^*$, one can first find the equilibrium surface shape $h(r,\phi)$ at any given moment of time, and then determine the pressure and the flow fields for this fixed functional form of $h$, as was asserted in the main text.

\newpage

\addcontentsline{toc}{chapter}{References}

\begin{singlespace}

\end{singlespace}


\begin{thebibliography}{99}

\bibitem{pre1} N.D.~Denkov, O.D.~Velev, P.A.~Kralchevsky, I.B.~Ivanov, H.~Yoshimura, K.~Nagayama, {\em Langmuir\/} {\bf 8}, 3183 (1992).

\bibitem{pre2} A.S.~Dimitrov, C.D.~Dushkin, H.~Yoshimura, K~Nagayama, {\em Langmuir\/} {\bf 10}, 432 (1994).

\bibitem{pre3} T.~Ondarcuhu, C.~Joachim, {\em Europhys.\ Lett.} {\bf 42}, 215 (1998).

\bibitem{pre4} J.~Boneberg, F.~Burmeister, C.~Shafle, P.~Leiderer, D.~Reim, A.~Fery, S.~Herminghaus, {\em Langmuir\/} {\bf 13}, 7080 (1997).

\bibitem{jpcb2} R.G.~Larson, T.T.~Petkins, D.E.~Smith, S.~Chu, {\em Phys.\ Rev.\ E\/} {\bf 55}, 1794 (1997).

\bibitem{jpcb1} J.P.~Jing, J.~Reed, J.~Huang, X.~Hu, V.~Clarke, J.~Edington, D.~Housman, T.S.~Anantharaman, E.J.~Huff, B.~Mishra, B.~Porter, A.~Shenkeer, E.~Wolfson, C.~Hiort, R.~Kantor, C.~Aston, D.C.~Schwartz, {\em Proc.\ Natl.\ Acad.\ Sci.\ U.S.A.} {\bf 95}, 8046 (1998).

\bibitem{pre5} A.B.~El~Bediwi, W.J.~Kulnis, Y.~Luo, D.~Woodland, W.N.~Unertl, {\em Mater.\ Res.\ Soc.\ Symp.\ Proc.} {\bf 372}, 277 (1995).

\bibitem{pre6} F.~Parisse, C.~Allain, {\em J.\ Phys. II\/} {\bf 6}, 1111 (1996).

\bibitem{pre7} F.~Parisse, C.~Allain, {\em Langmuir\/} {\bf 13}, 3598 (1996).

\bibitem{pre8} E.~Adachi, A.S.~Dimitro, K.~Nagayama, in {\em Film Formation in Waterborne Coatings,} edited by T.~Provder, M.A.~Winnik, M.W.~Urban, p. 419 (American Chemical Society, Washington DC, 1996).

\bibitem{pre9} E.~Adachi, A.S.~Dimitro, K.~Nagayama, {\em Langmuir\/} {\bf 11}, 1057 (1995).

\bibitem{pre0} J.~Conway, H.~Korns, M.R.~Fisch, {\em Langmuir\/} {\bf 13}, 426 (1997).

\bibitem{jpcb3} K.S.~Birdi, D.T.~Vu, A.~Winter, {\em J.\ Phys.\ Chem.} {\bf 93}, 3702 (1989).

\bibitem{jpcb4} K.S.~Birdi, D.T.~Vu, {\em J.\ Adhes.\ Sci.\ Technol.} {\bf 7}, 485 (1993).

\bibitem{jpcb5} M.E.R.~Shanahan, C.~Bourges, {\em Int.\ J.\ Adhes.} {\bf 14}, 201 (1994).

\bibitem{jpcb6} C.~Bourges, M.E.R.~Shanahan, {\em Langmuir\/} {\bf 11}, 2820 (1995).

\bibitem{jpcb7} S.M.~Rowan, G.~McHale, M.I.~Newton, {\em J.\ Phys.\ Chem.\ B\/} {\bf 99}, 13268 (1995).

\bibitem{jpcb8} S.M.~Rowan, G.~McHale, M.I.~Newton, M.~Toorneman, {\em J.\ Phys.\ Chem.\ B\/} {\bf 101}, 1265 (1997).

\bibitem{deegan1} R.D.~Deegan, O.~Bakajin, T.F.~Dupont, G.~Huber, S.R.~Nagel, T.A.~Witten, {\em Phys.\ Rev.\ E\/} {\bf 62}, 756 (2000).

\bibitem{deegan2} R.D.~Deegan, O.~Bakajin, T.F.~Dupont, G.~Huber, S.R.~Nagel, T.A.~Witten, {\em Nature\/} {\bf 389}, 827 (1997).

\bibitem{deegan3} R.D.~Deegan, {\em Phys.\ Rev.\ E\/} {\bf 61}, 475 (2000).

\bibitem{deegan4} R.D.~Deegan, {\em Ph.D.\ thesis\/} (University of Chicago, Dept.\ of Physics, 1998).

\bibitem{lebedev} N.N.~Lebedev, {\em Special Functions and Their Applications,} Revised English ed., Chapters 7 and 8 (Prentice-Hall, Englewood Cliffs, 1965).

\bibitem{jpcb0} R.G.~Picknett, R.~Bexon, {\em J.\ Colloid Interface Sci.} {\bf 61}, 336 (1977).

\bibitem{hu} H.~Hu, R.G.~Larson, {\em J.\ Phys.\ Chem.\ B\/} {\bf 106}, 1334 (2002).

\bibitem{popov1} Y.O.~Popov, T.A.~Witten, {\em Eur.\ Phys.\ J.\ E\/} {\bf 6}, 211 (2001).

\bibitem{popov2} Y.O.~Popov, T.A.~Witten, {\em Phys.\ Rev.\ E\/} {\bf 68}, 036306 (2003).

\bibitem{greenspan} H.P.~Greenspan, {\em J.\ Fluid Mech.} {\bf 84}, 125 (1978).

\bibitem{cameron} A.~Cameron, {\em Principles of Lubrication\/} (Longmans and Green, London, 1966).

\bibitem{brenner} M.P.~Brenner, {\em Ph.D.\ thesis\/} (University of Chicago, Dept.\ of Physics, 1994).

\bibitem{bensimon} D.~Bensimon {\em et al.}, {\em Rev.\ Mod.\ Phys.} {\bf 58}, 977 (1986).

\bibitem{prudnikov} A.P.~Prudnikov, Y.A.~Brychkov, O.I.~Marichev, {\em Integrals and Series,} Vol. 3 (Gordon and Breach, London, 1986).

\bibitem{davies} J.T.~Davies, E.K.~Rideal, {\em Interfacial Phenomenon\/} (Academic Press, New York, 1963).

\bibitem{jackson} J.D.~Jackson, {\em Classical Electrodynamics,} 2nd ed., Chapters 2 and 3 (Wiley, New York, 1975).

\bibitem{kraus} L.~Kraus, L.M.~Levine, {\em Comm.\ Pure Appl.\ Math.} {\bf 14}, 49 (1961).

\bibitem{blume1} S.~Blume, M.~Kirchner, {\em Optik\/} {\bf 29}, 185 (1969).

\bibitem{blume2} S.~Blume, G.~Kahl, {\em Optik\/} {\bf 70}, 170 (1985).

\bibitem{desmedt} R.~De~Smedt, J.G.~Van~Bladel, {\em IEE Proc.} {\bf 134}, 694 (1987).

\bibitem{dupont} T.~Dupont, private communication.

\bibitem{popov3} Y.O.~Popov, {\em J.\ Colloid Interface Sci.} {\bf 252}, 320 (2002).

\end{thebibliography}
\end{document}